%% file: paper.tex
\pdfoutput=1

\documentclass[ppnp,12pt]{elsarticle}

\usepackage{amssymb}

\usepackage{aas_macros}
\usepackage{sidecap}
\usepackage{dsfont}
\usepackage{url}

\newcommand{\dd}{{\rm d}}
\newcommand{\la}{\langle }
\newcommand{\ra}{\rangle}

\newcommand{\lr}{\left( }
\newcommand{\rr}{\right) }
\newcommand{\lee}{\left[ }
\newcommand{\re}{\right] }
\newcommand{\lgg}{\left\{ }
\newcommand{\rg}{\right\} }
\newcommand{\beq}{\begin{equation}}
\newcommand{\eeq}{\end{equation}}
\newcommand{\bea}{\begin{eqnarray}}
\newcommand{\eea}{\end{eqnarray}}

\journal{Progress in Particle and Nuclear Physics}

\begin{document}

\begin{frontmatter}



\title{Indirect and direct search for dark matter}


\author[a1]{M.\ Klasen}
\address[a1]{Institut f\"ur Theoretische Physik, Westf\"alische
 Wilhelms-Universit\"at M\"unster, Wilhelm-Klemm-Stra\ss{}e 9, D-48149
 M\"unster, Germany}
\author[a2,a3]{M.\ Pohl}
\address[a2]{DESY, D-15738 Zeuthen, Germany}
\address[a3]{Institut f\"ur Physik und Astronomie, Universit\"at Potsdam,
 D-14476 Potsdam, Germany}
\author[a4]{G.\ Sigl}
\address[a4]{II.\ Institut f\"ur Theoretische Physik, Universit\"at Hamburg,
 Luruper Chaussee 149, D-22761 Hamburg, Germany}

\begin{abstract}
The majority of the matter in the universe is still unidentified and under
investigation by both direct and indirect means. Many experiments searching
for the recoil of dark-matter particles off target nuclei in underground
laboratories have established increasingly strong constraints on the mass and
scattering cross sections of weakly interacting particles, and some have even
seen hints at a possible signal. Other experiments search for a possible mixing
of photons with light scalar or pseudo-scalar particles that could also
constitute dark matter. Furthermore, annihilation or decay of dark matter can
contribute to charged cosmic rays, photons at all energies, and neutrinos. Many
existing and future ground-based and satellite experiments are sensitive to such
signals. Finally, data from the Large Hadron Collider at CERN are scrutinized for
missing energy as a signature of new weakly interacting particles that may be
related to dark matter. In this review article we summarize
the status of the field with an emphasis on the complementarity between direct
detection in dedicated laboratory experiments, indirect detection in the cosmic
radiation, and searches at particle accelerators.

\end{abstract}


\end{frontmatter}


\newpage
\tableofcontents
\newpage


\input{section1.tex}
\input{section2.tex}
\input{section3.tex}
\input{section4.tex}
\input{section5.tex}

\input{section6.tex}

\section*{References}

\bibliographystyle{elsarticle-num} 

\end{document}

%% file: section1.tex
\section{Introduction}
\label{sec:1}

\subsection{Objective of this review article}

Only about 5\% of the mass-energy content of the universe is composed of ordinary baryonic matter, the bulk of which is diffuse gas rather than stars and galaxies. Nearly 70\% is carried by the so-called dark energy, that is mostly observed through its accelerating effect on the expansion of the universe as observed, e.g., by comparing the luminosity distance and the redshift of distant supernovae. The 2011 Nobel Prize in physics was awarded to Saul Perlmutter, Brian P. Schmidt, and Adam G. Riess for their groundbreaking work that led to the discovery of dark energy.

In this review we shall discuss the current understanding of the remaining 25\% of mass-energy in the universe, the so-called dark matter. Evidence for excess gravitational acceleration that cannot be explained by observable matter has been found on both small, galactic scales and large, cosmological scales. If Newton's law of gravity, or general relativity, is valid, then the universe must contain a constituent of unknown nature that betrays its presence only though gravitation. This dark matter may or may not be composed of particles. If so, the absence of a radiative signal or excess scattering of baryonic matter requires that those particles be uncharged and at most weakly interacting. Numerous candidate particles have been proposed over the years, and a hunt is on to detect them directly via elastic scattering in laboratory devices or indirectly through an astronomical decay or annihilation signal.

Beginning with a historical account, we present the current status of theoretical ideas and experimental constraints with an emphasis on complementarity. Other recent reviews include global accounts of dark matter \cite{Agashe:2014kda,Peter:2012rz,Bertone:2004pz} and more focussed publications, e.g., on structure formation \cite{DelPopolo:2008mr}, baryogenesis \cite{Racker:2014yfa}, or indirect searches \cite{He:2009ra,Cirelli:2012tf,Conrad:2014tla,Khlopov:2014nva}. Relevant topics for direct dark-matter searches are annual modulation \cite{Freese:2012xd} and elastic nuclear recoil \cite{Lewin:1995rx,Akimov:2001nr,Akimov:2011za}. Reviews of specific models are available for neutrinos \cite{RiemerSorensen:2013ih}, supersymmetry \cite{Jungman:1995df}, gauginos \cite{Cassel:2011zd,Mahmoudi:2014cya}, axinos \cite{Choi:2013lwa}, scalar condensates \cite{Magana:2012ph,Suarez:2013iw}, and asymmetric dark matter \cite{Petraki:2013wwa}, to only name a few.

\subsection{Observational evidence and alternative explanations}

In 1932, Jan Oort studied the motions of nearby stars and concluded that there must be more mass
in the local galactic plane than is seen as bright stars \cite{Oort:1932}. Although he interpreted
these observations in terms of dim stars, they were later taken to be the first indication of
the presence of dark matter.
Only a year later, Fritz Zwicky made the same claim, based on far better data. He assumed that galaxies in the Coma cluster would be virialized and thus found evidence of the total mass of the cluster being more than a hundred times that of the stars in the individual galaxies in it 
\cite{1937ApJ....86..217Z}. Today's value of the mass excess, based in particular on the now much lower value of the Hubble constant
of $H_0=67.8\pm0.9$ km/s/Mpc \cite{Planck:2015xua} instead of Zwicky's 558 in the same
units, is about 160 within 5 Mpc \cite{Fusco:1994}.

\begin{figure}[t]
 \includegraphics[width=\textwidth]{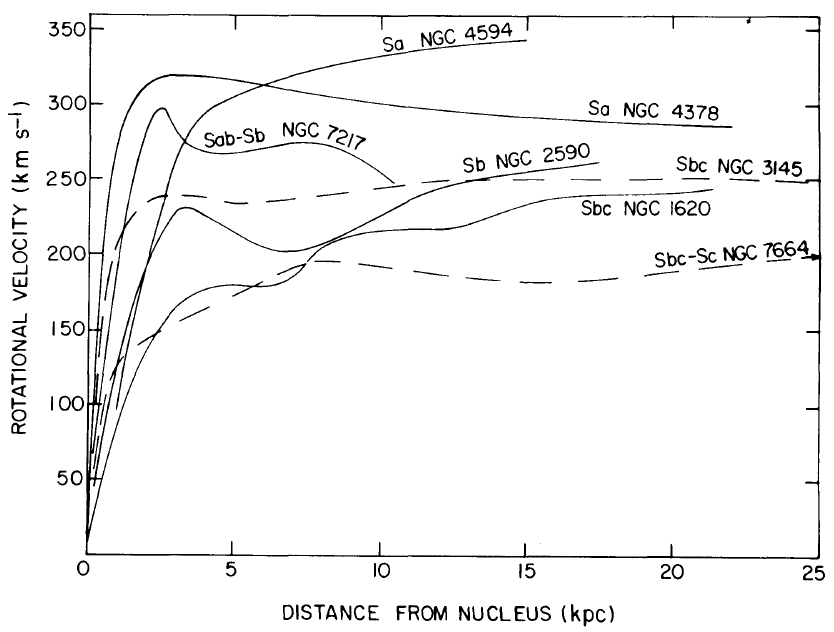}
 \caption{Rotational velocity curves of seven high-luminosity spiral galaxies of different Hubble type.
 Early-type galaxies have higher peak velocities than later types, but all curves remain constant out to
 large distances (taken from Rubin et al. \cite{Rubin:1978}).}
 \label{fig:rubin1978}
\end{figure}

In the 1960s, measurements of the rotation velocity of edge-on spiral galaxies became possible. The surprising finding was that most stars and the atomic hydrogen gas have the same orbit velocity irrespective of their distance from the center of the galaxy \cite{1970ApJ...159..379R,1978PhDT.......195B,1980ApJ...238..471R} and that this extends beyond the optical disk. By equating the centrifugal force $F=mv^2/r$ with the attractive gravitational force
$F=G_{\rm N}M(<r)m/r^2$, where $G_{\rm N}$ is Newton's gravitational constant and $M(<r)$ is the enclosed mass, one expects a
so-called ``Keplerian fall-off'' of the circular velocity, $v\propto 1/\sqrt{r}$, for tracers outside the central luminous disk. The constant circular velocities shown, e.g., in Fig.~\ref{fig:rubin1978} imply that the contained mass must increase with radius of the galaxy even in the outer regions, where few stars and little gas are found. Generally, the total mass is about five times that seen as baryonic matter.

Rubin's observations fit very well to early numerical simulations of galaxies composed
of 150 to 500 mass points \cite{Ostriker:1973}. They demonstrated that cold flat disks
such as the Milky Way could not survive due to a large-scale, bar-like instability beyond a ratio of
rotational kinetic energy over potential energy of $t=0.14\pm0.02$, unless one postulated
a spherical halo with halo-to-disk mass ratio between 1 and 2.5. To remain finite, the halo
mass can of course at some point no longer increase with $r$, and the density profile must fall off
faster than that of an isothermal sphere, $\rho\propto r^{-2}$. It is still unknown where this happens,
and the exact form of the dark-matter density profile is heavily debated. 

The mass distribution in galaxy clusters can also be determined with gravitational lensing. So-called strong lensing, which results in multiple images of a background object, is rare. Weak lensing is based on a statistical analysis of the distortions of the shapes of background objects \cite{2003ARA&A..41..645R}. The consistency of gravitational-lensing results with other types of measurements provides convincing evidence that galaxy clusters contain five times as much dark matter than baryonic matter, most of which is in the form of hot, dilute gas.

The most direct evidence for dark matter presented to date comes from a system known as the Bullet Cluster \cite{2007Natur.445..286M}. Two individual clusters appear to have collided some time ago. Most of the mass, determined using weak lensing, is located in the two individual clusters, whereas the intensity distribution of thermal X-ray emission indicates that the baryonic gas is concentrated in the collision region, as if it was slowed down when the two clusters collided. Hot gas interacts by Coulomb collisions, and so there cannot be an unimpeded penetration of the gas haloes of the two clusters. Stars and apparently also dark matter are collisionless and responsive only to collective gravitational forces. The dark matter components of the two clusters can thus pass through each other.

Also on cosmological scales one finds evidence for dark matter. It contributes not only to the total mass-energy content of the universe, but also impacts the growth of density fluctuations. 
Early in the cosmological evolution, before photons decoupled from baryonic matter, radiation provided the bulk of the pressure, but interacted only with ordinary baryonic matter. Density fluctuations in the baryonic matter arise from the interplay of gravity and pressure. Dark matter does not react to radiative pressure and thus modifies such acoustic oscillations. 

The earliest acoustic fluctuations are measured in the cosmological microwave background (CMB), whose temperature anisotropies ,
$\delta T/T\,(\theta,\phi)=\sum_{l,m}a_{lm}Y_{lm}(\theta,\phi)$, have been measured with unprecedented precision with the Wilkinson Microwave Anisotropy Probe
(WMAP) \cite{Nolta:2008ih}. The first peak in the angular power spectrum of temperature fluctuations is indicative of the amount of baryonic matter, whereas the other peaks carry a signature of the non-baryonic mass density. More specifically, 
the position of the first and the relative heights of the
second and third ``acoustic'' peaks, indicate the universe to be flat, i.e.\ to have a ratio $\Omega=\rho/\rho_c$ of its total density $\rho$ over the
``critical value'' $\rho_c=3H^2/(8\pi G)$ very close to unity, and to be dominated by cold dark matter (CDM) with a density ratio, $\Omega_{\rm c}$, about five times larger than that of ordinary baryonic
matter, $\Omega_{\rm b}$. Data of the subsequent Planck mission, when
fitted with a standard spatially-flat, six-parameter $\Lambda$CDM cosmological model, yield CDM and baryonic contents of
$\Omega_{\rm c}h^2=0.1186\pm0.0020$ and $\Omega_{\rm b}h^2=0.02226\pm0.00023$, respectively,
where $h=0.678\pm0.009$ denotes the present-day Hubble constant in units of 100 km/s/Mpc. The baryon density
inferred from CMB observations is in excellent agreement with the one obtained by comparing measured primordial
abundances of light elements with the predictions of big-bang nucleosynthesis which at 95\% confidence level is given by
$0.021\lesssim\Omega_{\rm b}h^2\lesssim0.025$~\cite{particle-data}.

After decoupling, baryonic density fluctuations can grow, and the uneven distribution of galaxies and clusters, known as baryonic acoustic oscillations (BAO), is another indicator of dark matter. The Sloan Digital Sky Survey (SDSS) performed a
survey of 205,443 galaxies at mean redshift of $z\simeq0.1$ (but also of
quasars as distant as $z\geq5$), covering an effective area of 2417 square degrees
\cite{Tegmark:2003uf}. The three-dimensional data permit studies of BAOs, help breaking degeneracies of the two-dimensional CMB, and also provide an independent measurement of the total matter content, $\Omega_m h=(\Omega_b+\Omega_c)h=0.213\pm0.023$,
in agreement with the CMB measurements (note the different powers of $h$).
Results from measurements of the CMB and BAO agree in the total matter density in the universe being about six times that in baryonic matter \cite{2014A&A...571A..16P}.

Type-Ia supernova surveys such as the Supernova Legacy Survey (SNLS) and that of SDSS
provide complementary constraints on the dark-energy content of the universe, $\Omega_\Lambda=\Lambda/(3H_0^2)$. When
interpreted within the standard flat $\Lambda$CDM cosmology ($1=\Omega_{\rm m}+\Omega_\Lambda$),
their combined data imply $\Omega_{\rm m}=0.295\pm0.034$, which perfectly agrees with the
CMB and BAO results cited above (note again the different powers of $h$) \cite{Betoule:2014frx}.

It should be noted that the standard $\Lambda$CDM cosmology, involving cold dark matter (CDM) and dark energy with similar impact as a cosmological constant $\Lambda$, is likewise not without problems. One issue is the missing-satellite problem, so named because far fewer dwarf galaxies are observed around Milky Way-like galaxies~\cite{2010AdAst2010E...8K} than are predicted by dark-matter N-body simulations. A possible solution could be that only a few dark-matter subhaloes experience star formation and evolve to dwarf galaxies on account of, e.g., expulsion of baryonic gas by ram pressure~\cite{Arraki:2012bu}. N-body simulations, however, point in the opposite direction: most of the massive subhaloes in their data are so dense and compact, that they are \emph{too big to fail} to form stars and thus become observable. The velocity dispersion should be significantly larger than that observed in real dwarf galaxies. 

Variations of the laws of gravity have been proposed as a means to explain all observational findings without dark matter. Developed in 1983, modified Newtonian dynamics (MOND) posits that below a certain level of gravitational acceleration, $a_0 \simeq 2\cdot 10^{-10}\ \mathrm{m}\,\mathrm{s}^{-2}$, the gravitational force is no longer proportional to the acceleration, but the acceleration squared \cite{1983ApJ...270..365M}. Originally a purely phenomenological nonrelativistic model designed to reproduce flat rotation curves, a relativistic formulation of MOND, called TeVeS, is now available \cite{2004PhRvD..70h3509B}, that also predicts gravitational lensing and furthermore evades the problem of
superluminal scalar-wave propagation of previous attempts. TeVeS is built on the notion that
the dynamics of matter fields is governed by a metric tensor $\tilde{g}_{\mu\nu}=e^{-2\phi}
g_{\mu\nu}-2u_\mu u_\nu \sinh(2\phi )$ that is different from the metric $g_{\mu\nu}$
appearing in the Einstein-Hilbert action. In particular, a timelike vector field $u_\mu$ satisfying
$g^{\mu\nu} u_\mu u_\nu=-1$ and a scalar field $\phi$ were introduced to exponentially stretch
and shrink the Einstein metric in the spacetime directions orthogonal and parallel to $u^\mu$,
respectively. The scalar field action is governed by a dimensionless function and a free
length scale, that can be related to the parameters $\mu$ and $a_0$ of MOND in the quasistatic limit.
In addition, the theory contains a scalar and a vector coupling constant as free parameters. 

Although most physicists prefer dark matter over MOND as an explanation of the excess gravitational acceleration in galaxies and the universe, the debate continues \cite{2015CaJPh..93..250M}. As an example, objections based on the rapid instability of spherical systems such as stars \cite{Seifert:2007fr} were alleviated by a generalized version of TeVeS 
\cite{Skordis:2008pq}. On the other hand, it is worth noting that the Bullet Cluster provides evidence for dark matter independent of the particulars of the laws of gravity \cite{2006ApJ...648L.109C}. MOND, or TeVeS, therefore, certainly has trouble accounting for the Bullet Cluster, but its predictions must nevertheless be tested, for example by measuring the orbits of binary stars of large separation, for which the gravitational acceleration is below $a_0$.

\subsection{Fundamental requirements for dark matter particles}
Let us suppose that dark matter is composed of individual particles with rest mass $m_0$. For hot dark matter, $m_0\,c^2$ is less than the energy scale of the universe shortly before recombination, and the particle was relativistic while structures started to grow, suppressing growth on small scales by smearing out such density fluctuations. Warm dark matter has a mass, and hence a velocity, just large enough that structure formation on the scale of a proto-galaxy is possible. Only cold dark matter is massive enough to permit growth of small-scale density fluctuations.

Simulations can be used to follow the evolution of structures after recombination, when radiative pressure is negligible. Structure formation in this phase is hierarchical, meaning that small structures form first and larger ones later. The matter distribution at redshift $z=0$ still carries imprints of the fluctuation spectrum at the time of recombination. Cold dark matter seems to fit best, and hot dark matter as dominant factor is virtually excluded \cite{2005Natur.435..629S}. Measurements of the Lyman-$\alpha$ forest, small-scale gas structures at redshifts of a few, are incompatible with a particle mass below the keV scale \cite{2009JCAP...05..012B}.

Very massive cold objects, for example planets, brown dwarfs, or primordial black holes, which are colloquially subsumed under the name Massive Compact Halo Objects (MACHOs), have been extensively searched for with microlensing experiments and found too few in number to account for dark matter \cite{1996ApJ...471..774A,2007A&A...469..387T,2011MNRAS.416.2949W}.

Dark matter cannot be baryonic, because that would be in conflict with the yield of big-bang nucleosynthesis \cite{1995Sci...267..192C} and results of a CMB fluctuation analysis. The mass range for which primordial black holes can constitute dark matter is also
quite restricted to roughly the mass range between $10^{17}\,M_\odot$ and $10^{25}\,M_\odot$~\cite{2010PhRvD..81j4019C}.
This makes it very likely that dark matter is a new type of particle.

A consequence of non-baryonicity is color neutrality, meaning dark matter particles may not engage in strong interactions, otherwise that particle would just act as baryonic matter. In its lightest form, it has to be electrically neutral, lest it be easily visible instead of being dark. Note that this does not preclude the existence of heavier charged siblings that may have existed in the early universe.

Weak interactions are not excluded. A coupling to the electroweak gauge
bosons $W^\pm$ and $Z$ must be somewhat weaker than that of other
Standard-Model particles, though, otherwise direct-detection experiments would already have seen such interactions.

The stability of at least the lightest dark-matter particles must be high or, conversely, the self-interactions rather weak. The lack of impact on the dark-matter distribution of cosmic events that do influence the baryonic-matter distribution, such as the merger of the Bullet Cluster imply that the dark-matter particles have to be highly collisionless, much more so than the intracluster gas that is itself collisionless in the usual definition of a preponderance of collective interactions over two-body collisions. In fact, these observations have been used to put
upper limits on the dark matter self-interaction scattering cross section per dark-matter mass $m_X$ due to the resulting
drag force per mass unit, $F_d/m_X\sim(\sigma_s/m_X)v^2\rho_X$, with $v$ and $\rho_X$ the relative
velocity and mass density, respectively. These constraints are currently of the order
$\sigma_s/m_X\lesssim1\,{\rm cm}^2/$g~\cite{Harvey:2015hha}. Interestingly, a significant offset between the center of the dark-matter halo
and the star distribution of one of the central galaxies in the galaxy cluster Abell 3827 has been claimed and interpreted in terms
of a dark-matter self-interaction scattering cross section of the order of $\sigma_s/m_X\sim10^{-4}\,{\rm cm}^2/$g~\cite{Massey:2015dkw},
but this is currently controversial~\cite{Kahlhoefer:2015vua}.

Depending on the nature of the decay products, astrophysical lower limits on the decay timescale are $\gtrsim 10^{26}\ \mathrm{s}$ \cite[e.g.][]{2010ApJ...712L..53P}. Self-annihilations must also be rare. Assuming that the particles were in thermodynamic equilibrium very early in the universe, the present-day average density of dark matter requires some loss in particle number on account of, e.g., annihilation. If $S$-wave processes dominate, i.e. the product of velocity and annihilation cross section, $\langle\sigma_{X\bar X}v\rangle$, is constant, then
the $X-$particle mass density in units of the critical density today is given by
\begin{eqnarray}\label{eq1}
  \Omega_X h^2&\simeq&835\,\frac{g_f^{1/2}}{H_0^2\langle\sigma_{X\bar X} v\rangle}\left(\frac{T_0}{M_{\rm Pl}}\right)^3
  \simeq\frac{10^{-37}\,{\rm cm}^2}{\langle\sigma_{X\bar X} v\rangle}\lesssim\Omega_ch^2\,,\\
  \langle\sigma_{X\bar X} v\rangle&\gtrsim&\langle\sigma_{\rm th}\,v\rangle=3\cdot 10^{-26}\ \mathrm{cm^3\,s^{-1}}\,,\nonumber
\end{eqnarray}
where $10\lesssim g_f\lesssim100$ is the number of relativistic degrees of freedom at the freeze-out temperature,
$T_f\simeq m_X/20$, below which the dark-matter particles of mass $m_X$ cease to be in thermal equilibrium, as
can be shown by solving the appropriate Boltzmann equation. Furthermore, $H_0$ and $T_0$ are the Hubble rate and
CMB temperature today and $M_{\rm Pl}=1.22\times10^{19}\,$GeV is the Planck mass.
This thermal freeze-out process and its scaling with the annihilation cross section can qualitatively
be understood as follows: The rate of change of the dark-matter number density, $n_X$, due to expansion of the universe
equals $-3H(T)n_X$ with $H(T)$ the Hubble rate at temperature $T$, whereas the one due to annihilation
and creation from inverse processes is proportional to $\langle\sigma_{X\bar X} v\rangle n_X^2$. For $T\gtrsim m_X$
the number density, $n_X$, is given by the thermal equilibrium abundance, and for $T\ll m_X$ annihilation and inverse processes become
inefficient so that $n_X\propto T^3$ due to the expansion of the universe. In the transition region $T\sim m_X$
both rates become comparable. Equating these rates and using the Friedmann equation for $H(T)$ then
leads to the scaling in Eq.~(\ref{eq1}). This
also implies that the annihilation cross section has to be equal to or larger than the
thermal cross section \cite{1990eaun.book.....K,1993NYASA.688..390G} if the particle is to constitute all or
part of the dark matter, respectively.

In principle, the particles constituting CDM could have any spin. If they happen to be fermionic and non-degenerate, an interesting lower bound on the fermion mass, known as the
Tremaine-Gunn bound \cite{1979PhRvL..42..407T}, can be derived. 
Fermions satisfy Pauli's principle, and so the fermionic occupation numbers
are always below the Boltzmann limit, $f_{\rm
  eq}(E)\leq g\exp\left[(\mu-E)/T\right]$. With $\mu=m$ one finds
\begin{equation}
\rho\simeq mn\lesssim gm\left(\frac{mT}{2\pi}\right)^{3/2}\simeq
gm\left(\frac{m^2\sigma_v^2}{6\pi}\right)^{3/2}\ ,
\label{eq2}
\end{equation}
where in the last step we have assumed a Gaussian distribution to compute
the velocity dispersion $\sigma_v$ in one direction, 
$\frac{1}{2}m\langle v^2\rangle\simeq\frac{1}{2}m\sigma_v^2=T/2$.
We know that individual dark matter haloes exist. In the simplest form they are isothermal, $\sigma_v=\mathrm{const.}$, for which the hydrostatic equilibrium implies $\rho\propto r^{-2}$. Inserting these expressions one finds
\begin{equation}\label{eq:Tremaine_Gunn}
  m\gtrsim\frac{(2\pi)^{1/8}}{(gG_{\rm N}\sigma_v r^2)^{1/4}}\simeq1.5\,
  g^{-1/4}\,\left(\frac{1000\,{\rm km}\,{\rm s}^{-1}}{\sigma_v}\right)^{1/4}
  \left(\frac{{\rm Mpc}}{r}\right)^{1/2}\,{\rm eV}\,.
\end{equation}
Some variation in the pre-factors is to be expected if not all assumptions are met, but Eq.~(\ref{eq:Tremaine_Gunn}) is accurate to within a factor
of a few. This limit is most severe for small haloes with low velocity dispersion. For dwarf satellite galaxies of the Milky Way we may estimate $r\approx 500$~pc and $\sigma_v\simeq 50$~km/s, resulting in $m\gtrsim g^{-1/4}\,(150\ \mathrm{eV})$. Obviously, very light fermionic particles constituting hot or warm dark matter appear
inconsistent with the existence of small dark matter haloes.

\begin{SCfigure}[0.85]
\centering
\caption{Summary of dark matter candidates in double-logarithmic display of mass $m_X$
and typical scattering cross section with ordinary matter. Hot, warm,
and cold dark matter are indicated in red, pink and blue, respectively. 
The gravitino is the supersymmetric partner of the graviton, whereas neutralinos are candidates for the
LSP. ADM denotes asymmetric dark matter, to be briefly discussed in Sect.~\ref{sec:2.4},
and ordinary as well as sterile neutrinos are candidates for hot and warm dark matter. Finally, SIMP stands for strongly interacting massive particles and WIMPZILLAS consitute supermassive dark matter that can only be created at GUT energy scales.
Taken from Baer et al. \cite{Baer:2014eja}.}
\includegraphics[width=0.51\textwidth]{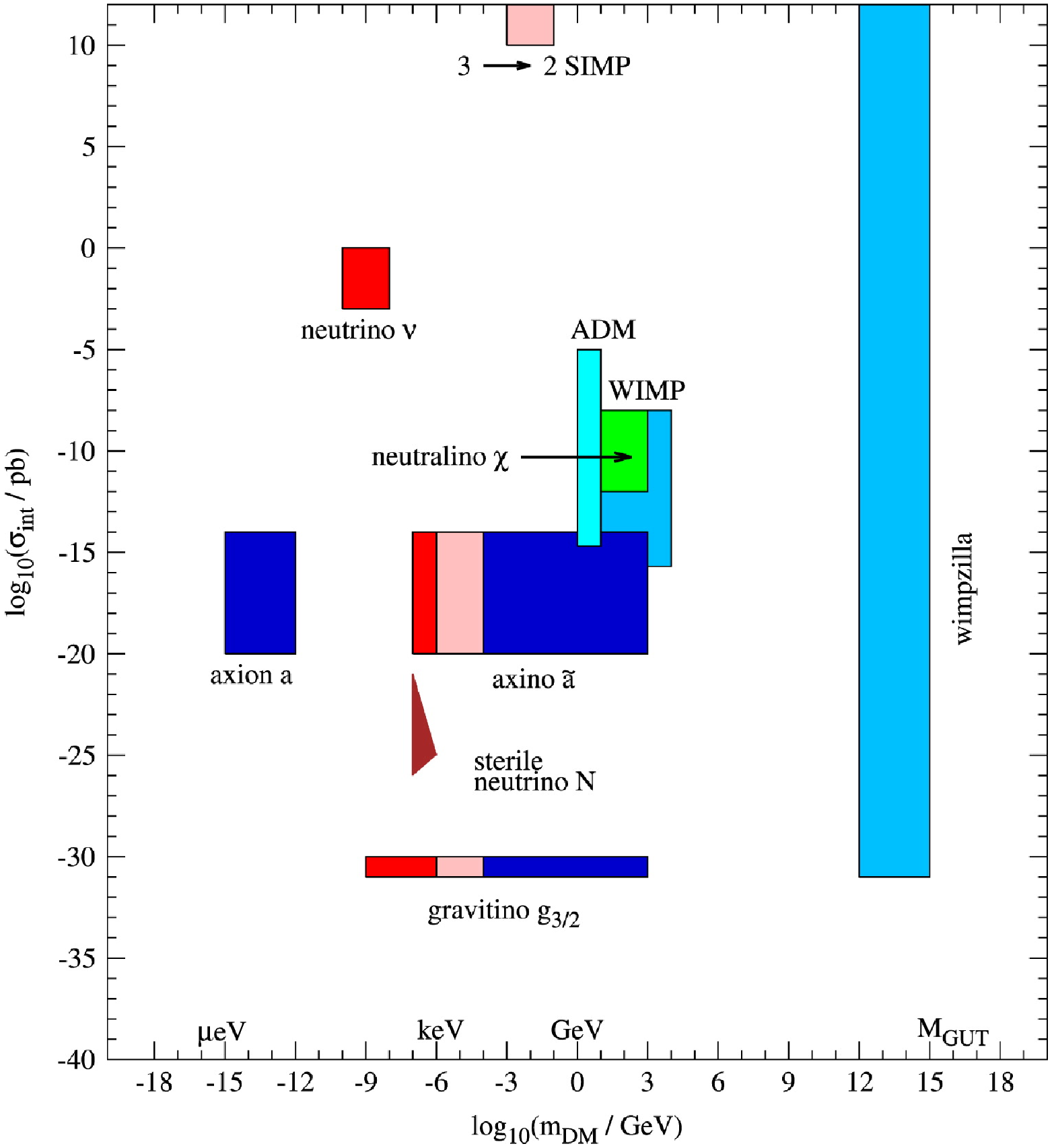}
\label{fig:dm-candidates}
\end{SCfigure}

The Standard Model of particle physics does not contain any
particle that meets all these requirements. In the following we list
three popular types of extensions to the Standard Model in which dark-matter candidates are discussed. An overview over these and other candidates is given in fig.~\ref{fig:dm-candidates}.

\begin{enumerate}

\item In supersymmetric extensions of the Standard Model the lightest supersymmetric particle (LSP) is stable on account of a specific symmetry. Such particles typically have masses above a few GeV and are thus candidates for CDM. Among these so-called  {\it weakly interacting massive particles} (WIMPs), one finds the supersymmetric partners of the gauge bosons and the Higgs boson, i.e. a neutralino, of a neutrino, i.e. a sneutrino, or even of the graviton, i.e. the gravitino.

\item Also subsumed as WIMPS are {\it Kaluza-Klein excitations} found in higher-dimensional extensions. Again, the lightest particle of this type can be a dark-matter
candidate. 

\item Axion-like Particles (ALPs) are generally very light, with masses below the MeV scale and typically much less, down to $\sim10^{-9}\,$eV. In contrast to WIMPs, they are not produced through thermal freeze-out, but form a {\it Bose-Einstein condensate} with very high occupation numbers.

\end{enumerate}

Of particular interest and subject to controversies is the distribution function of dark-matter particles. This is not an academic question, because the velocity distribution of particle in the barycenter of the solar system is the decisive input parameter for the interpretation of direct-detection experiments. The density profiles of dark-matter haloes and clumping on small scales determine the yield of dark-matter annihilation for indirect-detection experiments \cite[e.g.][]{2009PhRvD..79d1301P}, as the rate depends on the square of the density, $\rho$. For any fluctuating quantity the volume average of the density squared is amplified, 
\begin{equation}
\rho=\rho_0+\delta\rho \quad\rightarrow\quad
\langle\rho^2\rangle=\langle\rho_0^2\rangle+\langle(\delta\rho)^2\rangle\ge 
\langle\rho_0^2\rangle\ .
\label{eq2a}
\end{equation}
Simulations suggest that the mass distribution of dark matter clumps is well represented by a power law
\begin{equation}
\frac{dn(M)}{dM} = n_0\,M^{-f}
\label{densspec}
\end{equation}
whose index is approximately $f=2$ between $10^{-6}\,M_\odot$ and $10\,M_\odot$ \cite{diemand05},
and somewhat flatter, $f\simeq 1.8$, around $10^{6}\,M_\odot$ \cite[e.g.][]{helmi}. We note that the smallest clump
mass is determined by kinetic decoupling of the WIMPs and thus by their microphysics and can range between
$\sim10^{-11}\,M_\odot$ and $\sim10^{-3}\,M_\odot$~\cite{Bringmann:2009vf}. Apart from the low-mass cut-off of the
dark-matter clump distribution, the total number density of clumps also depends on the shape of the mass function which
in turn depends on the poorly known survival probability of such microhaloes \cite[e.g.][]{goerdt}.

For a dark matter clump of characteristic size, $r_s$, and density, $\rho_s$, the annihilation rate can be written as \cite{springel,kuhlen}
\begin{equation}
Q_\mathrm{ann}\propto \rho_s^2\,r_s^3 \propto M^d \ .
\label{sourcespec}
\end{equation}
For a cuspy density profile $\rho_s\,r_s^3$ should also be proportional to the mass of the clump, $M$. Then, 
$Q_\mathrm{ann}$ may increase faster than linearly with mass
\cite{koush, tasit}, i.e. $d >1$, and the annihilation rate per mass interval
peaks at the high-mass end of the distribution,
\begin{equation}
\frac{dQ}{d\ln M}= Q_c\, M\,\frac{dn(M)}{dM}\propto M^{d+1-f}\ .
\label{powerspec}
\end{equation}
If the power-law index in equation (\ref{powerspec}) is negative, $d+1-f < 0$, then the
dark matter annihilation is dominated by the numerous low-mass clouds, and we can treat clumping with a simple boosting factor. On the other hand, if $d+1-f > 0$,
then the few massive clumps dominate electron production, and the spatial distribution of annihilation events will no longer reflect the global dark-matter density profile.

On large scales, the density profiles of dark-matter haloes are likewise
not well known. Whereas various mathematical expressions can be adapted to fit the density profiles in the outer regions of the haloes, the behavior in the central regions is subject to a controversy known as the cusp/core problem \cite{1994Natur.370..629M}. The rotation curves of the innermost regions of dwarf galaxies suggest that the density profile levels off and forms a core. Observations of galaxy clusters likewise indicate a shallow mass profile.
If the rotation velocity falls off
faster than $r$ for $r\to0$, then the mass density,
$\rho(r)$, cannot increase for $r\to0$. Such a profile is well approximated as
\begin{equation}\label{eq:quasi_isothermal}
\rho_{\rm ISO}(r) = \frac{\rho_s}{1+\left(\frac{r}{r_s}\right)^2}\,,
\end{equation}
which turns into an isothermal profile for $r\gtrsim r_s$.
In contrast, N-body dark matter simulations predict dark
matter profiles that behave as $\rho(r)\propto r^{-1}$ or as $\rho(r)\propto r^{-1.5}$ for small $r$. The most widely used density profile, the Navarro-Frenk-While (NFW) profile~\cite{1996ApJ...462..563N}, is an adaption to simulation results and given by
\begin{equation}\label{eq:NFW}
\rho_{\rm NFW}(r) = \frac{r_s\,\rho_s}{r\,\left(1+\frac{r}{r_s}\right)^2}\,,
\end{equation}
where $r_s$ is a scale radius and $\rho_s$ a characteristic density.
More recent simulations suggest a slight modification in the innermost
regions, which is known as an Einasto density profile \cite{1965TrAlm...5...87E},
\begin{equation}\label{eq:Einasto}
\rho_{\rm Einasto}(r) = \rho_s\exp\left(\frac{2}{\alpha}\right)\exp\left[-\frac{2}{\alpha}\left(\frac{r}{r_s}\right)^\alpha\right]\quad \mathrm{with}\ \alpha\simeq 0.16\ .
\end{equation}
The slope of this profile depends on the radius $r$ as
\begin{equation}
\frac{d\ln\rho(r)}{d\ln r}=-2\left(\frac{r}{r_s}\right)^\alpha\ . 
\end{equation}
Warm dark matter would lead to a flattening of the central density profiles of dark-matter haloes compared to the ones discussed
above for CDM. A class of explanations for the cusp/core problem focusses on the neglect of baryons in dark-matter
simulations~\cite{2011ApJ...741L..29K}. Star
formation and supernova explosions may remove low-angular-momentum gas \cite{2012MNRAS.422.1231G}. Alternatively, 
angular-momentum transfer between baryons and dark matter may lead to heating of dark matter. One also needs to verify that residual collisionality in N-body simulations does not artificially shape the central density profile \cite{2003MNRAS.338...14P,2015APh....62...47B,2015arXiv150101959H}.  

%% file: section2.tex
\section{Candidates}
\label{sec:2}

\subsection{Neutrinos}
\label{sec:2.1}
Standard-Model neutrinos can not constitute a significant part of dark matter because current upper
limits on their mass imply that they would freely stream on scales of many Mpc and hence wash out the density
fluctuations observed at these scales. Many extensions of the Standard Model contain one neutrino state
per lepton generation which are singlets under the electroweak gauge group. Such states are generally termed
{\it sterile neutrinos}, although they are not sterile in the strict sense because they may be subject to other,
``hidden'' interactions, and the mass eigenstates are in general mixtures of these electroweak singlet states
and the active neutrinos. If the mass of the heaviest neutrino state is on the order of the Grand-Unification scale,
then one can show that the diagonalization of the $6\times6$ mass matrix of active and sterile neutrinos can naturally
provide the small, sub-eV mass scale of the known active neutrinos. This concept is known as the {\it seesaw mechanism}.

\begin{figure}[t]
 \centering
 \includegraphics[width=\textwidth]{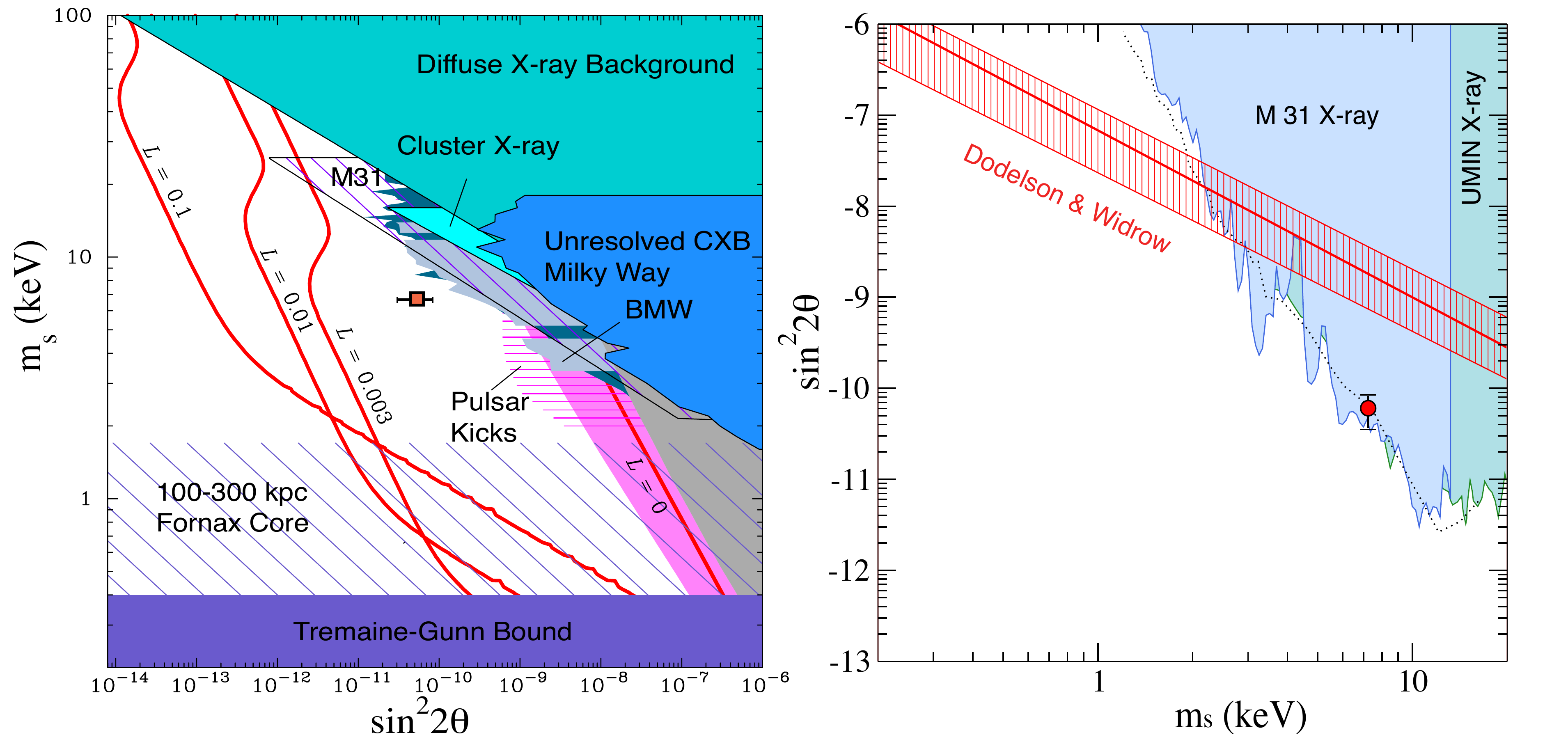}
 \caption{Summary of current constraints on sterile-neutrino mass and mixing angle with
active neutrinos, from Ref.~\cite{Bulbul:2014sua}. Red squares mark the values implied if
the putative 3.55-keV line were due to the decay of a sterile neutrino, constituting the entire dark matter, into
an active neutrino and a photon through a one-loop diagram involving a $W^-$.
Left panel: The colored area on the upper right is ruled out by various observations of diffuse $X-$ray
emissions. The red hatched area marked ``Pulsar Kicks'' would give the observed kicks to newly-born neutron stars
due to asymmetric emission of sterile neutrinos. Red lines denote parameter combinations, for
various lepton numbers as indicated, for which sterile neutrinos would be produced from active neutrinos through oscillations
and would constitute all dark matter, known as the Dodelson-Widrow scenario~\cite{Dodelson:1993je}.
The general lower bound on fermion dark-matter mass marked ``Tremaine-Gunn'' is discussed in Sect.~\ref{sec:1},
see Eq.~(\ref{eq:Tremaine_Gunn}) and also Ref.~\cite{Abazajian-sterile-nu} for
more information. Right panel: Recent $X-$ray constraints based
on observations of M31 by Chandra and XMM-Newton and of Ursa Minor by Suzaku. The Dodelson-Widrow band is the same as on the left panel ($L=0$). See also Ref.~\cite{Horiuchi:2013noa} for more information.}
 \label{fig:st2014}
\end{figure}

If the lightest sterile neutrino has a keV mass scale, it is a warm dark-matter
candidate~\cite{Merle:2013gea}. The couplings of
sterile neutrinos are usually too small to have been in thermal equilibrium in the early universe. However, if they mix
with active neutrinos, they can be produced through oscillations with active neutrinos which in turn have been
in thermal equilibrium at temperatures above an MeV in the early universe. If the total lepton number in the neutrino
sector vanishes, this process is known as the Dodelson-Widrow scenario~\cite{Dodelson:1993je}, whereas the case of non-vanishing lepton number
is often called the Shi-Fuller scenario~\cite{Shi:1998km}. One-loop diagrams involving a $W^-$ give rise to decays
into an active neutrino and a photon. If sterile neutrinos significantly contribute to dark matter, these decays can lead
to detectable diffuse $X-$ray emission, measurements of which rule out a considerable part
of the parameter space of sterile neutrinos. Figure \ref{fig:st2014} summarizes constraints on the mass, $m_s$, and the sterile-active mixing angle, $\theta$. Also shown in Fig.~\ref{fig:st2014} are the implications of interpreting in the
context of sterile-neutrino dark matter the recent putative observation of a 3.55-keV line.

\subsection{Weakly interacting sub-eV (or slim) particles (WISPs)}
\label{sec:2.2}
Many extensions of the Standard Model of particle physics, in particular scenarios
based on supergravity or superstrings, predict a {\it hidden sector} of new particles
interacting only very weakly with Standard-Model particles. Such scenarios do not
necessarily only contain Weakly Interacting Massive Particles (WIMPs\index{WIMP}), new heavy
states at the TeV scale and above, some of which are candidates for the dark matter,
but often also predict Weakly Interacting Sub-eV (or Slim) Particles (WISPs)
that can couple to the photon field, $A_\mu$~\cite{Jaeckel:2010ni}. Well-known examples are pseudo-scalar axions and axion-like
particles (ALPs), $a$, and hidden photons that kinetically mix with
photons. At the high end of the mass spectrum, various particle-physics models also
predict the existence of non-elementary particle states that can be either one-dimensional
topological defects such as monopoles or non-topological solitons such as condensations of bosonic
states and so-called {\it Q-balls}. In the following we briefly discuss the motivation for and constraints
on some of the lighter dark-matter candidates.

\subsubsection{Axion-like particles (ALPs)}
\label{sec:2.2.1}
In addition to the standard QCD Lagrange density,
QCD allows a term violating $CP$ symmetry that is of the form
\begin{equation}\label{eq:L_s_CP}
  {\cal L}_{\theta}=\frac{\alpha_{\rm s}}{4\pi}\,\theta\,G^\alpha_{\mu\nu}\tilde G^{\mu\nu}_\alpha
  =\frac{\alpha_{\rm s}}{4\pi}\,\theta\,\frac{1}{2}\epsilon^{\mu\nu\lambda\sigma}G^\alpha_{\mu\nu}G_{\lambda\sigma}^\alpha\,,
\end{equation}
where $G^\alpha_{\mu\nu}$ is the gluonic field-strength tensor with
$\alpha=1,\cdots,8$, and $\tilde G^\alpha_{\mu\nu}$ as its dual, and $\theta$ is a dimensionless number whose size has to
be experimentally determined. Experimental limits on the $CP$ symmetry-violating electric-dipole moment
of the neutron requires the parameter $\tilde\theta\equiv\theta+\arg\det M$ to be smaller than $\sim10^{-10}$, where $M$ is the quark mass matrix. There is a priori no theoretical reason why this
number should not be of order unity, and so this small value is known as the {\it strong $CP$ problem}.

A possible solution to this problem is to promote $\tilde\theta$ to a
dynamical pseudo-scalar {\it axion field}, $a$, whose expectation value
can be dynamically driven to zero. This behavior can be achieved, if the low-energy effective Lagrange density for the axion has the form
\begin{equation}\label{eq:L_a}
  {\cal L}_a=\frac{1}{2}\partial_\mu a\partial^\mu a+\frac{s\alpha_{\rm s}
    }{4\pi f_a}\,a\,G^\alpha_{\mu\nu}\tilde G^{\mu\nu}_\alpha+
    \frac{\alpha_{\rm em}}{8\pi f_a}\,a\,F_{\mu\nu}\tilde F^{\mu\nu}-\frac{1}{2}m_a^2\,a^2\,,
\end{equation}
where $s$ is a model-dependent number, $f_a$ is the {\it Peccei-Quinn
energy scale}~\cite{Peccei:1977hh}, and $F_{\mu\nu}$ is the electromagnetic-field tensor with its dual $\tilde F^{\mu\nu}=\epsilon^{\mu\nu\lambda\sigma}F_{\lambda\sigma}/2$. If 
the effective axion mass, $m_a$, vanishes, the effective action based on the Lagrange density (cf. Eq.~\ref{eq:L_a})
is invariant under the {\it Peccei-Quinn shift symmetry}, $a\to a+$const. This
invariance follows from the fact that the parity-odd terms involving field strengths in Eq.~(\ref{eq:L_a}) can be written as
divergence of a chiral fermion current. The divergence can then be transferred to the axion field by partial
integration, and only derivatives of the axion field remain.
Furthermore, by comparing Eqs.~(\ref{eq:L_s_CP}) and~(\ref{eq:L_a}) one sees that $\theta^\prime\equiv\theta+sa/f_a$
plays the role of an angle, that in the early universe within one causal volume will take
a random value of order unity. This aspect will play a role for the axion relic density that we will
discuss in Sect.~\ref{sec:2.2.3}.

\begin{figure}[ht]
\center{\includegraphics[width=0.8\textwidth]{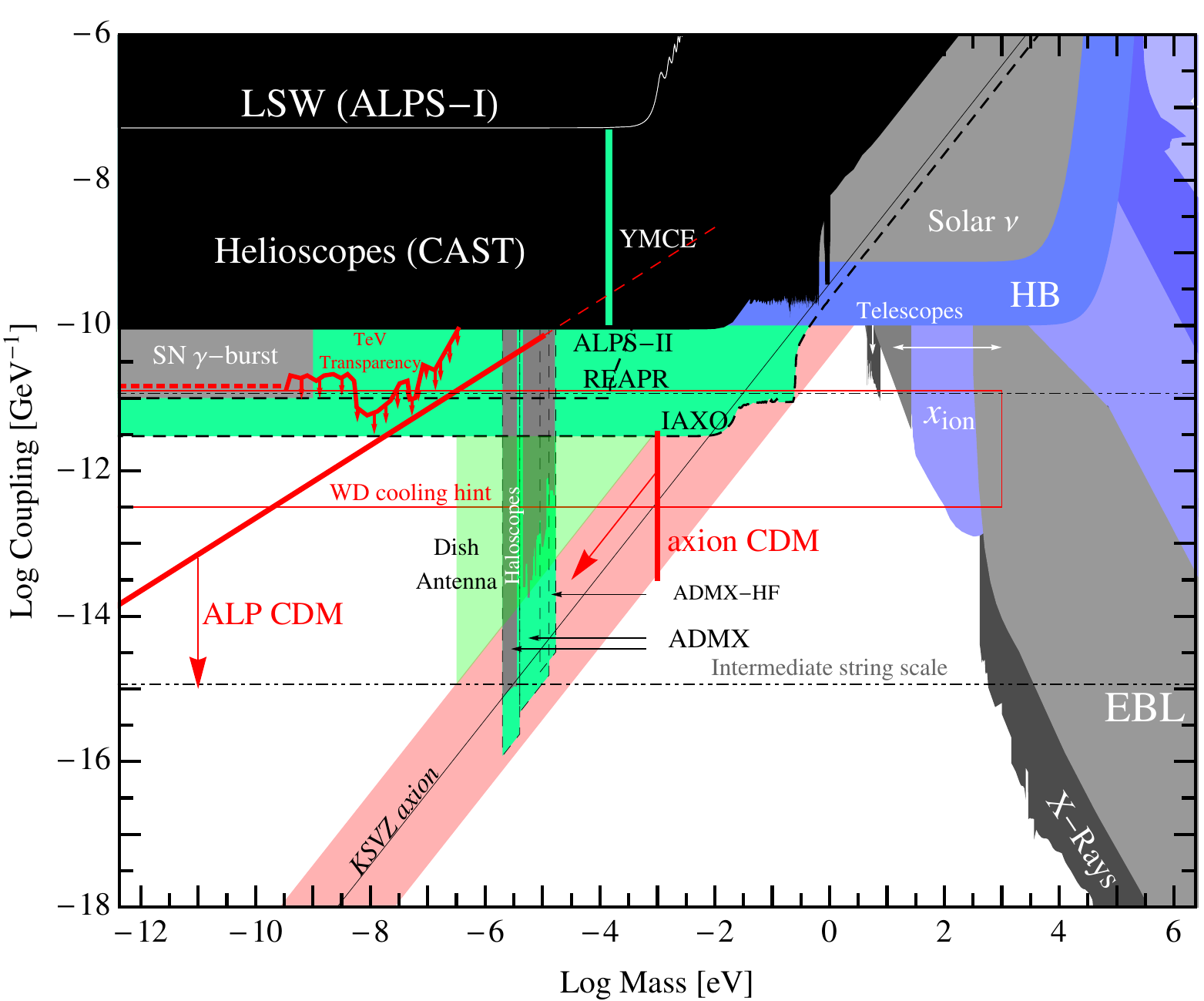}}
\caption[...]{Constrained and still testable regions of photon-ALP mixing in the $m_a-g_{a\gamma}$ plane. Experimentally excluded
regions are shown in black, constraints from astronomical observations in gray, and
constraints from astrophysical and cosmological arguments in blue; the sensitivity of planned experiments
is indicated in light green. Plotted in red are the parameter regions in which ALPs could account for all dark matter, see Sect.~\ref{sec:2.2.3}.
The {\it KSVZ axion} is one of the original
QCD\index{quantum chromodynamics (QCD)} axion models which roughly exhibit the
scaling in Eq.~(\ref{eq:m_a}). From Ref.~\cite{Hewett:2012ns}.}
\label{fig:ALPs}
\end{figure}

The coupling of Axion-Like Particles (ALPs) to photons can now be put into the form
\begin{equation}\label{eq:axion}
  {\cal L}_{a\gamma}=\frac{\alpha_{\rm em}}{8\pi f_a}\,a\,F_{\mu\nu}\tilde F^{\mu\nu}
  =-\frac{\alpha_{\rm em}}{2\pi f_a}\,a\,{\bf E}\cdot{\bf B}\,,
\end{equation}
where in the last expression we have rephrased $F_{\mu\nu}\tilde F^{\mu\nu}$ in terms of
the electric and magnetic field strengths ${\bf E}$ and ${\bf B}$, respectively, and the often-used convention
\begin{equation}\label{eq:g_a}
g_{a\gamma}\equiv\frac{\alpha_{\rm em}}{2\pi f_a}\equiv\frac{1}{M_a}
\end{equation}
was applied. Eq.~(\ref{eq:axion}) is the basis to derive many oscillation phenomena between
photons and ALPs, as we shall see in Sect.~\ref{sec:4.6}.

The ALPs couplings to gluons give rise to mixing between ALPs and neutral pions,
$\pi^0$. In the original QCD axion scenario, the chiral symmetry breaking of the strong interactions which gives rise
to the neutral-pion mass, $m_\pi$, also leads to an axion potential which at zero temperature has the form
\begin{equation}\label{eq:V_a0}
V(a)\simeq\Lambda_{\rm QCD}^4\left(1-\cos\frac{a}{f_a}\right)\,,
\end{equation}
where $\Lambda_{\rm QCD}\simeq215\,$MeV is the confinement scale.
Expansion for $a/f_a\ll1$ then yields an axion mass
\begin{equation}\label{eq:m_a}
m_a\simeq0.6\,\left(\frac{10^{10}\,{\rm GeV}}{f_a}\right)\,{\rm meV}\,,
\end{equation}
which is thus inversely proportional to the Peccei-Quinn scale\index{Peccei-Quinn scale}. We mention in passing that in so-called {\it natural inflation scenarios}
approximately massless (pseudo) Nambu-Goldstone bosons\index{Nambu-Goldstone boson} such as the axion can also play the role of the inflaton
with a potential of the form Eq.~(\ref{eq:V_a0}), with $\Lambda_{\rm QCD}$ replaced by a more general energy scale.

Fig.~\ref{fig:ALPs} shows a compilation of current constraints on and sensitivities of planned experiments
to photon-ALP mixing in the $m_a-g_{a\gamma}$ plane.

In the laboratory photon-ALPs mixing can be
probed with so-called {\it shining light through a wall} experiments. Photons of a laser beam are
partly converted to ALPs in a strong magnetic field in front of a wall and then reconverted behind the wall by a similar magnetic field in an optical cavity of high $Q$ value. One such experiment, the
Axion-Like Particle Search\footnote{Alternatively called Any Light Particle Search (ALPS), \url{https://alps.desy.de}}, is operated at DESY and
uses a 5-T magnetic field and an optical cavity of 8.4m length.

\subsubsection{Hidden photons and other WISPs}
\label{sec:2.2.2}
A hidden-photon field, $X_\mu$, describes a hidden $U(1)$-symmetry group
and mixes with the photon through a Lagrange density of the form
\begin{equation}\label{eq:hidden_photon}
{\cal L}_{X\gamma}=-\frac{1}{4}F_{\mu \nu}F^{\mu \nu}
 - \frac{1}{4}X_{\mu \nu}X^{\mu \nu}
+\frac{\sin\chi}{2}X_{\mu \nu} F^{\mu \nu}
+ \frac{\cos^2\chi}{2}m_{\gamma^\prime}^2 X_{\mu}X^\mu + j^\mu_{\rm em} A_{\mu}\,,
\end{equation}
where $X_{\mu\nu}$ is the hidden-photon field-strength tensor, $m_{\gamma^\prime}$ the
hidden-photon mass, and $\chi$ a dimensionless mixing parameter;
$j_{\rm em}^\mu$ is the electromagnetic current. Typical values for
the mixing parameter range from $\sim10^{-2}$ down to $10^{-16}$.

\begin{figure}[ht]
\center{\includegraphics[width=\textwidth]{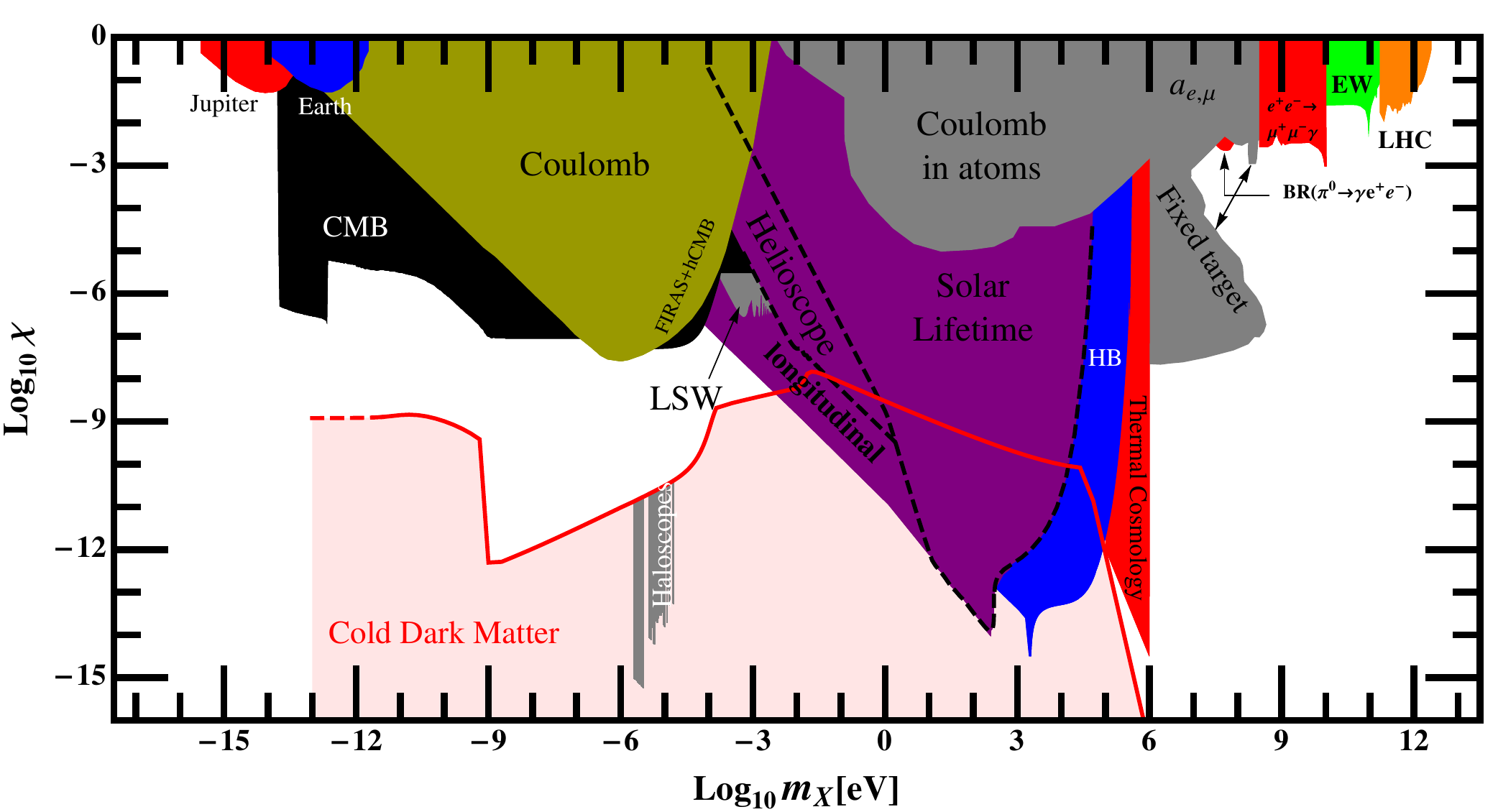}}
\caption[...]{Constrained and still testable regions of photon/hidden-photon mixing in the $m_{\gamma^\prime}-\chi$ plane. Experimentally excluded regions are shown in green and gray, the other colors indicate constraints from astronomical and
astrophysical observations, as indicated. Constraints from CMB distortions are shown in black.
Below the thin red line the hidden photon would account for all the dark matter
observed by non-thermal production in the early universe. From Ref.~\cite{Hewett:2012ns}.}
\label{fig:hidden_photons}
\end{figure}

Fig.~\ref{fig:hidden_photons} shows a compilation of current constraints on, and sensitivities of, planned experiments
to photon/hidden-photon mixing in the $m_{\gamma^\prime}-\chi$ plane.

The ALPS experiment can also be used to probe photon/hidden-photon mixing when operated without magnetic 
field. The resulting constraints are included in Fig.~\ref{fig:hidden_photons}.

\subsubsection{WISPs as dark matter}
\label{sec:2.2.3}
WISPs are typically very weakly coupled and very light. If they have ever been in thermal equilibrium
in the early universe, they would have thermally decoupled already at very high temperatures. In this case,
as long as they have not formed a Bose-Einstein condensate, they would contribute to hot dark matter, and
their mass density today would be far
too small to explain dark matter. Therefore, if WISPs significantly contribute to dark matter today, very likely they
have been produced non-thermally in the early universe and/or formed a Bose-Einstein condensate.

Let us discuss non-thermal WISP production modes on the specific example of the QCD\index{quantum chromodynamics (QCD)} axion. There are
essentially two potential contributions to non-thermal relic axions, and both are related to breaking of the
Peccei-Quinn symmetry\index{Peccei-Quinn symmetry}: First, once this symmetry is broken, there will be
a potential energy whose zero-temperature limit we encountered in Eq.~(\ref{eq:V_a0}). In general, the
axion field, $a(t,{\bf r})$, will not be at the minimum of this potential, resulting in oscillations around this
minimum associated with kinetic and potential energy. Oscillations
of the homogeneous mode of scalar or pseudo-scalar fields manifest themselves as non-relativistic dark matter with
an initial energy density given by
\begin{equation}\label{eq:rho_axion}
  \rho_{a,0}(T)\simeq\frac{1}{2}m^2_a(T)a_0^2\simeq\frac{[f_am_a(T)]^2}{2s^2}\theta_{a,0}^2\,,
\end{equation}
where $m_a(T)$ is the temperature-dependent axion mass, $a_0$ is the value of the axion field at the time of
symmetry breaking, $s$ is a dimensionless parameter that depends on the axion model, and $\theta_{a,0}\equiv sa_0/f_a$ the corresponding angle. Since the resulting energy density
is proportional to the square of the deviation from the equilibrium position, $\theta_{a,0}$, this process is also known as
the {\it axion misalignment mechanism}. The resulting energy density today
can be obtained by solving an equation of motion analogous to that for the inflaton field,
\begin{equation}\label{eq:eom_a}
  \ddot{a}+3H(t)\dot{a}+m_a^2[T(t)]a=0\,,
\end{equation}
which is a damped-oscillator equation with a time-dependent mass $m_a[T(t)]$. In terms of the critical energy density, one obtains
\begin{equation}\label{eq:Omega_a_ma}
  \Omega^{\rm ma}_ah^2\simeq0.11\left(\frac{f_a}{5\times10^{11}\,\theta_{a,0}^2\,{\rm GeV}}\right)^{1.184}
  \simeq0.11\,\theta_{a,0}^2\,\left(\frac{12\,\mu{\rm eV}}{m_a}\right)^{1.184}\,,
\end{equation}
where in the second expression we have used the relation Eq.~(\ref{eq:m_a}) between the zero-temperature mass, $m_a$, and $f_a$
for QCD axions. A generalization of Eq.~(\ref{eq:Omega_a_ma}) to ALPs gives
\begin{equation}\label{eq:Omega_ALP}
  \Omega_ah^2\simeq0.16\left(\frac{m_a}{{\rm eV}}\right)^{1/2}
  \left(\frac{f_a}{10^{11}\,{\rm GeV}}\right)^2\,\left(\frac{\theta_{a,0}}{\pi}\right)^2\,.
\end{equation}
If the ALP or axion field becomes massive before inflation, the initial misalignment angle, $\theta_{a,0}$, is the same as that in the visible
universe today, whereas if the field becomes massive only after the end of inflation, the misalignment angle will be
different in different Hubble volumes at the end of inflation and must be averaged to obtain today's energy density,
$\left\langle\theta_{a,0}^2\right\rangle=\pi^2/3$.

A second contribution to relic axions arises because PQ-symmetry breaking also gives rise
to {\it axionic strings}, since breaking of a $U(1)$ symmetry always gives rise to strings.
The Higgs-Kibble mechanism predicts that about one string will be produced per
causal volume. If PQ-symmetry breaking occurs after inflation, such that the strings are not
inflated away, the decay of these strings will then give rise to the contribution
\begin{equation}\label{eq:Omega_a_s}
  \Omega^{\rm s}_ah^2\simeq0.11\left(\frac{400\,\mu{\rm eV}}{m_a}\right)^{1.184}\,.
\end{equation}
It is interesting to note that if the inflation scale is close to the upper limit given by the tensor-to-scalar
ratio\index{tensor-to-scalar ratio}, then most likely the ALP scale $f_a\lesssim H_i\simeq10^{14}\,$GeV
because otherwise axion isocurvature fluctuations\index{isocurvature fluatuations} would be
produced during inflation, inconsistent with upper limits from CMB observations. From Eq.~(\ref{eq:Omega_a_ma}) with
$(\theta_{a,0}/\pi)^2\to1/3$ and Eq.~(\ref{eq:Omega_a_s}) one would deduce a rather narrow window of $10^9\,{\rm GeV}\lesssim f_a\lesssim 10^{12}\,{\rm GeV}$
in which ALPs or axions could represent the bulk of dark matter. According to Eq.~(\ref{eq:m_a}) the axion mass would then be in the
range $10^{-6}\,{\rm eV}\lesssim m_a\lesssim 10^{-5}\,{\rm eV}$, whereas ALP masses could be in the eV range.

If WISPs make up a significant part of dark matter, they can also be experimentally detected with
so-called {\it haloscopes}\index{haloscopes} which typically consist of a resonant cavity that can be tuned to the photon
energy corresponding to the WISP mass. The leading
Axion Dark Matter eXperiment\footnote{\url{http://www.phys.washington.edu/groups/admx/home.html}} (ADMX) uses a cryogenically cooled high-$Q$ value tunable
microwave cavity that is immersed in a magnetic field of about $8\,$~T. The resulting constraints on ALP
parameters are shown in Fig.~\ref{fig:ALPs}. An extension to higher frequencies, ADMX-HF, is planned.

Very recent ideas focus on dish antennas to detect the conversion into radio photons of WISP dark matter in the
mass interval $10^{-6}\,{\rm eV}\lesssim m_{\rm WISP}\lesssim 1\,$eV~\cite{Horns:2012jf}.

Finally, due to their, albeit very weak, coupling to photons and, possibly, other Standard-Model particles, WISPs
can also provide potential indirect signatures via their decays in Standard-Model particles. For example, the ALP
coupling to two photons (cf. Eq.~\ref{eq:axion}) leads to a lifetime
\begin{equation}\label{eq:tau_a}
  \tau_a=\frac{64\pi}{g^2_{a\gamma}m_a^3}\simeq5.1\times10^{26}\,\left(\frac{f_a}{10^{10}\,{\rm GeV}}\right)^2
  \left(\frac{{\rm eV}}{m_a}\right)^3\,{\rm s}\,.
\end{equation}
For the preferred parameters discussed above this is obviously much larger than the age of the universe. If ALPs
constitute a fraction of dark matter, for $m_a\simeq7.1\,$keV and $10^{11}\,{\rm GeV}\lesssim f_a\lesssim 10^{16}\,{\rm GeV}$, the lifetime may be sufficiently short
to explain, for example, the recent indications for a 3.55-keV photon
line that we discuss in Sect.~\ref{sec:4.2}.

\subsection{Weakly interacting massive particles (WIMPs)}
\label{sec:2.3}
Supersymmetry (SUSY) relates fermions to bosons whose spins differ by $1/2$, and they appear as so-called super-partners in super-multiplets.
Furthermore, SUSY commutes with the energy-momentum operator. Among other things this implies that,
for unbroken supersymmetry, the rest masses of all members of a super-multiplet must be equal. Since such pairings
are obviously not realized in nature, SUSY must be broken. An important feature, and one of its main motivations,
is that even broken SUSY stabilizes scalar masses which would otherwise be unstable to radiative corrections,
because SUSY links the scalar mass to the mass of the corresponding
fermionic partner which is protected by gauge symmetry. Furthermore, supersymmetric extensions of the
Standard Model typically have an additional discrete {\it R-symmetry}, also called R-parity. Since supersymmetric
partners of ordinary Standard Model particles are odd, the lightest SUSY particle, known as the
{\it lightest supersymmetric particle (LSP)}, is thus stable against decay and constitutes a
candidate for dark matter, called a {\it weakly interacting massive particle (WIMP)}.
In particular, the super-partners of the neutral electroweak gauge bosons, the photon and $Z^0$, and of the two neutral
Higgs-boson states present in SUSY models, or linear combinations of these four states, are promising WIMP candidates
because of their weak interactions. They are known as {\it neutralinos}.

WIMP candidates also occur in extensions of the Standard Model containing compactified extra dimensions.
Such scenarios are motivated by string-theory models which require extra dimensions to be anomaly-free, and also
by the fact that in a flat space of $n$ extra dimensions and a volume $V_n$ the fundamental, higher-dimensional gravity
scale, $M_*$, is related to the Planck mass, $M_{\rm Pl}$, observed in four dimensions by the relation
\begin{equation}\label{eq:m_pl_n}
M_{\rm Pl}^2=M_*^{n+2}V_n\,.
\end{equation}
This allows to ``tune'' $V_n$ such that the fundamental gravity scale, $M_*$, can be around a TeV, thus avoiding the
huge and unexplained difference between the electroweak scale and gravity. Assuming that the wave function
factorizes into an ordinary three-dimensional part and a part that only depends on the coordinates of the extra dimensions, $s_i$,
the latter must be single-valued and thus must have the form
$$\exp\left(i\sum_{i=1}^n\frac{2\pi n_is_i}{r_n}\right)\,,$$
with $n_i\in\mathds{Z}$. The dispersion relation of a particle with momentum ${\bf p}_3$ in the ordinary (infinitely extended)
three spatial dimensions then reads
\begin{equation}\label{eq:E_disp_n}
  E^2({\bf p})=M^2+{\bf p}_3^2+\sum_{i=1}^n\left(\frac{2\pi n_i}{r_n}\right)^2\,.
\end{equation}
From the point of view of ordinary three-dimensional space, there may be so-called {\it Kaluza-Klein excitations} of the
quantum fields describing elementary particles, with mass contributions equal to multiples of $2\pi/r_n$,
\begin{equation}
  \frac{2\pi}{r_n}\simeq6\times10^{3}\left(\frac{M_*}{{\rm TeV}}\right)
  \left(\frac{M_*}{M_{\rm Pl}}\right)^{2/n}\,{\rm GeV}\,.\label{rextra2}
\end{equation}
If the compact space of extra dimensions has a reflection symmetry, as is the case in the example above,
the so-called Kaluza-Klein parity, $(-1)^{\sum_{i=1}^nn_i}$, is conserved at tree level.
The lightest Kaluza-Klein state is thus stable or long lived which makes it a dark-matter candidate.

\subsection{Alternatives}
\label{sec:2.4}
The abundance of WIMPs discussed above is governed by the thermal freeze-out paradigm, in which
WIMPs drop out of thermal equilibrium at temperatures $T\lesssim m_X/20$, as we have discussed in
Sect.~\ref{sec:1}. However, there are also dark-matter candidates which have never been in thermal equilibrium
in the early universe. This is typically the case for sterile neutrinos and WISPs. But whereas sterile neutrinos and WISPs tend to be
relatively light and are produced by mixing with Standard-Model particles or through the dynamics of
a scalar or pseudo-scalar field, so-called {\it feebly interacting massive
particles} (FIMPs) with mass comparable to the electroweak scale, alternatively also called
{\it frozen-in massive particles}, could be produced by collisions or
decays of Standard-Model particles. If such particles start from a vanishing initial abundance, they are said
to be produced by {\it freeze-in}, in contrast to the freeze-out scenario
discussed above, in which dark-matter particles have interacted frequently enough with
Standard-Model particles to have initially been in thermal equilibrium.
Let us briefly develop the general properties of such freeze-in scenarios, following Ref.~\cite{Hall:2009bx}.

Suppose that the coupling of FIMPs, $X$, to Standard-Model particles is characterized by a dimensionless
number, $\lambda\lesssim1$, and the mass of the heaviest particle is denoted by $m$. Then the FIMPs
are produced by interactions of the thermal bath of Standard-Model particles with a characteristic cross
section that is on the order $\sigma_X\sim\lambda^2/T^2$ for $T\gtrsim m$. Since the density of Standard-Model particles is $\sim T^3$, during one Hubble time, $t_\mathrm{H}\simeq M_{\rm Pl}/T^2$, the physical density of FIMPs
increases by $\Delta n_X\sim\sigma_X T^6 t_\mathrm{H}\sim\lambda^2M_{\rm Pl}T^2$. Number densities, $n$, are often expressed
in terms of the yield, $Y\equiv n/s$, where $s\sim T^3$ is the entropy density which is itself on the order of the
photon number density. Yields are only changed by interactions and are unaffected by the expansion of the
universe. In our case we thus have $\Delta Y_X\sim\lambda^2M_{\rm Pl}/T$, which is largest at low
temperatures and applies as long as the abundances are not Boltzmann-suppressed, and thus for $T\gtrsim m$.
One thus finds
\begin{equation}
  Y_X\sim\lambda^2\frac{M_{\rm Pl}}{m}\label{eq:Y_FIMP}
\end{equation}
for the yield today. For comparison, the standard yield of thermal WIMPs is $Y_X\simeq\rho_c\Omega_X/(m_Xs_0)$, where
$\rho_c$ and $s_0$ are the critical density and entropy density today, respectively. Using Eq.~(\ref{eq1})
this gives
\begin{equation}
  Y_X\sim\frac{1}{M_{\rm Pl}m_X\langle\sigma_{X\bar X} v\rangle}\simeq
  \frac{1}{\lambda^2}\frac{m_X}{M_{\rm Pl}}\,,\label{eq:Y_WIMP}
\end{equation}
where in the last expression we have phrased the cross section as
$\langle\sigma_{X\bar X} v\rangle\simeq\lambda^2/m_X^2$. Identifying $m$ with $m_X$, one now
realizes that Eqs.~(\ref{eq:Y_FIMP}) and~(\ref{eq:Y_WIMP}) are just the inverse of each other !
Thus, whereas the WIMP abundance decreases with increasing coupling, $\lambda$, the FIMP abundance
increases with $\lambda$. Candidates for FIMPs include moduli or their supersymmetric partners in
string theories, Dirac neutrinos within weak-scale supersymmetry, massive gauge bosons of an extra $U(1)$ group
that mix with photons, as already encountered in Sect.~\ref{sec:2.2.2}, and weakly coupled Kaluza-Klein states
from extra dimensions as discussed in Sect.~\ref{sec:2.3}.

Another alternative discussed in the literature is known as {\it asymmetric dark matter}. Usually it is assumed
that the initial abundance of dark-matter particles, $X$, is equal to that of their anti-particles, $\bar X$. One could, however, also imagine a situation in which $X$ and $\bar X$ are distinguishable and initially
not of equal abundance. In this case, in thermal freeze-out, for example, the more numerous component
would always survive so that the relic abundance (cf. Eq.~\ref{eq1}) would only describe the symmetric component.
The annihilation cross section could then be much larger, potentially permitting complete annihilation of the symmetric partner. In this situation the final dark-matter abundance would be determined by the initial
asymmetry, analogous to the baryon number in certain scenarios of baryo- and leptogenesis. It has
been speculated that unknown physics beyond the Standard Model could render comparable the final number
densities for dark matter and baryons, so that $\Omega_X/\Omega_b\simeq m_X/m_N$, suggesting a dark-matter mass $m_X$ of a few GeV.
If today dark matter would be constituted only by the symmetric part there would essentially be no remaining
annihilations today, and thus the indirect signatures due to dark-matter annihilation that we will discuss in Sect.~\ref{sec:4}
would be moot. The remaining dark matter could, however, still be trapped in astrophysical bodies on account of
scattering with ordinary matter. 

%% file: section3.tex
\section{Direct searches}
\label{sec:3}

\subsection{Detection principles}

Direct-detection experiments rely on the scattering of dark-matter particles from the halo of the
Milky Way in a detector on Earth. The latter is usually set up deep underground, e.g.\
at Gran Sasso in Italy, Sanford in South Dakota or in the future Jin-Ping in China, to shield it
from cosmic radiation, which, together with natural radioactivity in the rock and the detector
material, constitutes the most important background. To discriminate the dark-matter signal from
background processes and thermal noise, one exploits usually two out of three possible
signals: energy deposition (phonons) in (often cryogenic) calorimeters, scintillation light
(photons), and ionisation (electrons). Localization of the
scattering event allows one to define a fiducial volume inside the detector, which is then effectively
shielded by the detector material itself. For the latter, current experiments use CaWO$_4$
(CRESST \cite{Angloher:2011uu,Brown:2011dp,Angloher:2014myn}), CsI(Ti) (KIMS \cite{Lee.:2007qn,%
Kim:2012rza}), Ge (CDMS \cite{Ahmed:2008eu,Ahmed:2009ht,Ahmed:2010wy,Agnese:2013rvf,%
Agnese:2013jaa,Agnese:2014aze}, CoGeNT \cite{Aalseth:2008rx,Aalseth:2012if}, EDELWEISS
\cite{Armengaud:2012pfa,Armengaud:2013rta}), NaI(Ti) (DAMA/LIBRA \cite{Bernabei:2013xsa}) or Si
(CDMS Si \cite{Agnese:2013rvf}) crystals, but also liquid Ar (DarkSide \cite{Bossa:2014cfa}),
Ne (DEAP \cite{Boulay:2012hq}) or Xe (LUX \cite{Akerib:2013tjd}, XENON \cite{Angle:2007uj,%
Angle:2008we,Aprile:2012nq,Aprile:2013doa}, XMASS \cite{Abe:2012ut}, ZEPLIN
\cite{Akimov:2011tj}). For many of these experiments,
upgrades are in preparation (e.g.\ DARWIN \cite{Baudis:2012bc}, EURECA \cite{Kraus:2011zz},
LZ \cite{Malling:2011va}, PANDA-X \cite{Xiao:2014xyn}, XENON1T/nT), in some cases
focussing on the conversion
of axions into photons (ADMX \cite{Sikivie:1983ip}). It is also possible to exploit
the metastability of a superheated liquid composed of carbon and fluorine (COUPP \cite{Behnke:2012ys},
PICASSO \cite{Archambault:2012pm}, PICO \cite{Amole:2015lsj}, SIMPLE \cite{Felizardo:2011uw}), the
annual modulation of the dark-matter signal as the Earth orbits around the
Sun (DAMA/LIBRA \cite{Bernabei:2013xsa}), or the direction of the dark matter ``wind'', blowing
from the Cygnus constellation towards the Sun as it orbits around the galactic center (DMTPC
\cite{Monroe:2011er}, DRIFT \cite{Battat:2014van}, MIMAC \cite{Riffard:2015rga}, NEWAGE
\cite{Miuchi:2010hn}).

\subsection{Local dark matter density and velocity}

The direct-detection rate depends on the local dark-matter density, currently estimated
to be $\rho\simeq0.39\pm0.03$ GeV cm$^{-3}$ \cite{Catena:2009mf}. The systematic uncertainty of this value is up to 40\%, on account of, e.g., effects of the galactic disk, whereas small halo substructures
(clumps or streams) are unlikely to influence it significantly. It is therefore commonly adopted
as a Standard Halo Model (SHM) value in halo-dependent comparisons of direct-detection experiments
\cite{Vogelsberger:2008qb}.

The direct-detection rate also depends on the local dark-matter velocity distribution
$f({\bf v})$. If we assume the velocity distribution at the solar circle to follow a Gaussian and take into account the orbit velocity of the Sun around the galactic center,
$v_0=220$ km s$^{-1}$, and the orbit speed of the Earth around the Sun, $v_e$,
we obtain the normalized one-dimensional Maxwell-Boltzmann
distribution
\bea
 f(v)\dd v&=& {v \dd v\over v_e v_0 \sqrt{\pi}}
 \lgg\exp\lee-{(v-v_e)^2\over v_0^2}\re
   -\exp\lee-{(v+v_e)^2\over v_0^2}\re\rg,
\eea
which must in addition be truncated at the local escape velocity from the Milky Way $v_{\rm esc}=
533^{+54}_{-41}$ km s$^{-1}$ \cite{Piffl:2013mla}. The dark-matter flux passing through a detector on
Earth can therefore be quite large, $nv=\rho v/m_X\simeq$ 10$^5$ cm$^{-2}$ s$^{-1}$ for
$m_X=100$ GeV.

\subsection{Recoil energy and rate}

The recoil energy induced by a dark matter particle of mass $m_X$ on a nucleus of mass $m_N$
is
\bea
 E_R&=&{q^2\over2m_N}~=~{\mu^2v^2\over m_N}(1-\cos\theta^*),
\eea
where $q$ is the momentum transfer, $\mu=m_Xm_N/(m_X+m_N)$ the reduced mass,
$v$ is now the mean dark-matter velocity relative to the target, and $\theta^*$ the scattering
angle in the center-of-mass system \cite{Lewin:1995rx}. If we allow for inelastic scattering,
i.e.\ that the dark-matter particle leaves the detector in an excited state, the minimal and maximal recoil energies
\bea
 E_R^{\rm min,max}&=&\frac{\mu^2}{m_N}
 \left[v^2-\frac{\delta}{\mu}\mp v\left(v^2-2\frac{\delta}{\mu}\right)^{1/2}\right]
\eea
acquire a dependency on the increment $\delta$ of the dark-matter rest mass, $m_X$
\cite{TuckerSmith:2001hy}.
In the elastic case ($\delta=0$) and for $m_X\leq m_N$, the maximum
\bea
 E_R^{\rm max}&\simeq&
 90\,\left(\frac{m_X}{100\,{\rm GeV}}\right)^2
 \left(\frac{100\,{\rm GeV}}{m_N}\right)\left(\frac{v}{200\,{\rm km}\,{\rm s}^{-1}}\right)^2˛{\rm keV}
\eea
ranges typically between 1 and ${\cal O}(100)$ keV, the upper limit imposed by $v_{\rm esc}$.
Dark matter with masses below 1 GeV is typically below the detection threshold for nuclear
recoil, but it could deposit enough energy to interact with electrons (1 to 10 eV). It could
then be detected via electron ionization, excitation or molecular dissociation. This option is
particularly interesting for candidates preferably interacting with leptons, so-called
leptophilic dark matter, that have received attention in the context of the observed positron
excess (see below).

In a detector composed of different target nuclei, the total recoil energy spectrum
\bea
 {\dd R\over\dd E_R}&=& 
 \sum_N N_N{\rho\over m_X}\int_{v_{\min}}^{v_{\rm esc}} {\dd\sigma_N\over\dd E_R}
 v f({\bf v})\dd^3 v
\eea
is obtained by summing the differential rates for all target nuclei $N$, weighted by their number
per unit mass of detector ($N_N$). The individual nuclear recoil rates depend on the number
density of the dark-matter particle, $\rho/m_X$, which must be rescaled if the considered
candidate is not the only contributor to the total relic density, and the differential scattering
cross section 
\bea
  \frac{\dd\sigma_N}{dE_R}&=&\frac{m_N}{2\mu^2v\sqrt{v^2-2\delta/\mu}}\,\sigma_N(E_R)
\eea
integrated over the velocity distribution with a lower limit of integration,
\bea
 v_{\min}&=&{m_NE_R^{\rm min}/\mu+\delta\over\sqrt{2m_NE_R^{\rm min}}}\,,
\eea
that depends on the minimal detectable recoil energy.
The recoil rates are usually small, of order 1 event/100 kg/day.

Experimentally, the recoil energy spectrum must still be convolved with an energy-resolution
function and multiplied with an energy-dependent counting efficiency or cut acceptance. The
probability that only a fraction of the true recoil energy is experimentally visible is
incorporated in a so-called ``Q-factor'',
 $Q(E_R)={E_{\rm vis}/E_R}$.
$E_{\rm vis}$ is
often quoted in keVee (keV electron-equivalent) or in photoelectrons. Since both background
and signal events mostly involve scattering off electrons and nuclei, respectively, the Q-factor is separately determined
from $\gamma$-ray (e.g.\ $^{57}$Co or $^{137}$Cs) and neutron (e.g.\ AmBe or $^{252}$Cf) calibration
sources. In crystals, where mostly phonons are produced, it ranges between 0.09 for iodine
and 0.3 for germanium. In noble gases, a similar factor measures the scintillation efficiency
of dark matter particles relative to photons. Both correction factors are subject to large
experimental uncertainties at low energies.

\subsection{Spin-independent cross sections}

Spin-independent (SI) cross sections
\bea
 \sigma^{\rm SI}_N(E_R)&=&[Z+(A-Z)(f_n/f_p)]^2\,\frac{\mu^2}{\mu_p^2}\, \sigma_p F^2_{\rm SI}(E_R)\,,
\eea
result from scalar or vector couplings. Here, $Z$ and $A$ are the proton and total nucleon numbers
of the target nucleus $N$, $f_{p(n)}$, $\mu_p$ and $\sigma_p$ are the dark-matter couplings,
reduced mass and total cross section for protons (neutrons), and $F_{SI}(E_R)$ is the
spin-independent nuclear form factor, i.e.\ the Fourier transform of the nucleon density in
the nucleus. It is often assumed to be similar for protons and neutrons and have the Helm form
\bea
 F^2_{\rm SI}(E_R)&=&\lr{3j_1(qr_0)\over qr_0}\rr^2e^{-s^2q^2}
\eea
with $q=\sqrt{2m_NE_R}$, the effective nuclear radius $r_0\simeq\sqrt{(1.2~{\rm fm}~A^{1/3})^2-5s^2}$,
and the nuclear skin thickness $s\simeq1$ fm \cite{Helm:1956zz}.
For isospin-conserving couplings ($f_n=f_p$), the nuclear cross section is
enhanced by a factor $A^2$ over the proton cross section, i.e.\ the whole nucleus interacts
coherently provided the inverse momentum transfer is smaller than the size of the nucleus.
Since heavier nuclei consist of more neutrons than protons, isospin-violating coupling with
$f_n/f_p\simeq-Z/(A-Z)$ can reduce the cross sections for particular nuclei, and thus weaken the
tension between conflicting experimental results \cite{Feng:2011vu}.
To adjust the sensitivity of Xe- and
Ge-based experiments to the signal regions claimed by CDMS-II, CoGeNT, and most
strikingly DAMA/LIBRA (see below),
one needs, e.g., $f_n/f_p=-0.7$ and $-0.8$, respectively.

If, e.g., dark matter scatters off nuclei via the exchange of a scalar
particle $\phi$ with mass $m_\phi$ and dimensionless coupling constants $\lambda_X$,
$\lambda_p$ and $\lambda_n$ to dark matter, protons and neutrons, respectively, the
differential cross section is
\bea
  \frac{\dd\sigma^{\rm SI}_N}{\dd E_R}&\simeq&\frac{m_N}{(q^2+m_\phi^2)^2}
  \,\frac{\lambda_X^2\left[Z\lambda_pF_p(E_R)+(A-Z)\lambda_nF_n(E_R)\right]^2}
  {2\pi v\left(v^2-2\delta/\mu\right)^{1/2}}.
\eea
Since  $E_R^{\rm max}-E_R^{\rm min}=2\mu^2v\left(v^2-2\delta/\mu\right)^{1/2}/m_N$,
the total spin-independent scattering cross section is then of the order
\bea
  \sigma^{\rm SI}_N(E_R)&\simeq&\frac{\lambda_X^2\lambda_p^2\mu^2A^2}
  {(2m_NE_R+m_\phi^2)^2}F^2_{\rm SI}(E_R).
\eea

\begin{figure}[h]
 \includegraphics[width=0.495\textwidth]{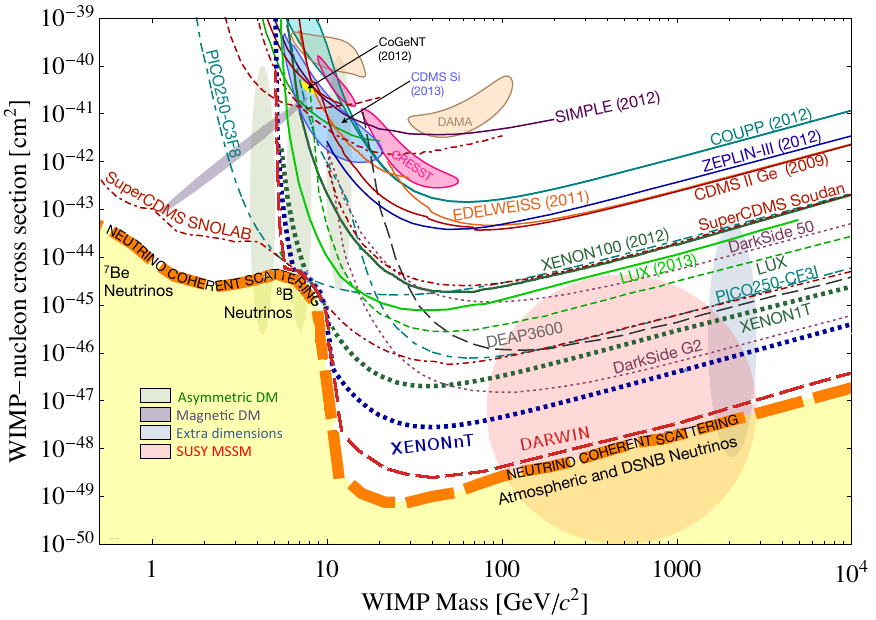}
 \includegraphics[width=0.505\textwidth]{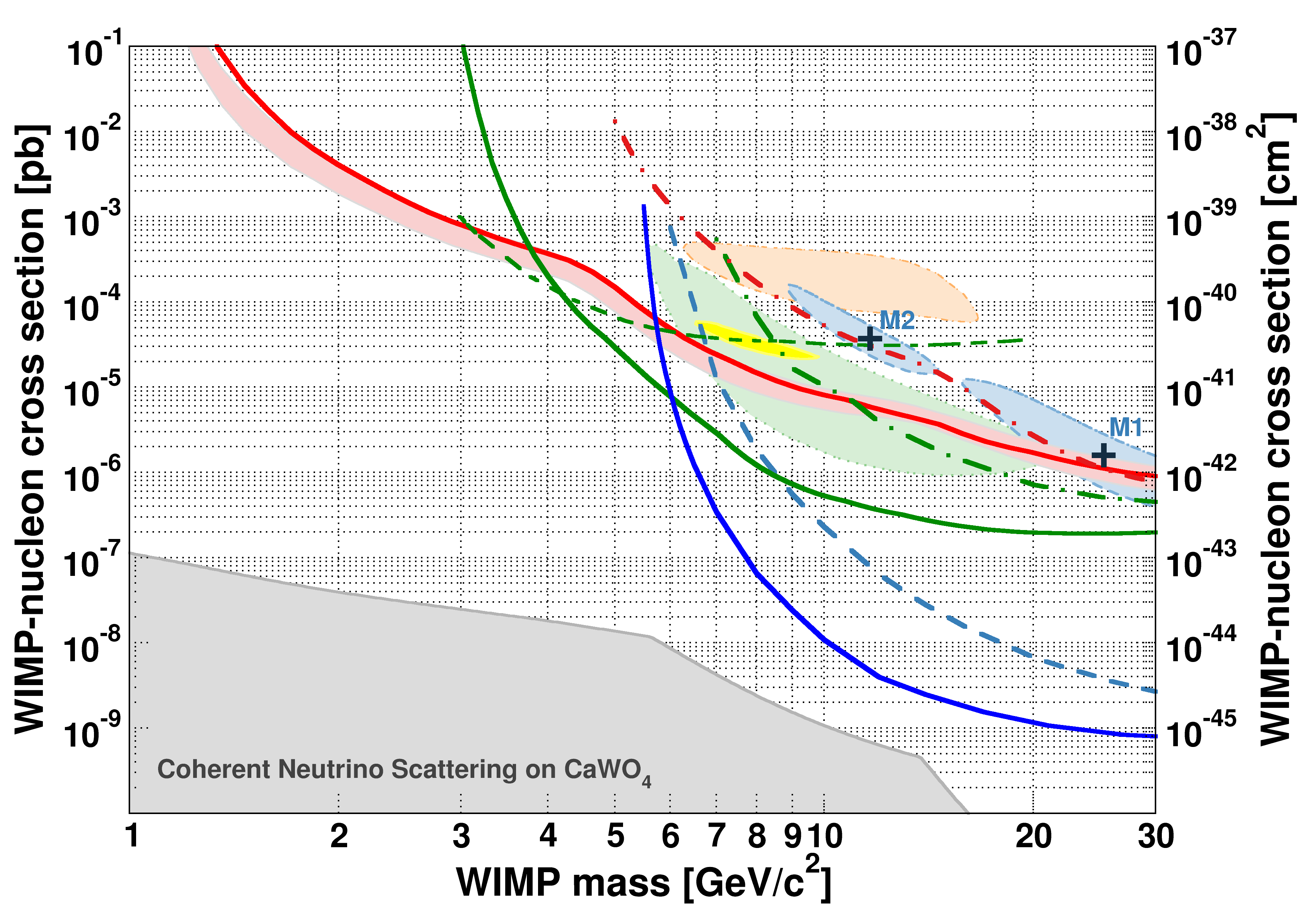}
 \caption{Left: Spin-independent WIMP-nucleon scattering results \cite{Cushman:2013zza,%
 Baudis:2014naa}. Claims from the crystal-based
 experiments CDMS Si \cite{Agnese:2013rvf}, CoGeNT \cite{Aalseth:2012if}, CRESST
 \cite{Angloher:2011uu} and DAMA/LIBRA \cite{Bernabei:2013xsa} are shown together with exclusion
 limits from the noble-gas experiments ZEPLIN-III~\cite{Akimov:2011tj},
 XENON10~\cite{Angle:2007uj}, XENON100~\cite{Aprile:2012nq}, and LUX~\cite{Akerib:2013tjd}.
 Also shown are the projected sensitivities of DarkSide-50 \cite{Bossa:2014cfa}, LUX
 \cite{Akerib:2013tjd}, DEAP3600 \cite{Boulay:2012hq}, XENON1T, DarkSide G2, XENONnT (similar
 sensitivity as LZ \cite{Malling:2011va}) and DARWIN \cite{Baudis:2012bc}. In the yellow shaded region
 coherent neutrino scattering becomes an important background.
 Right: Spin-independent WIMP-nucleon scattering results at low mass \cite{Angloher:2014myn}.
 The 90{\%} C.L. upper limit from CRESST-II (solid red) is depicted together with the expected
 sensitivity (1{$\sigma$} C.L.) from the background-only model (light red band).
 The CRESST-II 2{$\sigma$} contour reported for phase 1 \cite{Angloher:2011uu} is
 shown in light blue. The dash-dotted red line refers to the reanalyzed data from the
 CRESST-II commissioning run \cite{Brown:2011dp}. Shown in green are the limits (90{\%}
 C.L.) from Ge-based experiments SuperCDMS (solid) \cite{Agnese:2014aze}, CDMSlite (dashed)
 \cite{Agnese:2013jaa} and EDELWEISS (dash-dotted) \cite{Armengaud:2012pfa}. The parameter
 space favored by CDMS Si \cite{Agnese:2013rvf} is shown in light green (90{\%} C.L.),
 the one favored by CoGeNT (99{\%} C.L. \cite{Aalseth:2012if}) and DAMA/LIBRA
 \cite{Bernabei:2013xsa} in yellow and orange.
 The exclusion curves from liquid Xe experiments (90{\%} C.L.) are drawn in
 blue (solid for LUX \cite{Akerib:2013tjd}, dashed for XENON100 \cite{Aprile:2012nq}).
 Marked in grey is the limit for a background-free CaWO$_4$ experiment
 arising from coherent neutrino scattering, dominantly from solar neutrinos
 \cite{Gutlein:2014gma}.
 }
 \label{fig:si2014}
\end{figure}

Currently, three direct-detection experiments claim evidence for potential signals
of dark matter. At Gran Sasso, DAMA/LIBRA with 14 annual cycles and a total exposure of
1.33 ton-years finds with $9.3\sigma$ confidence an annual modulation of single-hit events
in the (2-6) keV energy interval with amplitude (0.0112 $\pm$ 0.0012) counts/kg/keV/day, a
measured phase of (144 $\pm$ 7) days and a period of (0.998 $\pm$ 0.002) years, all
well in agreement with those expected for dark-matter particles \cite{Bernabei:2013xsa}.
CoGeNT at the Soudan Underground Laboratory claims a 2$\sigma$ (now somewhat less) dark-matter excess and annual modulation \cite{Aalseth:2012if}, and CDMS-II at SNOLAB observes
three unexplained low-energy events in their Si-detector data sample. After their detector upgrade, CRESST-II no longer finds
a previously claimed excess \cite{Angloher:2011uu}
\cite{Angloher:2014myn}. These claims are shown in Fig.\ \ref{fig:si2014} (left) as colored
areas. They cluster in the mass region of tens of GeV and at
cross sections between 10$^{-42}$ and 10$^{-39}$ cm$^2$. All other direct detection
searches have set exclusion limits on the SI dark matter-nucleus cross section that
contradict the claims listed above. They are shown in Fig.\
\ref{fig:si2014} (left) as full lines. The currently best exclusion bounds come from
XENON100 (soon to be updated by XENON1T) \cite{Aprile:2012nq} and LUX
\cite{Akerib:2013tjd}. The signal claims also seem to be in conflict with the first PANDA-X
results \cite{Xiao:2014xyn}.

Asymmetric dark matter models (green ovals) have been
constructed with a view to resolve the apparent contradiction between claimed
signals and exclusion limits, whereas null results from the LHC have driven supersymmetric (big red circle) and
extra-dimensional models (blue oval) to high masses and low cross sections
(see below).
DARWIN is designed to probe the entire parameter region for dark-matter masses above
$\sim$6\,GeV, limited by the neutrino background (yellow region) \cite{Baudis:2014naa}. Experiments based on the mK cryogenic
technique, such as SuperCDMS \cite{Agnese:2014aze} and EURECA \cite{Kraus:2011zz}, have access
to lower WIMP masses.
The low-mass region is emphasized in Fig.\ \ref{fig:si2014} (right), where the
claims, including the area favored by CRESST-II \cite{Angloher:2011uu} (light blue), are shown together the exclusion curves. We include that from the CRESST-II low-threshold analysis
of a single upgraded detector module \cite{Angloher:2014myn} (red), that did not confirm the
previous excess.

\subsection{Spin-dependent cross sections}

Spin-dependent (SD) cross sections
\bea
 \sigma_N^{\rm SD}(E_R)&=&
 32\mu^2G_F^2[(J_N+1)/J_N][\la S_p\ra a_p+\la S_n\ra a_n]^2 F^2_{\rm SD}(E_R)
\eea
are due to axial vector coupling. They depend on the nuclear spin $J_N$, the dark-matter
coupling to protons and neutrons, $a_{p,n}$, the nuclear form factor $F_{\rm SD}(E_R)$ and
in particular on the expectation values $\la S_{p,n}\ra$ of the proton and neutron spin in
the target nucleus \cite{Ressell:1997kx,Bednyakov:2004xq,Toivanen:2009zza,Menendez:2012tm}.
Since these are of order unity,
SD cross sections are smaller than SI cross sections by a factor $A_N^2$, and the bounds on the
latter are thus considerably better than those on the former.

\begin{figure}[t]
 \includegraphics[width=0.5\textwidth]{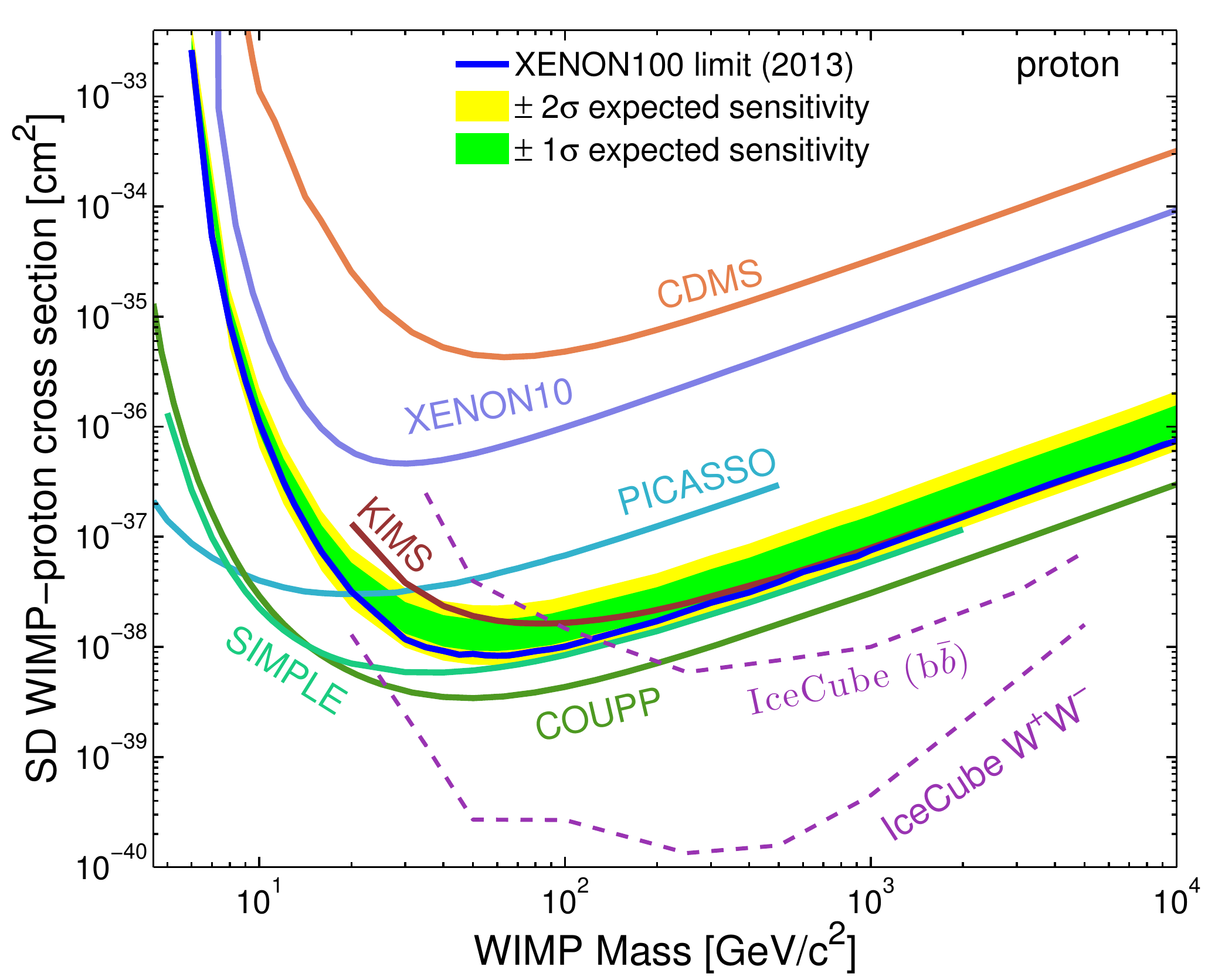}
 \includegraphics[width=0.5\textwidth]{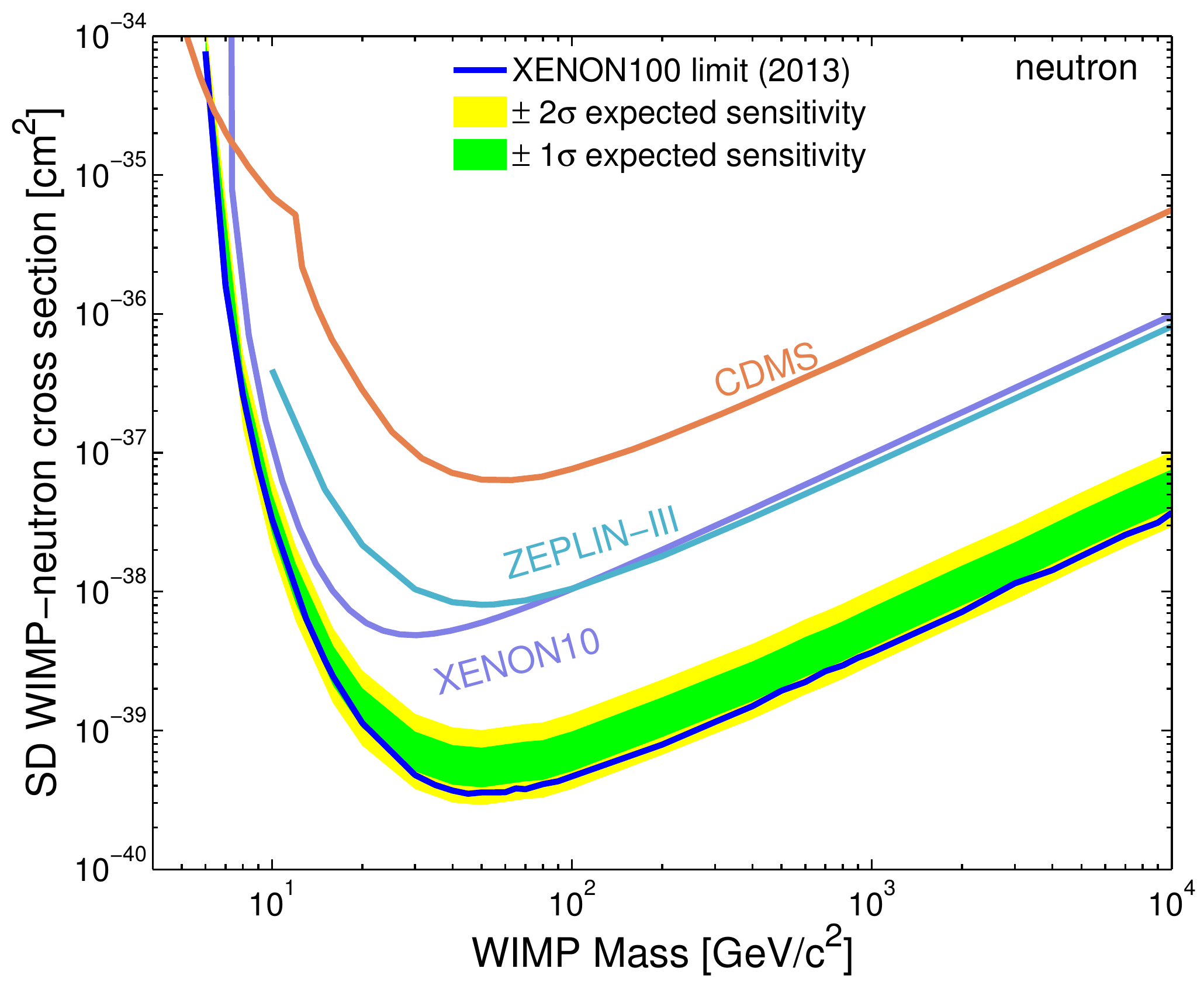}
 \caption{Spin-dependent WIMP-nucleon scattering results \cite{Aprile:2013doa}.
 XENON100 90\%\,C.L. upper limits on the cross sections on protons (left) and
 neutrons (right) based on Ref.\ \cite{Menendez:2012tm}
 (dark blue curve) are shown together with their 1$\sigma$ and 2$\sigma$ uncertainties on the
 expected sensitivity (green and yellow shaded bands) and results from XENON10
 \cite{Angle:2008we} based on Ref.\ \cite{Ressell:1997kx}, CDMS
 \cite{Ahmed:2008eu,Ahmed:2010wy}, ZEPLIN-III \cite{Akimov:2011tj} based on
 Ref.\ \cite{Toivanen:2009zza}, PICASSO \cite{Archambault:2012pm}, COUPP \cite{Behnke:2012ys},
 SIMPLE
 \cite{Felizardo:2011uw}, KIMS \cite{Kim:2012rza}, and results derived from indirect IceCube
 limits \cite{Aartsen:2012kia} in the hard ($W^+$$W^-$, $\tau^+\tau^-$) and soft ($b\bar{b}$)
 annihilation channels.}
 \label{fig:sd2013}
\end{figure}

To compare results obtained with different target materials, a common practice is to report
the cross section for the interaction with a single proton or neutron, assuming 
no coupling to the other type of nucleon. Calculations of the structure
functions, $S_{p,n}$, are traditionally based on the nuclear-shell model
\cite{Ressell:1997kx,Bednyakov:2004xq,Toivanen:2009zza}, but have
recently also been performed using chiral effective field theory, yielding much
better agreement between calculated and measured Xe spectra, both in terms of energy
and in ordering of the nuclear levels \cite{Menendez:2012tm}. The currently
best upper limits from XENON100 for protons and neutrons are shown as full
blue curves in Fig.\ \ref{fig:sd2013} (left and right) with their
1$\sigma$ and 2$\sigma$ uncertainty bands (green and yellow), together with results from other competing
experiments. The sensitivity to proton couplings is weaker, since both Xe isotopes
($^{129}$Xe and $^{131}$Xe) have an unpaired neutron, but an even number of protons
leading to $|\la S_p\ra|\ll|\la S_n\ra|$. The IceCube detector at the south pole
has searched for muon neutrinos from dark-matter annihilation at the center of
the Sun. This indirect-detection mode will be discussed in Sect.~\ref{sec:4.5}.
If one assumes a particular dark-matter candidate, such as the lightest neutralino of supersymmetry \cite{Wikstrom:2009kw}, the
indirect-detection limits from heavy quark ($b$), lepton ($\tau$) and weak gauge
boson ($W^\pm$) annihilation products can be translated into competitive
SD direct-detection limits \cite{Aartsen:2012kia}.

\subsection{Light dark matter}

As discussed in Sect.\ \ref{sec:2}, axions were introduced by Peccei and Quinn as a
solution of the strong $CP$ problem in the form of a pseudo-Nambu-Goldstone boson emerging from
the breaking of a global U(1) symmetry \cite{Peccei:1977hh}. Although the original model has been
ruled out, invisible axions arising from a higher symmetry-breaking scale are still allowed,
as explicitly demonstrated in the DFSZ and KSVZ models \cite{Dine:1981rt,Zhitnitsky:1980tq,%
Kim:1979if,Shifman:1979if}.

The Sun is believed to be an intense source of axions through
production via bremsstrahlung, Compton scattering, axio-recombination and axio-deexcitation
which are standard electrodynamic processes, in which the final-state photon is replaced by an
axion~\cite{Redondo:2013wwa}. Conversely, axion-like particles (ALPs) could have been
generated via a non-thermal production mechanism in the early universe, so that they would
now constitute the dark matter in our galaxy. Both types of particles give rise to
direct observable signatures through their coupling to photons ($g_{A\gamma}$), electrons
($g_{Ae}$), and nuclei ($g_{AN}$).

\begin{figure}[t]
 \includegraphics[width=0.54\textwidth]{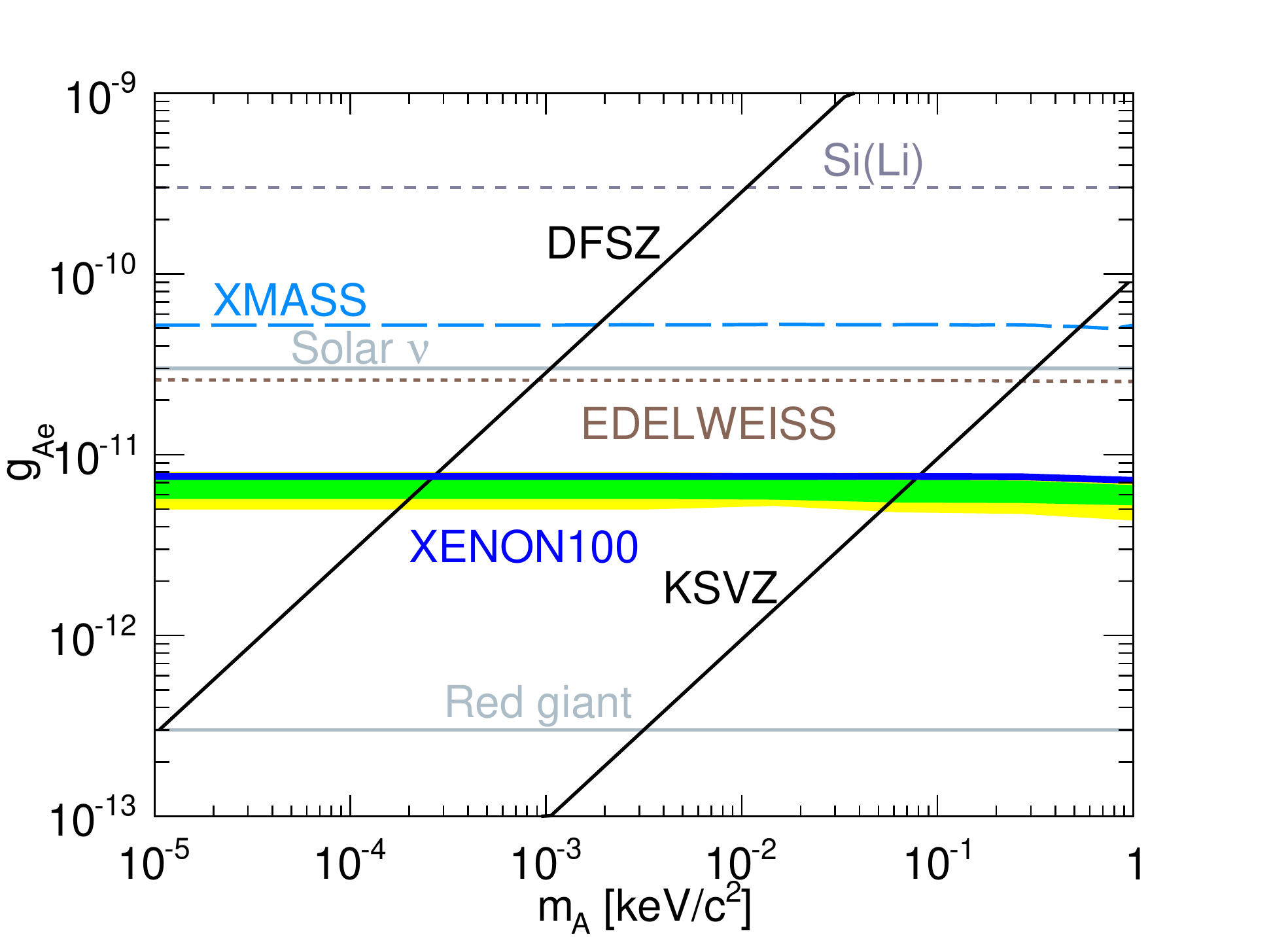}\hspace*{-5mm}
 \includegraphics[width=0.54\textwidth]{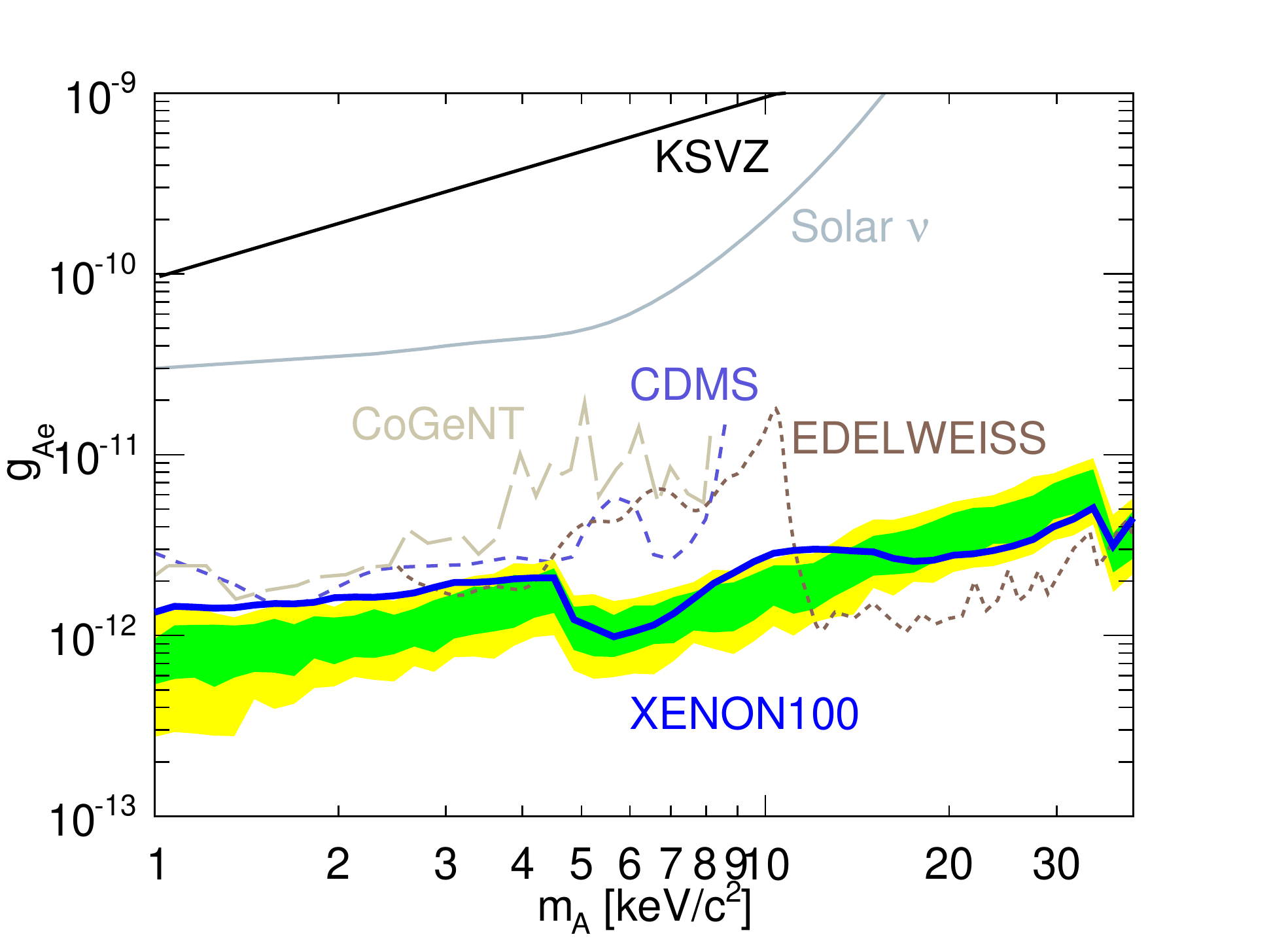}
 \caption{Left: Current limits on solar axions  \cite{Aprile:2014eoa}.
 The XENON100 limits (90\% C.L., blue line)
 are shown together with the expected sensitivity at 1$\sigma$ and 2$\sigma$ based on the
 background hypothesis (green and yellow bands). Also shown are
 limits by EDELWEISS-II~\cite{Armengaud:2013rta} and XMASS~\cite{Abe:2012ut}
 together with those from a Si(Li) detector \cite{Derbin:2012yk}.
 Indirect astrophysical bounds from solar neutrinos~\cite{Gondolo:2008dd} and red
 giants~\cite{Viaux:2013lha} are represented by dashed lines. The benchmark DFSZ and KSVZ models
 are represented by black solid lines \cite{Dine:1981rt,Zhitnitsky:1980tq,Kim:1979if,%
Shifman:1979if}.
 Right: Current limits on galactic axions \cite{Aprile:2014eoa}.
 The XENON100 limit (90\% C.L.) on ALPs coupling to electrons as a function of the
 mass is shown under the assumption that ALPs constitute all the dark matter in our galaxy
 (blue line). The expected sensitivity is shown by the green/yellow bands ($1 \sigma
 / 2 \sigma$). The other curves are constraints set by CoGeNT~\cite{Aalseth:2008rx}
 (brown dashed line), CDMS~\cite{Ahmed:2009ht} (grey continuous line), and
 EDELWEISS-II~\cite{Armengaud:2013rta} (red line, extending up to 40 keV).
 Indirect astrophysical bounds from solar neutrinos~\cite{Gondolo:2008dd} are
 represented as a dashed line. The benchmark KSVZ model is represented by a
 black solid line~\cite{Kim:1979if,Shifman:1979if}.}
 \label{fig:as2014}
\end{figure}

The coupling $g_{Ae}$ may be tested via scattering
of electrons through the axio-electric effect \cite{Dimopoulos:1985tm,Avignone:1986vm,%
Pospelov:2008jk,Derevianko:2010kz,Arisaka:2012pb}.
Fig.~\ref{fig:as2014} (left) shows the current exclusion limits on $g_{Ae}$.
The XENON100 results (90\% C.L., blue line)
are shown together with the expected sensitivity at 1$\sigma$ and 2$\sigma$ based on the
background hypothesis (green and yellow bands).
The range of validity of the light solar-axion flux limits the search to $m_A < 1$ keV
\cite{Redondo:2013wwa}. For comparison, other recent direct search limits and astrophysical
bounds are also shown, together with the theoretical benchmark models DFSZ and KSVZ. For solar
axions with masses below 1 keV, the electron coupling can be limited by the XENON100 data
to $g_{Ae}<7.7 \times 10^{-12}$ at 90\% C.L. Within the DFSZ and KSVZ models, this limit excludes
axion masses above 0.3 eV and 80 eV, respectively. For comparison,
the CAST experiment has tested the coupling
to photons $g_{A\gamma}$ and excluded axions within the KSVZ model in the mass range between
0.64 eV and 1.17 eV \cite{Arik:2013nya}.

The current exclusion limits on galactic ALPs from XENON100, competing
experiments and astrophysical bounds are shown in Fig.\ \ref{fig:as2014} (right)
together with the KSVZ benchmark model. The steps in the sensitivity around 5 and 35
keV reflect jumps in the photoelectric cross section reflecting atomic energy levels.
In the 5-10 keV mass range, XENON100 sets the best upper limit, excluding an axion-electron
coupling $g_{Ae}>1 \times 10^{-12}$ at the 90\% C.L., if one assumes that ALPs constitute all
of the galactic dark matter. This case has been discussed in Sect.\ \ref{sec:2.2.3}.

%% file: section4.tex
\section{Indirect searches}
\label{sec:4}
\subsection{General facts on indirect detection of WIMPs}
\label{sec:4.1}
Dark matter cannot only be detected directly in dedicated experiments searching for nuclear
recoils from the scattering of dark-matter particles or produced in particle accelerators such
as the LHC, but it can also reveal its existence indirectly. The total number of dark-matter
particles does not change significantly after freeze-out in the early universe, but their
spatial distribution changes considerably during structure formation. The very self-annihilation
that plays a central role in this freeze-out can give rise to a significant flux of $\gamma$-rays,
neutrinos, and even antimatter such as antiprotons and positrons, especially in regions with
large dark-matter density. The energy of the secondary particles can reach up to the
mass of the dark-matter particle, which typically would be a few hundred GeV. 

Positrons would eventually annihilate with the electrons of the interstellar plasma and give
rise to 511-keV line emission. In fact, such a line has already been observed with
INTEGRAL~\cite{Jean:2003ci}, in particular from the galactic center, but uncertainties remain
as positrons are copiously produced by $\beta$ decay of unstable isotopes~\cite{integral2010}.
Secondary electrons and positrons can give rise to synchrotron radiation that may be detected
in the radio band. Therefore, detectors for cosmic rays and $\gamma$-rays, neutrino telescopes,
and radio telescopes can be used for indirect dark-matter detection. 

Since dark-matter annihilation scales with the square of its density, indirect detection is
more sensitive to cosmological and astrophysical processes than is direct detection. Such
processes include the appearance of cusps around the central super-massive black holes in our
own as well as other galaxies and the formation of clumps that condense around intermediate-mass
black holes or that arise from initial density fluctuations.

As the fluxes and spectra of ordinary cosmic rays are modified during propagation to Earth, so
are those of the annihilation and decay products of dark matter. Charged particles undergo
diffusion in the galactic magnetic field, whereas photons and neutrinos
will propagate on straight trajectories and can therefore reveal the
spatial distribution of dark matter. Eventually, this might lead to a measurement of the
small-scale structure of dark matter, thus permitting an experimental test of our theoretical
understanding of the formation and survival of cusps and clumps.

The annihilation yield of a dark-matter particle $X$ is usually calculated under the assumption
that the particle in question is of Majorana type, i.e. it is its own antiparticle, $\bar X$.
Otherwise there would be an additional factor $1/2$. Then, the flux of neutral secondaries can
be written as
\begin{equation}
\Phi_i({\bf n},E)=\langle\sigma_{X\bar X}v\rangle \frac{dN_i}{dE} \frac{1}{8 \pi\, m_X^2}
\int_{{\mbox{l.o.s.}}}dl\ 
\rho_X^2\left[{\bf r}(l,{\bf n})\right],\label{eq:dm_flux}
\end{equation}
where the index $i$ denotes the secondary particle observed (we focus on $\gamma$-rays and
neutrinos) and the integration variable $l$ refers to the path length along the line of sight
(l.o.s.) in direction ${\bf n}$. Furthermore,  
$dN_i/dE$ is the multiplicity spectrum of secondary particle $i$, $m_X$
is the dark matter particle mass, and $\rho_X[{\bf r}(l,{\bf n})]$ is the dark-matter density
along the line of sight. 
One often assumes that the annihilation
cross section averaged over the dark-matter velocity distribution, $\langle\sigma_{X\bar X}v\rangle$, is the same as that relevant for thermal relic freeze-out, i.e. Eq.~\ref{eq1}. Note, however, that the particle velocities
relevant for indirect detection are very different from those pertaining to thermal freeze-out:
the typical WIMP\index{WIMP} velocity in the Milky Way is $v\simeq300\ {\rm km}\,{\rm s}^{-1}$,
whereas freeze-out happened when $T_f\simeq m_X/20$, implying  $v\simeq0.4\,c$.
Therefore, if the WIMP annihilation cross section has a strong velocity dependence, as is possible if
$S$-waves\index{s-wave} are suppressed and higher partial waves dominate, then constraints from freeze-out
and from indirect detection may not be directly comparable.

To separate the factors depending on the dark-matter profile 
from those depending only on particle physics, we introduce,
following~\cite{Bergstrom:1997fj}, the quantity
\begin{equation}
J\left({\bf n}\right) = \frac{1} {8.5\, \rm{kpc}} 
\left(\frac{1}{0.3\,{\rm GeV}\,{\rm cm}^{-3}}\right)^2
\int_{\mbox{l.o.s.}}dl\ \rho_X^2\left[{\bf r}(l,{\bf n})\right]\,,
\end{equation}
where we used galactic scales and $\rho_X\simeq0.3\,{\rm GeV}{\rm cm}^{-3}$ to render $J$ dimensionless. Note that various definitions of $J$ can be found in the literature.
We define $\overline{J}({\bf n},\Delta\Omega)$ as the average of $J({\bf n})$
over a circular region of solid angle $\Delta\Omega$, centered around ${\bf n}$.

The parameter $J$ describes the intensity distribution of neutral secondaries, whereas $\overline{J}$ is a measure of the flux from the region $\Delta\Omega$, given by
\begin{equation}
\Phi_{i}(\Delta\Omega, E)\simeq C\,\frac{dN_i}{dE} 
\left( \frac{\frac{1}{c}\langle\sigma_{X\bar X}v\rangle}
{\rm{pb}}\right)\left( \frac{1\,\rm{TeV}} 
{\rm{m_X}}\right)^2 \overline{J}\left(\Delta\Omega\right)
\,\Delta\Omega
\label{eq:dm_flux2}
\end{equation}
with $C=2.75\times10^{-12}\ \rm{cm}^{-2} \rm{s}^{-1}$.

The enhancement of annihilation rates due to small-scale structure
in the form of dark-matter clumps (cf. Eq.~\ref{eq2a}) is often expressed in terms of a so-called
boosting factor, $B\simeq\langle\rho^2\rangle/\langle\rho\rangle^2$, that measures the
annihilation yield relative to that computed using a smooth dark-matter distribution.

Another type of signal amplification can arise, if dark-matter interactions involve as a force carrier a new boson, $\phi$, of sufficiently low mass, $M_\phi\lesssim 10$~GeV. Originally envisioned by Sommerfeld in a different context \cite{1931AnP...403..257S}, the wave function of the incoming particle, conventionally approximated as a plane wave, can be significantly distorted, leading to strong upward or downward variations in the cross section. If the parameters are chosen appropriately, this so-called \emph{Sommerfeld enhancement} can lead to a substantial increase in the annihilation cross section in galaxies and clusters, while leaving the arguments about thermal freeze-out in the early universe intact \cite{2009PhRvD..79a5014A}. Furthermore, certain lowest-order cross sections
can be suppressed by symmetries or helicity effects, such as the annihilation of two non-relativistic Majorana fermions into a relativistic fermion/anti-fermion
pair. In this case \emph{internal bremsstrahlung}, the attachment of an additional photon to one of the charged final-state fermions,
can actually enhance the annihilation cross section, because the
larger available phase space alleviates suppression arising from symmetry or helicity effects.

If, instead of annihilation, dark matter dominantly decays with lifetime $\tau_X$
and a multiplicity spectrum of secondary particles $dN_i/dE$, the flux observed on
Earth would read
\begin{equation}
\Phi_i({\bf n},E)=\frac{dN_i}{dE} \frac{1}{4 \pi \tau_Xm_X}
\int_{\mbox{l.o.s.}}dl\ 
\rho_X\left[{\bf r}(l,{\bf n})\right]\,.\label{eq:dm_flux_decay}
\end{equation}
Note that in this case there is no boosting, because the flux depends only linearly on the dark-matter density.

\subsection{Photons}
\label{sec:4.2}
Let us begin with a possible signature of dark matter in the X-ray sky. Recent analyses of the X-ray spectrum observed toward the Andromeda galaxy, the Perseus galaxy cluster, and various other sources stacked together indicated a
3.55-keV line that is not expected as a thermal emission of an astrophysical plasma \cite{Bulbul:2014sua,Boyarsky:2014jta}. If interpreted in terms
of decaying sterile neutrinos, one derives a neutrino mass $m_{\nu_s}\simeq7.1\,$keV and a mixing
angle with active neutrinos $\sin^2\theta_s=(5.5-50)\times10^{-12}$. In Fig.~\ref{fig:st2014} of Sect.~\ref{sec:2} these values are indicated by a red square and found not in conflict with other constraints.

The detection of an X-ray line and its interpretation is the subject of an ongoing debate, though. The Suzaku satellite finds a line in the X-ray emission of the Perseus galaxy cluster, but not for other
galaxy clusters~\cite{Urban:2014yda}. Concerns have been raised that the intensity distribution of line emission doesn't correspond to that expected for dark-matter decay \cite{Carlson:2014lla}.
No X-ray lines are observed from dwarf galaxies~\cite{Malyshev:2014xqa}, for which confusion with thermal line emission should be absent, and other galaxies~\cite{Anderson:2014tza}.
Several theoretical pre- and postdictions are related to this possible X-ray line, involving axion-like particles~\cite{Cicoli:2014bfa},
axion-photon conversions~\cite{Conlon:2014xsa}, that we will discuss in general in Sect.~\ref{sec:4.6}, and scenarios
in which inelastic scattering of dark matter leads to an excited state that subsequently decays \cite{Cline:2014vsa}.
Among alternative astrophysical explanations, the well-known atomic lines K XVIII or Cl XVII have been proposed \cite{Jeltema:2014qfa} and criticized on account of large
systematic uncertainties \cite{Boyarsky:2014paa} or wrong transition probabilities in the case of Cl XVII \cite{Bulbul:2014ala}.

\begin{figure}[ht]
\includegraphics[width=\textwidth,clip=true,angle=0]{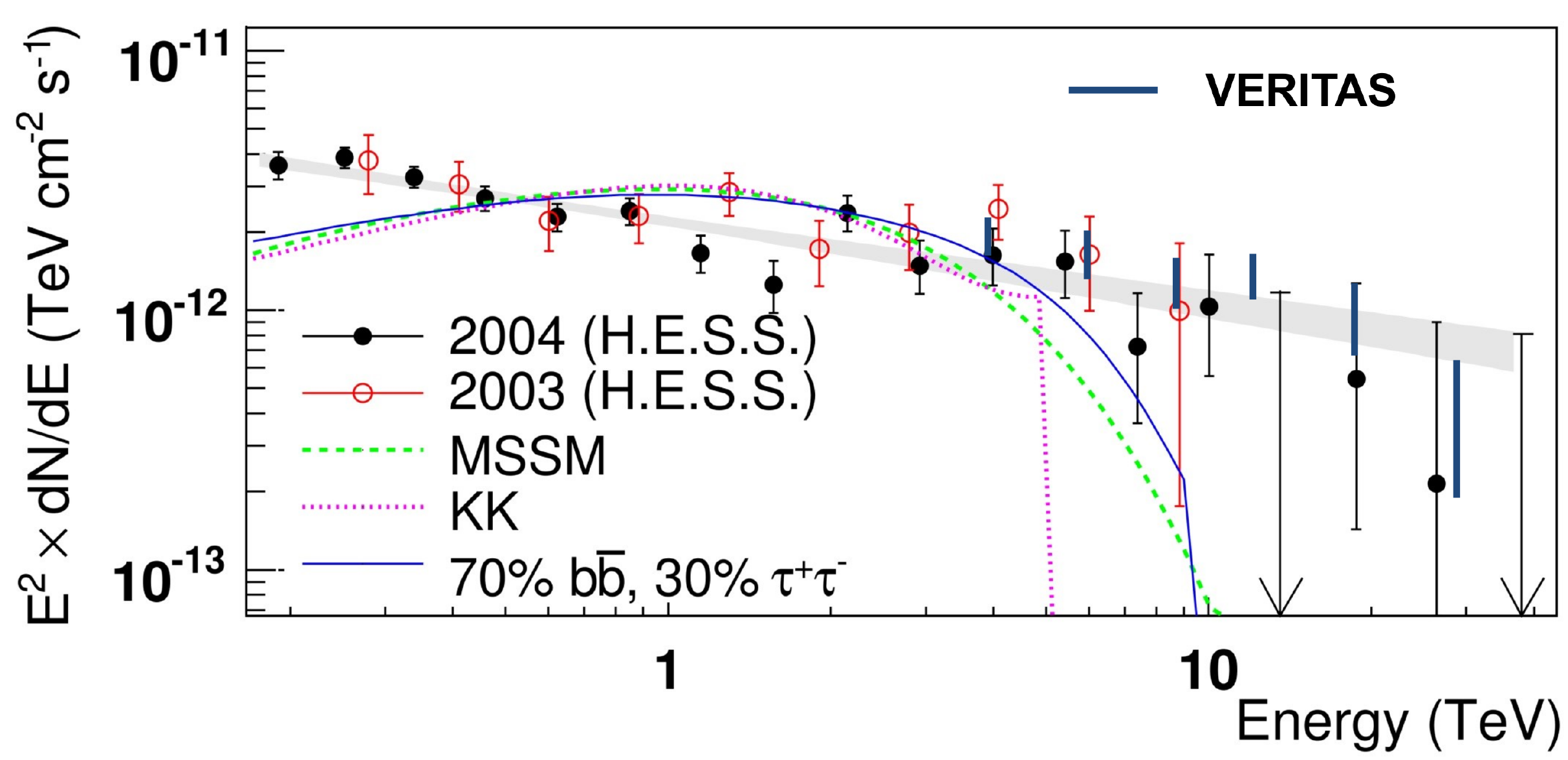}
\caption[...]{Spectral energy density, $E^2\times \mathrm{d}N/\mathrm{d}E$, of $\gamma$-rays
from the galactic-center source seen by H.E.S.S. \cite{Aharonian:2006wh}, complemented with data points obtained with VERITAS \cite{2014ApJ...790..149A}.
Upper limits are 95\% CL. The shaded area shows a power-law fit, $\mathrm{d}N/\mathrm{d}E \sim E^{-\Gamma}$.
The dashed line illustrates typical spectra of phenomenological minimal supersymmetric
standard-model annihilation for neutralino masses of 14~TeV.
The dotted line shows the distribution predicted for Kaluza-Klein dark matter \index{Kaluza-Klein dark matter}
with a mass of 5~TeV. The solid line gives the spectrum of a 10~TeV dark-matter
particle annihilating into $\tau^+\tau^-$ (30\%) and $b\bar{b}$ (70\%).}
\label{fig:hess-dm}
\end{figure}

\begin{figure}[ht]
\includegraphics[width=0.42\textwidth]{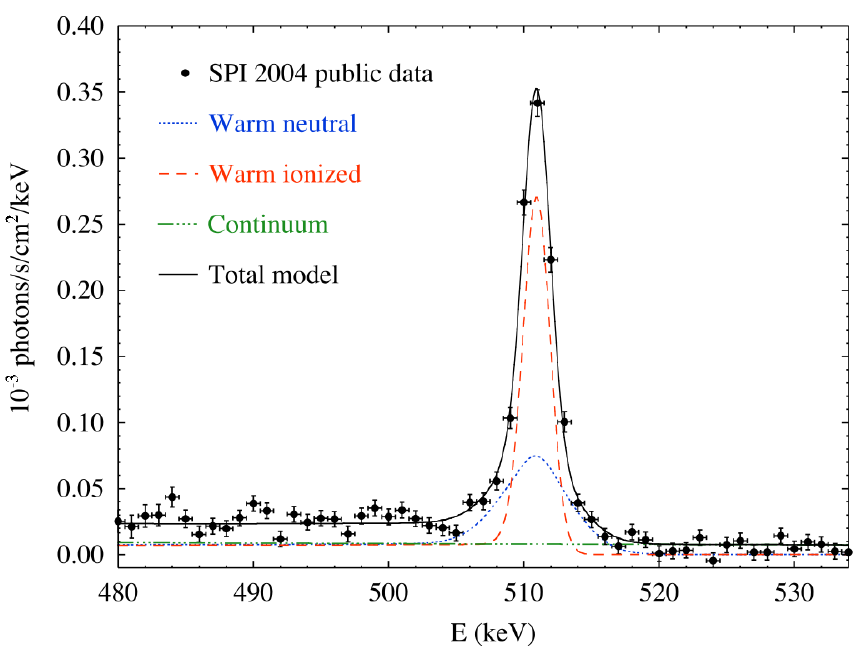}
\includegraphics[width=0.55\textwidth]{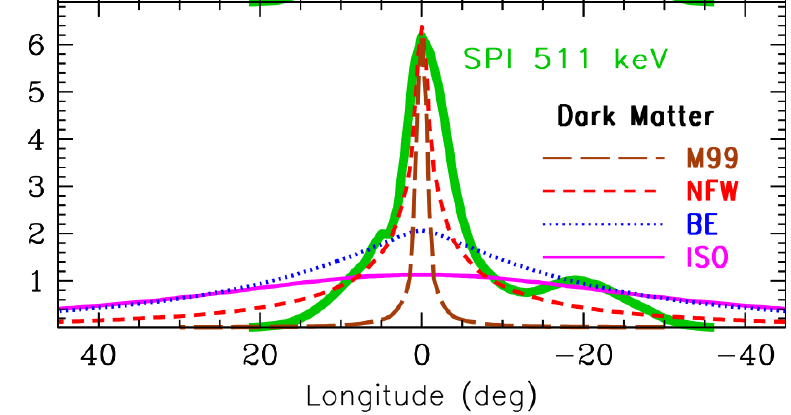}
\caption{The 511-keV emission seen with INTEGRAL/SPI. Left panel: The spectrum from the inner galaxy ($b<15^\circ$) compared to fits to a combination
of astrophysical components. Right panel: Longitudinal
distribution of line flux compared to predictions for dark-matter annihilation for various density profiles (right panel). Taken from
Ref.~\cite{integral2010}.}
\label{fig:integral-dm}
\end{figure}

Let us now turn to $\gamma$ rays. Indirect detection of WIMPs requires disentangling the flux of dark-matter annihilation products from that of conventional astrophysical contributions. Several observations have been interpreted as possible dark-matter annihilation signatures. These
include TeV-band $\gamma$-ray emission~\cite{Aharonian:2006wh} from the galactic center,
described in Fig.~\ref{fig:hess-dm}, and the INTEGRAL observation of a 511-keV $\gamma$-ray line from the galactic-center region~\cite{Jean:2003ci}, elucidated in Fig.~\ref{fig:integral-dm}.

However, with more and newer data, these signatures seem to find convincing explanations involving normal
astrophysical processes: The TeV $\gamma$-rays seen by H.E.S.S. and VERITAS \cite{2014ApJ...790..149A} display no spectral structure up to $>30\,$TeV which would
require an unnaturally heavy dark-matter primary. Instead, the spectrum appears compatible with predictions based on 
an accelerated primary cosmic-ray component~\cite{Aharonian:2009zk}. In addition, even for the cuspy NFW profile, the required
velocity-weighted annihilation cross section into $\gamma$ rays would be of the order $3\times 10^{-24}\,\mathrm{cm^3\,s^{-1}}$, higher by a factor $100$ than that needed to produce
the correct thermal relic abundance, cf. Eq.~(\ref{eq1}). Whatever the interpretation, the measurement provides an upper limit to the annihilation cross section~\cite{HESS:2015cda}. The 511-keV intensity distribution seen with INTEGRAL does not show the
spherical symmetry expected for the central region of a dark-matter halo, but looks more like the galactic bulge~\cite{Weidenspointner:2008zz}.

\begin{figure}[ht]
\includegraphics[width=0.47\textwidth,clip=true]{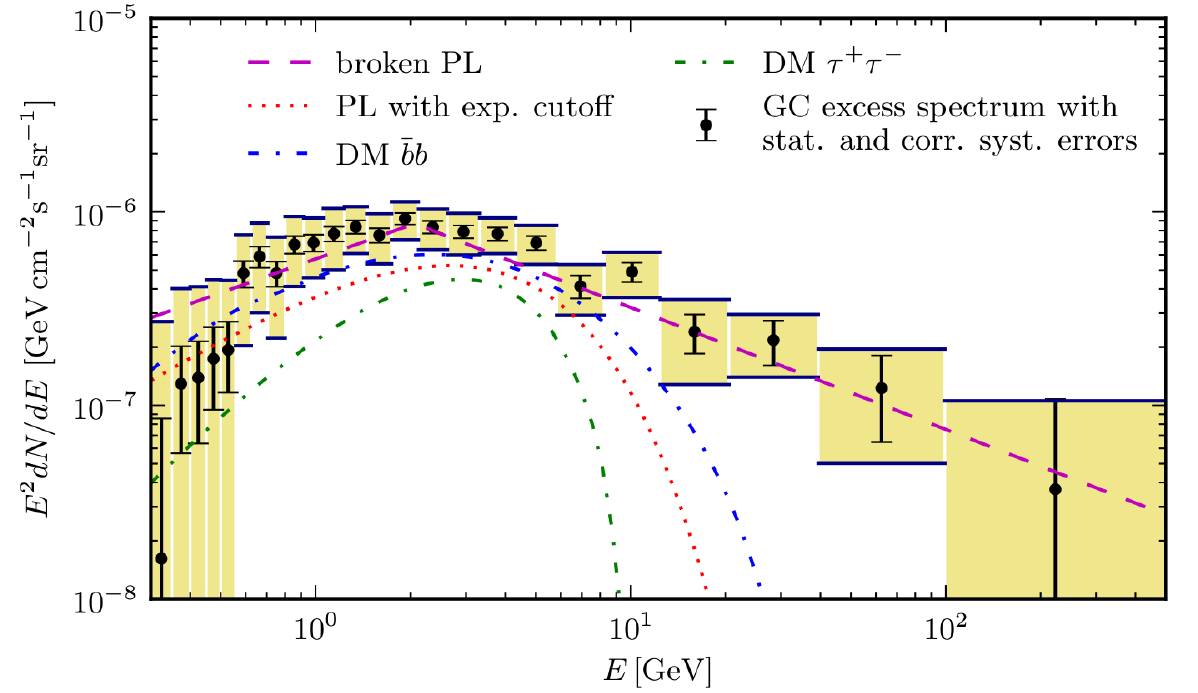}
\includegraphics[width=0.5\textwidth,clip=true]{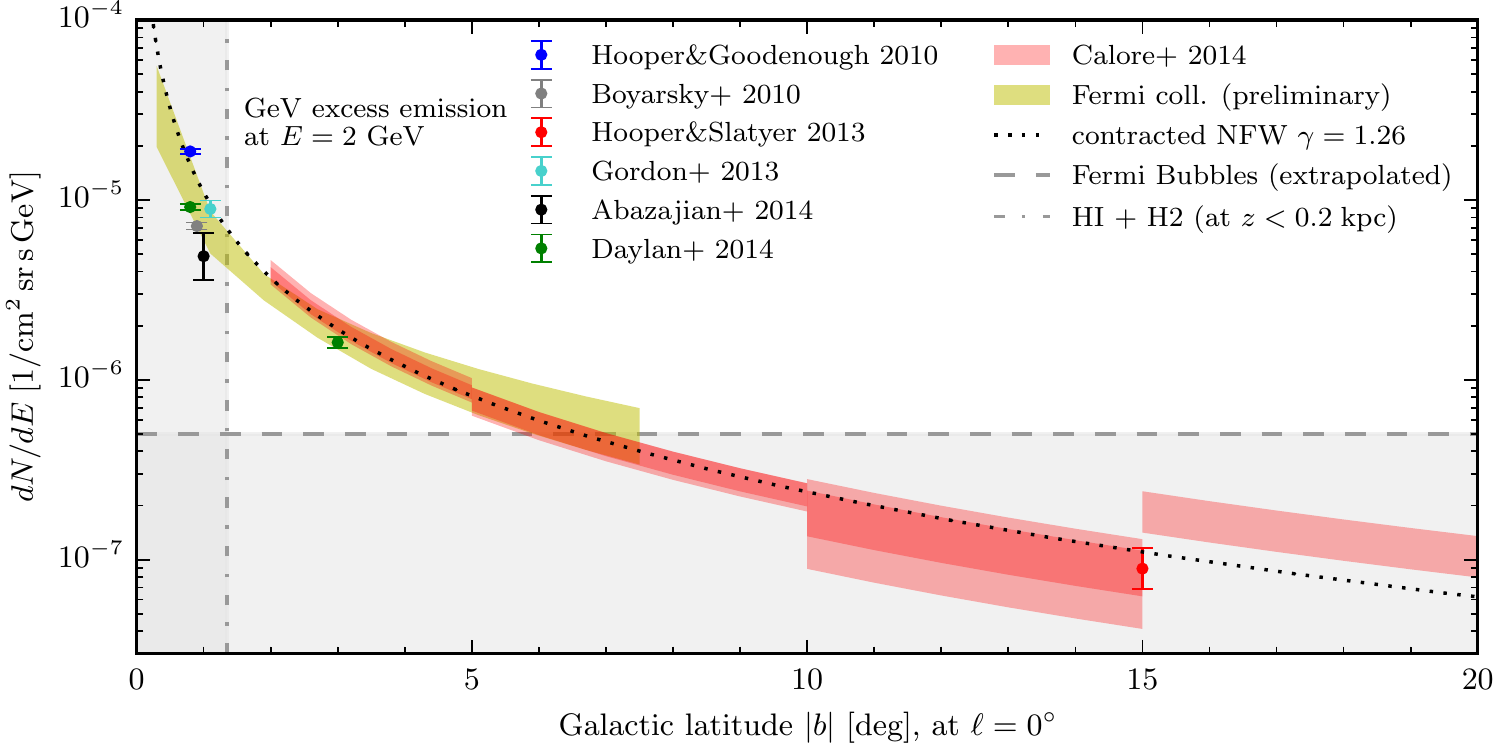}
\caption{The Fermi GeV $\gamma$-ray excess.
Left panel: The residual $\gamma$-ray spectrum after subtraction of astrophysical $\gamma$-ray emission averaged over the region $2^\circ\leq|b|\leq20^\circ$ and
$|l|\leq20^\circ$.
Also shown are best-fit model spectra of two dark-matter annihilation channels.
From Ref.~\cite{Calore:2014xka}. Right panel: Intensity profile of the excess at $E_\gamma=2\,$GeV compared
with a contracted NFW profile $\propto r^{-1.26}$. From Ref.~\cite{Calore:2014nla}.}
\label{fig:Fermi-excess}
\end{figure}

In recent years, the LAT detector aboard the Fermi satellite\footnote{\url{http://www-glast.stanford.edu/} and \url{http://fermi.gsfc.nasa.gov/}} has surveyed the
$\gamma$-ray emission from the inner galaxy. One result is evidence of excess emission extending up to $40^\circ$ above and
below the galactic plane, that has a spectrum roughly following an $E^{-2}$ power law,
significantly harder than that expected of emission from the galactic disk~\cite{Su:2010qj}. This structure is colloquially known as ``Fermi bubbles'' and roughly matches a radio
counterpart found earlier, the so-called ``WMAP haze''~\cite{Dobler:2009xz}
that was recently confirmed by the Planck experiment~\cite{Planck:2012rta}. The WMAP haze has a smaller vertical extent than the Fermi bubbles, possibly on account of the difference in the radiative energy-loss time of synchrotron-radiating and inverse-Compton scattering electrons. Whereas the Fermi bubbles may be interpreted in terms of outflows triggered by past black-hole activity or a central starburst, a second $\gamma$-ray excess on smaller scales has been claimed that may be related to dark matter~\cite{Hooper:2011ti}. In Fig.~\ref{fig:Fermi-excess} we show the spectrum and latitudinal distribution of this GeV excess.
If it is interpreted in terms of dark matter, one would require a particle  mass in the range of $7-12\,$GeV for leptonic channels or $25-45\,$GeV for hadronic final states. The cross section would be roughly consistent~\cite{Hooper:2013rwa} with that estimated from
thermal freeze-out, cf. Eq.~(\ref{eq1}), although its value would be uncertain by up to a factor 50 on account of the poorly known density profile of dark matter. In any case, conventional astrophysical scenarios cannot be excluded, and so one can, strictly speaking, only place an upper limit on the cross section of dark-matter annihilation.

A word on $\gamma$-ray producing annihilation channels is in order. Depending on the nature of the initial annihilation products, two processes are to be dinstinguished.
First, a pair of quarks may be created either directly or through electroweak gauge boson pairs. The quarks subsequently evolve into, e.g.,
neutral pions that decay into $\gamma$ rays. Second, charged annihilation products should emit bremsstrahlung. Both lead to a broad energy distribution of $\gamma$ rays that is easy to confuse with contributions from other astrophysical processes. Truly incontrovertible evidence for dark matter would be a $\gamma$-ray line because we do not know an astrophysical process that would produce a line at energies $\gtrsim 1$~GeV. Unfortunately, the process $X\bar{X}\rightarrow \gamma\gamma$ is not possible with tree-level processes
because dark matter is neutral, and it requires a loop of virtual charged particles. 

As CDM WIMPs are by definition slow, annihilation into two photons will inevitably give a line at $E_\gamma=m_X$, and so one would automatically measure the particle mass as well. The channel $X\bar X\to\gamma Z,H$ would also lead to a narrow $\gamma$-ray line, albeit at the energy $E_\gamma=m_X\left[1-m_{Z,H}^2/(4m_X^2)\right]$. Recent reports of a
$\gamma$-ray line around 130 GeV~\cite{Weniger:2012tx,Bringmann:2012ez} caused substantial excitement, but subsequent
analysis suggested that the feature is likely caused by either a statistical or an instrumental effect~\cite{Gustafsson:2013fca}.

Among all targets of $\gamma$-ray observations, dwarf spheroidal galaxies are of particular interest, because they are dark-matter dominated and thus less subject to uncertainties
from astrophysical flux contributions. Upper limits have been derived by both
ground- and space-based $\gamma$-ray detectors \cite{2012PhRvD..85f2001A,Abramowski:2014tra,2014JCAP...02..008A,Drlica-Wagner:2015xua}.
The coincidence of a $\gamma$-ray source with
the dwarf galaxy Reticulum 2 is now subject of further investigation~\cite{Geringer-Sameth:2015lua}. Other constraints have been derived from the galactic halo and subhaloes, galaxy clusters (for a recent summary see, e.g., \cite{Buckley:2013bha}), and from
the extragalactic $\gamma$-ray background~\cite{Ajello:2015mfa}.
For $m_X\lesssim 100\,$GeV, $\gamma$-ray observations challenge the thermal freeze-out scenario, cf. Eq.~\ref{eq1}, and the thermal-relic model is excluded for particle masses below 
$m_X\approx10\,$GeV. The upcoming CTA experiment\footnote{\url{http://www.cta-observatory.org/}} promises to provide the sensitivity to test the freeze-out model up to the TeV-scale.

\subsection{Electromagnetic cascades and their effects on the CMB}
\label{sec:4.3}

\begin{figure}[t]
 \centering
 \includegraphics[width=0.75\textwidth]{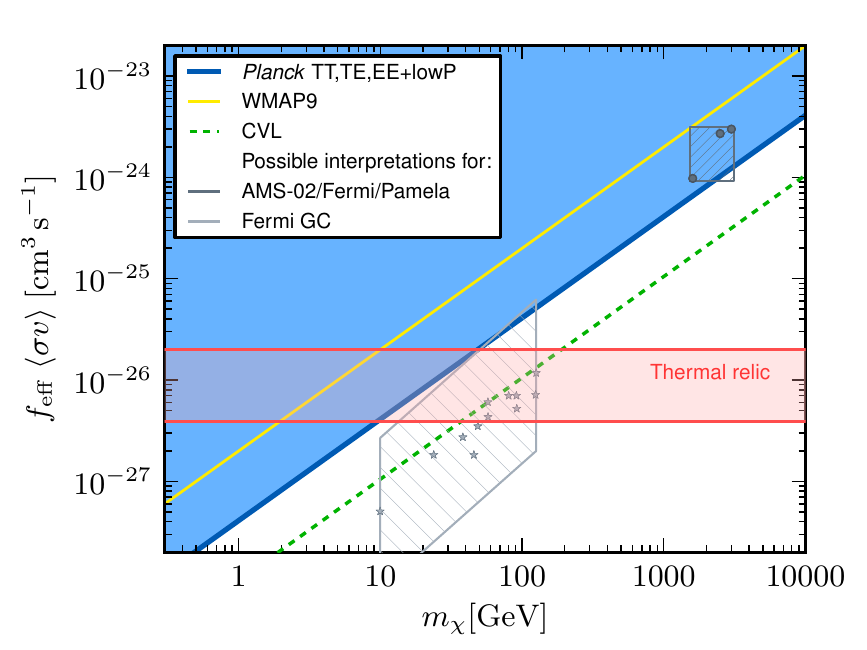}
 \caption{Constraints on the self-annihilation cross section at the time of recombination,
 $\langle \sigma v \rangle_{z_\ast}$, times an efficiency parameter, $f_{\rm eff}$.
 The blue area shows the parameter space excluded by the
 TT{,} TE{,} EE, and low-P data of Planck at 95\,\% CL. The yellow line indicates
 the constraint using WMAP9 data. The dashed green line delineates the region ultimately
 accessible by a cosmic-variance-limited experiment with angular resolution
 comparable to that of Planck. The horizontal red band denotes the
 thermal-relic cross section scaled with $f_{\rm eff}$ for
 different annihilation channels. The dark grey circles show the best-fit
 DM models for the cosmic-ray positron excess~\cite{Cholis:2013psa}. The light
 grey stars refer to best-fit models for the galactic-center
 gamma-ray excess~\cite{Calore:2014nla}, with the light grey area indicating the astrophysical uncertainties. Taken from \cite{Planck:2015xua}.}
 \label{fig:pl2015}
\end{figure}

Dark-matter annihilation with a non-vanishing branching ratio
into the electromagnetic channel leads to distortions of the CMB which can be probed with data of, e.g., WMAP or Planck. The analysis is simplified, if the annihilation products are relativistic. In that case, the rate of change in comoving energy density can be estimated as
\begin{equation}\label{eq:sigma_CMB0}
  \frac{d}{dt}\left[\frac{\rho}{(1+z)^4}\right]=m_X\left\langle\sigma_{X\bar X}v\right\rangle\frac{n^2_X(z)}{(1+z)^4}=
  \Omega_m^2\rho_{c,0}^2\frac{\left\langle\sigma_{X\bar X}v\right\rangle}{m_X}(1+z)^2\,.
\end{equation}
CMB observations are sensitive to the injection of additional energy, and so the absence of signatures of additional energy sources provides an upper limit on the annihilation cross section, for which the conversion efficiency of turning energy into CMB distortions is parametrized by a factor $f_{\rm eff}(z)$. Results of such an analysis are summarized in Fig.~\ref{fig:pl2015}~\cite{Planck:2015xua}. To be noted from the figure is the exclusion of the thermal freeze-out scenario for low WIMP masses.
The dependence of the upper limits on the WIMP mass is linear, as it is for the $\gamma$-ray limits, and the advantage of CMB-based limits lies in the absence of
astrophysical uncertainties. The efficiency parameter $f_{\rm eff}(z)$ has a relatively small uncertainty.

If we assume that the thermal-relic scenario is essentially correct, then the combination of the corresponding cross section (cf. Eq.~\ref{eq1}) with the constraints displayed in Fig.~\ref{fig:pl2015} provide a lower limit on the WIMP mass \cite[e.g.][]{Steigman:2015hda},
\begin{equation}\label{eq:m_X_CMB}
  m_X\gtrsim m_{\rm CMB}= (100\,{\rm GeV})\,f_{\rm eff}(z_{\rm rec})\,.
\end{equation}
On the other hand, the same assumption combined with cross section unitarity, $\langle\sigma_{X\bar X} v\rangle\lesssim1/m_X^2$, implies $m_X\lesssim20\,$TeV. One would then conclude that thermal-relic WIMPs with significant branching to electromagnetic annihilation products must have a mass between a few tens of GeV and about $20\,$TeV.

If the WIMP ($X$) carries only part of the dark matter density, $\Omega_X<\Omega_m$, then the constraint must be adjusted. If we denote with $\langle\sigma_{\rm th}v\rangle$ the thermal-relic cross section
saturating $\Omega_m$, and $\langle\sigma_{\rm CMB}v\rangle$ is the CMB constraint (the thick blue line in Fig.~\ref{fig:pl2015}), then the
cross section limits
change to $\left\langle\sigma_{X\bar X}v\right\rangle\gtrsim\langle\sigma_{\rm th}v\rangle$ for $m_X\gtrsim m_{\rm CMB}$
and $\left\langle\sigma_{X\bar X}v\right\rangle\gtrsim\langle\sigma_{\rm th}v\rangle^2/\langle\sigma_{\rm CMB}v\rangle$
for $m_X\lesssim m_{\rm CMB}$. 

\subsection{Antimatter as a product of dark-matter interactions}
\label{sec:4.4}

\begin{figure}[ht]
\includegraphics[width=0.49\textwidth,clip=true,angle=0]{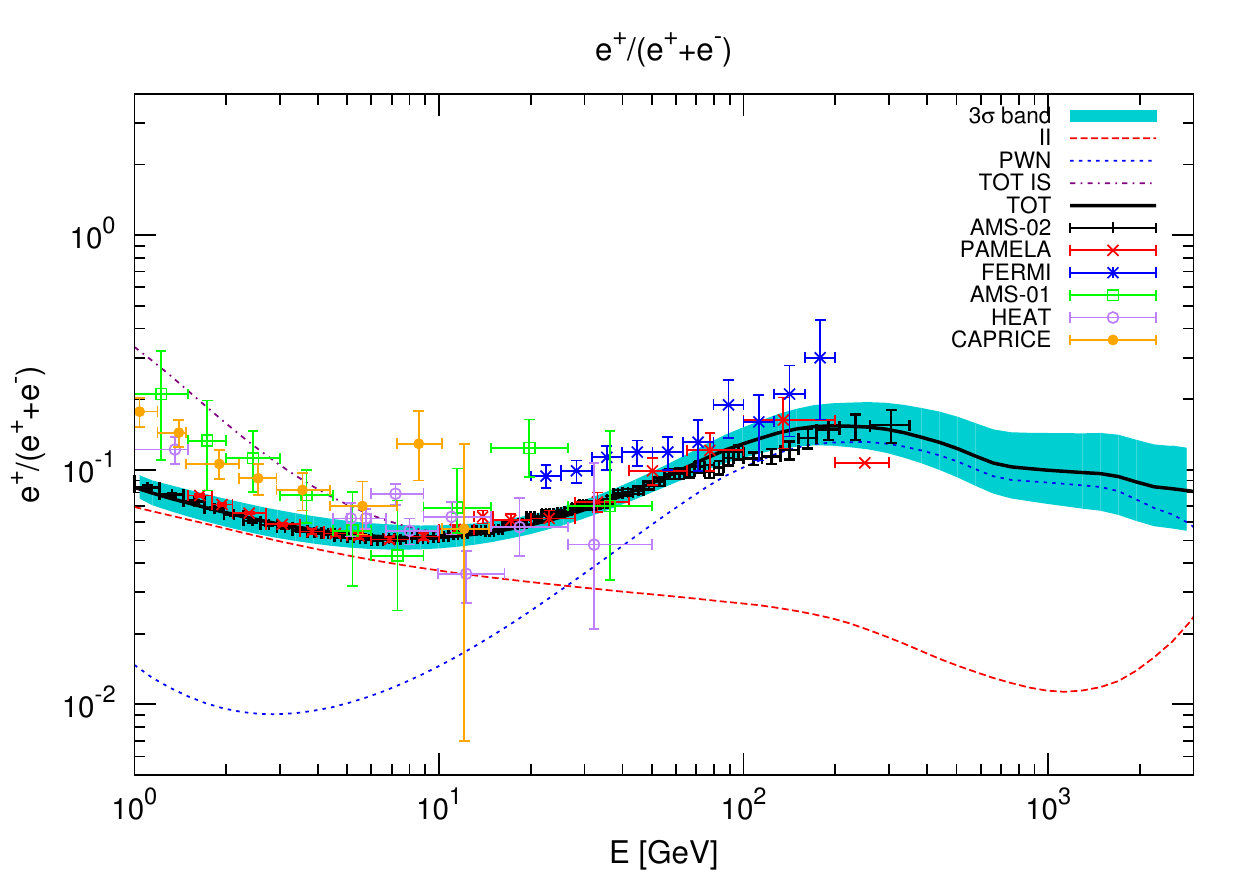}
\includegraphics[width=0.49\textwidth,clip=true,angle=0]{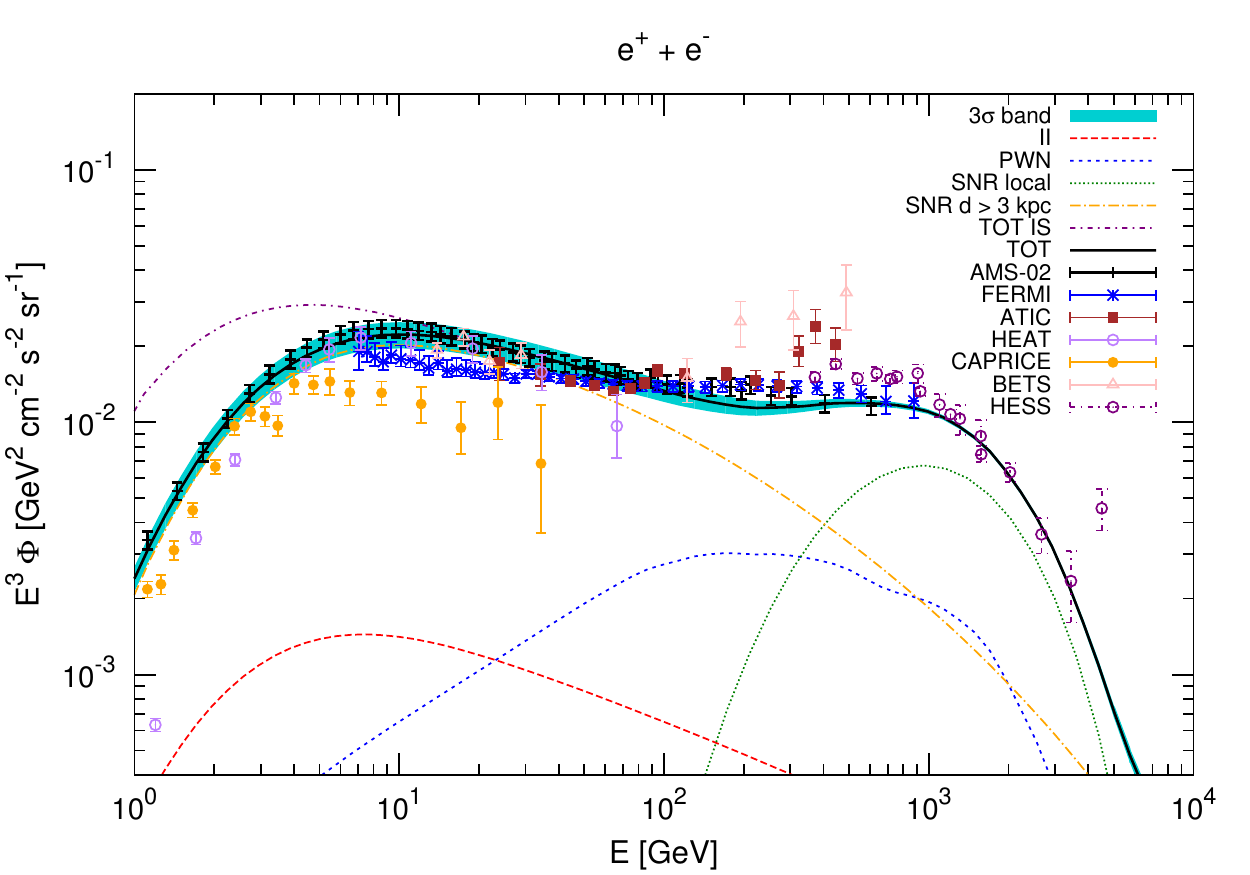}
\caption[...]{Left panel: Compilation of data on the GeV ``excess'' of galactic positrons. Expectations for interstellar production and speculations on, e.g., a contribution from pulsar-wind nebulae are added. 
Right panel: The electron plus positron flux as measured with various detectors. The shaded bands indicate the uncertainties arising from modeling contributions of primary electrons and positrons from supernova remnants, pulsar wind
nebulae, and from secondary electrons and positrons produced during propagation. Taken from Di Mauro et al. ~\cite{DiMauro:2014iia}.}
\label{fig:electrons}
\end{figure}

There is now strong evidence for an increase with energy of the cosmic-ray positron fraction, 
$N_e^+/(N_e^++N_e^-)$. In Fig.~\ref{fig:electrons} we summarize results obtained with the PAMELA satellite~\cite{Adriani:2008zr}, Fermi LAT~\cite{FermiLAT:2011ab}, and AMS-02~\cite{Aguilar:2013qda}, that together clearly demonstrate an excess between a few GeV and a few hundred GeV, compared to the inevitable contribution from cosmic-ray interactions in the interstellar medium.
This strengthened the case for a positron fraction in the electron plus positron flux that
is increasing between $\sim10\,$GeV and $\sim100\,$GeV (see Fig.~\ref{fig:electrons},
left panel). Conventional wisdom is that all positrons are secondary particles produced through collisions of cosmic rays with gas in the interstellar medium. As a function of energy, the interstellar contribution to the positron fraction should fall off. The observed positron excess therefore calls for additional sources of positrons, a candidate for which would be pulsars or pulsar-wind nebulae. Both primary and secondary positrons can be produced within the sources by electromagnetic
cascades induced by accelerated electrons~\cite{Hooper:2008kg}. Interactions of accelerated primary
particles~\cite{Blasi:2009hv} are possible, but imply similar structures in the spectra of secondary cosmic-ray elements~\cite{Mertsch}. In any case, 
fits of such astrophysical models can reproduce the rising positron fraction.

PAMELA has not
detected any significant enhancement of the antiproton flux beyond what is expected from interactions
of cosmic rays with the gas in our galaxy~\cite{Adriani:2008zq}. Any interpretation of the positron excess in terms of annihilating or decaying dark matter thus 
requires so-called leptophilic dark matter, i.e. predominantly leptons as end products. In addition, the thermal-relic cross section (cf. Eq.~\ref{eq1}) is too small by a factor $\gtrsim 100$ to provide the observed positron flux. Correspondingly strong boosting would be required by either dark-matter clumping or Sommerfeld enhancement. In an interpretation involving dark-matter decay, the limits on the particle lifetime would be weak.

The newest results by AMS-02~\cite{Accardo:2014lma,Aguilar:2014mma} extend up to 500 GeV in positron energy and show that the positron fraction levels off and remains approximately constant at
energies above 200 GeV. If the positron excess is indeed due to dark matter, a particle mass of a few hundred GeV would be most likely. The implied extragalactic photon emission from electrons and positron severely challenge both scenarios, dark-matter decay \cite{2010ApJ...712L..53P} and annihilation \cite{2009JCAP...07..020P}. Galactic photon emission constrains the excess production of electrons and positrons as well. Besides $\gamma$-ray emission, synchrotron radiation in the radio band provides constraints on the branching of dark-matter annihilation into electrons and positrons. In particular for the galactic-center region, they are comparable to upper limits based on $\gamma$-ray emission or possibly
stronger, depending on the scenario~\cite{Regis:2008ij}. Care must be exercised, though, because the synchrotron emissivity does not only depend on the uncertain density profile of dark matter in the innermost regions of the galaxy, as does the $\gamma$-ray emissivity, it  
also hinges on the poorly known magnetic-field strength
close to the galactic center.

In principle, individual anti-nucleons produced by dark-matter annihilation or decay can coalesce to
heavier anti-nuclei. Very little astrophysical background is expected
for mass number $A\geq3$. In addition, the background from cosmic-ray interactions should also quickly fall off with increasing energy, whereas, e.g., the anti-deuteron flux from cold dark matter has a flat spectrum.
No significant limits on the fluxes of anti-nuclei have been placed to date. The planned
General Antiparticle Spectrometer experiment\footnote{\url{http://gamma0.astro.ucla.edu/gaps/}} (GAPS) \cite{Mognet:2013dxa} promises to provide constraints at kinetic energies up to 0.3 GeV per nucleon, covering anti-protons as well, and the energy range around 1 GeV per nucleon should be accessible with AMS-02.

In the right panel of Fig.~\ref{fig:electrons}, we have already displayed the total electron-plus-positron spectrum observed with Fermi LAT~\cite{Abdo:2009zk}. Measurements with 
H.E.S.S. indicate a cut-off or spectral break at about $1\,$TeV~\cite{Aharonian:2009ah}.
Between $\sim100\,$GeV and $\sim1000\,$GeV, an excess is visible compared to that expected for primary electrons from sources with uniform power-law spectra and homogeneous spatial distribution. Both assumptions are not well justified. In fact, the observed spectrum can
be well explained by combining contributions from individual supernova remnants and pulsars. Future experiments, that will measure the combined electron and positron flux up to several TeV, include the balloon-based Cosmic Ray Electron Synchrotron
Telescope (CREST)~\cite{Coutu:2011zz} and the CALorimetric Electron Telescope
(CALET)~\cite{Adriani:2015wga} on the International Space Station.

Model predictions of the electron and positron fluxes are obtained
by solving a diffusion-energy loss equation that can be derived from a Fokker-Planck description of particle transport. Among the input parameters are the diffusion coefficient and the relevant distributions of both astrophysical sources and sites of dark-matter annihilation or decay.
Numerical solutions can be obtained with dedicated software packages
such as GalProp\footnote{\url{http://galprop.stanford.edu/}} and DRAGON\footnote{\url{http://www.desy.de/~maccione/DRAGON/index.html}}, whereas analytical methods typically involve integrating the Green's function for the problem.

The anisotropy of the galactic electrons and positrons can also serve as a probe for the origin of the excess positrons. The positron source rate from dark matter is expected to be rather smoothly distributed, unless strong boosting is dominated by high-mass dark-matter clumps, and thus a relatively small anisotropy is expected~\cite{Borriello:2010qh}. More detailed investigations
predict an upper limit on the dipolar anisotropy given by
\begin{equation}
  |\hbox{\boldmath$\delta$}(E_e)|\lesssim3\times10^{-3}\,\left(\frac{E_e}{100\,{\rm GeV}}\right)\quad
  \mbox{for $30\,{\rm GeV}\lesssim E_e\lesssim2\times10^3\,{\rm GeV}$}\,.\label{eq:e_p_dm_aniso}
\end{equation}
This is consistent with current upper limits which are at the level of a few percent in this energy range~\cite{e_aniso_fermi}. The electron anisotropy for conventional sources such as SNRs is highly dependent on the actual location and activity period of the sources, but on average larger than that quoted in Eq.~\ref{eq:e_p_dm_aniso} by a factor of a few~\cite{2013ApJ...766....4P}. Future detectors
will have sufficient sensitivity to detect anisotropies of that order.

Distinguishing dark-matter contributions from ordinary astrophysical signatures in the spectra of cosmic rays, antimatter, radiation fields from the radio to the $\gamma$-ray band,
and also neutrinos requires a multi-messenger approach that includes accurate modeling of the acceleration and propagation of cosmic rays, as well as their production of secondary $\gamma$-rays and neutrinos.

\begin{figure}[ht]
\centering
\includegraphics[width=0.7\textwidth,clip=true,angle=0]{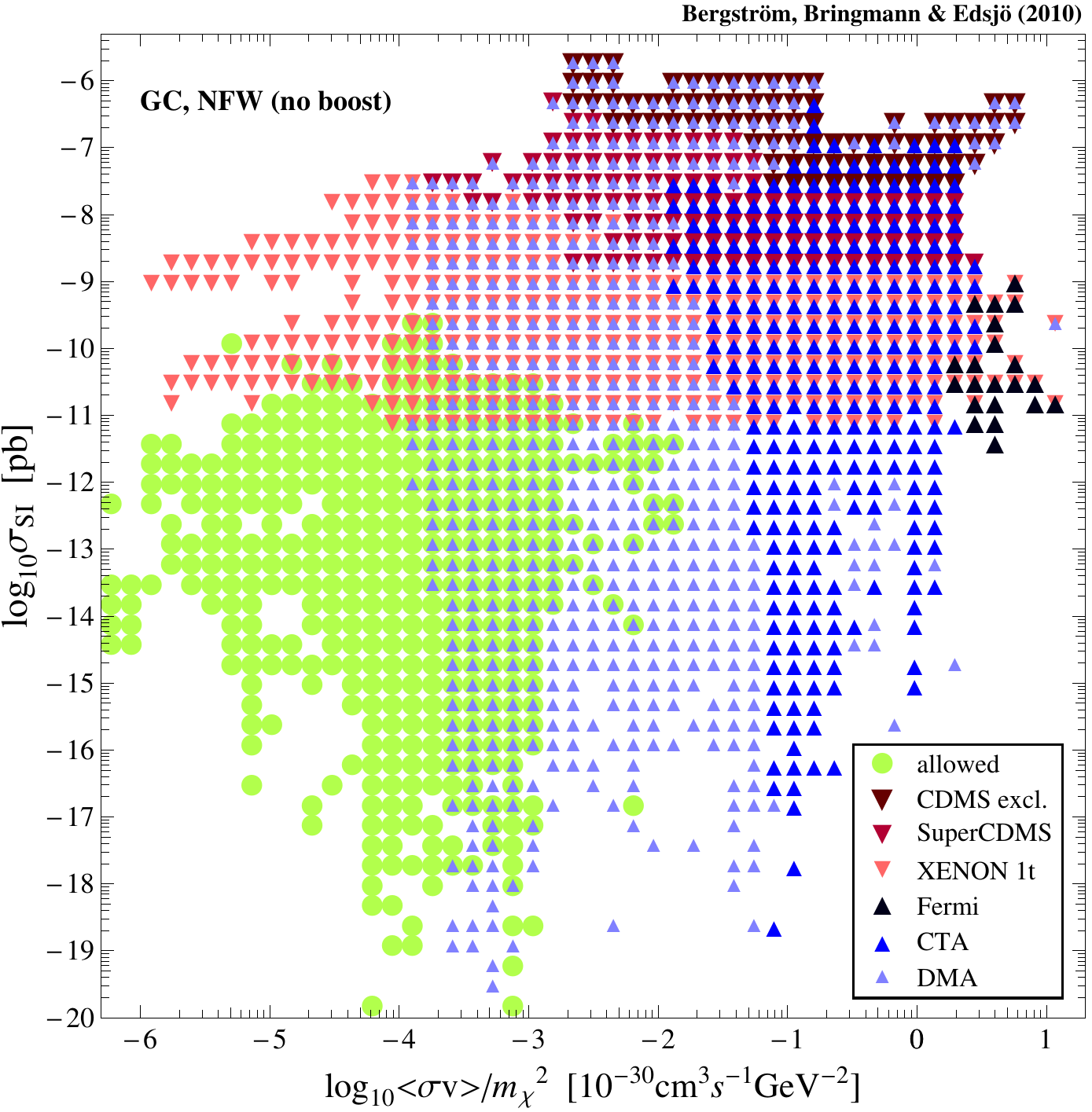}
\caption{Experimental coverage of the cross-section plane of spin-independent WIMP-nucleon scattering and WIMP self-annihilation. As tracer of the latter we plot on the $x-$axis
$\left\langle\sigma_{X\bar X}v\right\rangle/m_X^2$ (cf. Eq.~\ref{eq:dm_flux}) and restrict to
$\gamma$-ray observations of the 
galactic center. The density of dark matter profile is assumed to follow an NFW profile without
additional boosting. All cross section combinations permitted in typical MSSM and mSUGRA models are shown. DMA
denotes a possible dedicated dark matter array. From Ref.~\cite{Bergstrom:2010gh}.}
\label{fig:dm_complementarity}
\end{figure}

To be noted is the complementarity of direct and indirect dark-matter searches. In Fig.~\ref{fig:dm_complementarity} we indicate the reach of current and future experiments in constraining MSSM and
mSUGRA dark-matter models through measurements of spin-independent WIMP-nucleon scattering and
WIMP annihilation, the latter on the example of $\gamma$-rays from the galactic center. The top half of the plot refers to large
scattering cross sections addressed by the direct-detection experiments, whereas indirect detection probes the right half of the plot, where annihilation cross sections are large.
It is interesting to note that Fermi LAT in combination with CTA will probe WIMP annihilation into $\gamma$ rays down to very small values of the annihilation cross section~\cite{Funk:2013gxa}, which includes the region of small scattering cross section, and so CTA may reach deeper than the largest future direct-detection experiments. The mass range $m_X\lesssim10\,$GeV
will be difficult to probe both directly, on account of the small nuclear recoil, and indirectly, because low-energy secondary
$\gamma$-rays and neutrinos would fall below the energy threshold of ground-based detectors. Such low-mass WIMPs are easier to probe at accelerator experiments, as we shall discuss in Sect.~\ref{sec:5}.

\subsection{WIMP capture and neutrinos}
\label{sec:4.5}
WIMP dark matter
can be trapped in the gravitational well of celestial bodies such as stars and planets, resulting in an enhanced dark-matter
density near the center of these objects. An inevitable consequence is a large
annihilation rate that depends on the WIMP-proton scattering cross section, as the following considerations illustrate.

The rate of change of the number of WIMPs, $N_X$, is given by
\begin{equation}
\dot N_X=\Gamma_c-\Gamma_a=\Gamma_c-\frac{\langle\sigma_{X\bar X}v\rangle}{V_X}\, N_X^2\,,\label{eq:capture_rate}
\end{equation}
where $\Gamma_c$ is the WIMP capture rate, $\Gamma_a\equiv C_a N_X^2$ is the WIMP
annihilation rate in the compact body, and $V_X$ is the effective volume in which most of the WIMPs reside, typically the central core of the object in question. The steady-state  solution to Eq.~\ref{eq:capture_rate} and the relaxation timescale toward the equilibrium situation are
\begin{equation}
N_\mathrm{eq}=\sqrt{\frac{\Gamma_c\,V_X}{\langle\sigma_{X\bar X}v\rangle}}\ ,\qquad\qquad
\tau_\mathrm{eq}=\sqrt{\frac{V_X}{\Gamma_c\,\langle\sigma_{X\bar X}v\rangle}}\ .
\label{eq:C_a2}
\end{equation}
Obviously, knowledge of the
capture rate, $\Gamma_c$, is required for estimating either quantity. A WIMP will be captured if it is scattered to a velocity smaller than the
escape velocity of the astronomical body in question. Typical WIMP velocities in the Milky Way,
$v\simeq300\,{\rm km}\,{\rm s}^{-1}$, are smaller than the escape velocity of the Sun,
$v_\odot\simeq620\,{\rm km}\,{\rm s}^{-1}$, and so capturing WIMPs in the Sun is an efficient process, that is 
dominated by axial (spin-dependent) WIMP scattering on hydrogen, the most common
element in the Sun. The rate can be estimated as~\cite{Jungman:1995df}
\begin{equation}
\Gamma_{c,\odot}\simeq(10^{23}\,{\rm s}^{-1})\,\left[\frac{\rho_X}{0.3\,{\rm GeV}\,{\rm cm}^{-3}}\right]
\left[\frac{100\,{\rm GeV}}{m_X}\right]\left[\frac{\sigma^{\rm SD}_{p}}{10^{-40}\,{\rm cm}^2}\right]
\left[\frac{270\,{\rm km}\,{\rm s}^{-1}}{v}\right]\,,\label{eq:capture_rate2}
\end{equation}
where $\sigma^{\rm SD}_{p}$ is the spin-dependent WIMP-proton scattering cross section and
the typical WIMP velocity $v$ is comparable to the
escape velocity of the galaxy. The escape velocity of the Earth, $v_\oplus\simeq11\,{\rm km}\,{\rm s}^{-1}$,
is much smaller than the typical WIMP velocity, and so capture in the Earth has a very low efficiency. 

Using the parameter scales given in Eq.~\ref{eq:capture_rate2} and further
$$\langle\sigma_{X\bar X}v\rangle\simeq 10^{-26}\,\mathrm{cm^3\,s^{-1}}\qquad
\mathrm{and}\qquad V_X\approx 10^{27}\,\mathrm{cm^3}\ ,$$
the relaxation timescale would be
\begin{equation}
\tau_\mathrm{eq}\approx 10^{15}\,\mathrm{s}\simeq 3\cdot 10^7\,\mathrm{yr}
\,. \label{eq:capture_rate3}
\end{equation}
For a wide range of WIMP masses and cross sections one would therefore expect a dark-matter density in the core of the Sun that is close to the steady-state value, whose ratio to the local dark-matter density, $\rho_X/m_X$, can be written as
\begin{equation}
\frac{N_\mathrm{eq}\,m_X}{V_X\,\rho_X}\approx 10^{12}\,.\label{eq:capture_rate4}
\end{equation}
In the steady state, the total rate of annihilation events by definition equals the capture rate. Therefore, in this limit any secondary particle flux resulting from WIMP annihilation
in the star depends on the same cross section that is probed by direct-detection experiments. In that respect, indirect detection by capture in the Sun is complementary to direct dark-matter detection.

All final-state particles except neutrinos would be thermalized inside the star. The
neutrinos would escape and can thus potentially be detected with large 
high-energy neutrino detectors such as IceCube. Of particular interest is the mass range $m_X>m_{W^\pm}$, for which
the annihilation channel $X+X\to W^++W^-$ is open and the subsequent $W^\pm$ decays provide a peak in the neutrino spectrum near the energy $ m_X/2$. IceCube
and its DeepCore subarray have placed severe constraints on the cross section of both spin-independent and
spin-dependent scattering~\cite{Aartsen:2012kia}. Whereas for the spin-independent WIMP-nucleon cross section the IceCube limit at WIMP masses of a few hundred GeV and is not competitive with
that of XENON100 limits, the spin-dependent cross section is constrained to
$\sigma^{\rm SD}_N\lesssim 10^{-40}\,{\rm cm}^2$ at WIMP masses of a few hundred GeV, which is the best limit published to date
for $m_X\gtrsim35\,$GeV. It is also of the same order as cross sections predicted within the MSSM, so that the IceCube measurements permit to test that scenario. 

The Super-Kamiokande experiment has a much lower energy threshold compared to IceCube, which permits addressing the mass range $5\,{\rm GeV}\lesssim m_X\lesssim50\,$GeV. Preliminary upper limits on $\sigma^{\rm SI}_N$ are on the order of
$10^{-39}\,{\rm cm}^2$ and $10^{-40}\,{\rm cm}^2$ for annihilation into $b\bar b$ and $\tau\bar\tau$, respectively.
The future Hyper-Kamiokande and IceCube-PINGU experiments promise to reach sensitivities down to
$\sigma^{\rm SD}_N\simeq10^{-41}\,{\rm cm}^2$ between 10 GeV and a few hundred GeV.

\subsection{Astrophysical and cosmological signatures of WISPs}
\label{sec:4.6}
In this section we consider signatures of WISPs in astrophysical and cosmological observations that arise from the coupling of
WISPs to photons. The physical concepts behind WISPs were described in Sect.~\ref{sec:2.2}.
The
most well-known examples of WISPs are pseudo-scalar axions, ALPs, and hidden photons. 
In the presence of electromagnetic fields, in
particular of magnetic fields, photons can
oscillate into axions or ALPs and vice-versa, an effect known as the
Primakoff effect~\cite{Pirmakoff:1951pj}. 

If the wavelength of the photon and the
ALP are much smaller than the length scales on which the properties of the medium
and the magnetic field vary, the coupling of photons and ALPs
is mathematically 
analogous to that of neutrino oscillations. The length scale of oscillations is typically much larger than those accessible with laboratory experiments, but orders of magnitude smaller than astrophysical and cosmological scales. Astronomical signals travel a long distance, $L$, and they thus correspond to an average over many oscillations. Instead of an oscillatory variation of photon number $\propto \sin^2 (k_\mathrm{osc}L/2)$, we would expect a constant reduction to a level given by the effective mixing angle in the dilute plasma between galaxies. 

The mixing of photons with hidden photons can occur in vacuum, but might be
modified in the presence of
a plasma that gives the photons an effective mass, whereas the WISP mass is
essentially unchanged. This can give rise to matter oscillations reminiscent
of the Mikheyev-Smirnov-Wolfenstein (MSW) effect for neutrino 
oscillations~\cite{Wolfenstein:1977ue,Mikheev:1986gs}. In particular, resonant conversion of
photons into WISPs can occur within a plasma, i.e. both within astrophysical sources and during propagation
of photons from the source to the observer.

Once produced inside an astrophysical source, the
weak coupling of WISPs to ordinary matter permits their escape without significant reabsorption, opening an additional channel of radiative
cooling. The few neutrinos observed from the
cooling phase of SN1987A have been used to place limits on the mass and coupling of axions by requiring that core-collapse supernovae not
cool significantly faster than by neutrino
emission alone~\cite{Keil:1996ju}.

Probing photon-WISP
oscillations requires a detailed understanding of the photon-emission
process. One of the best-understood
radiation sources in the universe is the CMB. Apart from dipole distortion, its spectrum deviates from a perfect blackbody by less than about
$10^{-4}$, and the amplitude of anisotropy
is on the order of $10^{-5}$. The photons of the CMB have travelled a distance
commensurate with the present-day Hubble radius. Any photon-WISP mixing at a level of
$10^{-4}$ would induce a spectral distortion or an anisotropy in
conflict with CMB measurements. Since photon-ALP mixing
requires the presence of a magnetic field, the absence of a significant
modification of the CMB imposes an upper limit on the combination of the root mean square of the large-scale extragalactic
magnetic-field strength and a parameter of the ALP model. 

In contrast to
the case of ALPs, oscillations between photons and hidden photons do not require a magnetic field. The absence of CMB distortions thus directly constrains 
the vacuum mixing
angle to be $\chi\lesssim 10^{-7}-10^{-5}$ for
hidden-photon masses of $10^{-14}\,{\rm eV}\lesssim
m_{\gamma^\prime}\,c^2\lesssim 10^{-7}\,$eV~\cite{Mirizzi:2009iz}. 

The physics of non-thermal astrophysical sources is not well understood, resulting in a substantial uncertainty concerning the flux and spectrum of X-rays
and $\gamma$-rays. The observed photon spectra from these objects are often well represented 
by power laws, suggesting that the photons emissivities inside the sources are power laws as well. Then, photon-ALP mixing can induce
steps in the spectra that may be detectable, and the absence of such spectral features can be used to place limits on ALP models. One of the free parameters of the
theory is the Peccei-Quinn scale or axion decay constant $f_a$~\cite{2008LNP...741....3P}. Spectral modifications on the keV to TeV scale are expected for ALP masses of
$m_a\sim10^{-6}\,$eV and $f_a\lesssim 10^{13}\,$GeV~\cite{Hooper:2007bq,Hochmuth:2007hk}, somewhat depending on the magnetic-field strength. Exploiting these 
effects is complementary to, and potentially more sensitive than, other
experimental avenues. 

The currently best constraints have been derived from solar observations. Photons from the
Sun would convert to ALPs in the solar magnetic field and then turn back into photons while passing through the field of an artificial magnet in front of a telescope. The
CERN Axion Solar Telescope (CAST) experiment provided the strongest
constraint to date, $f_a\gtrsim 10^{10}\,$GeV for $m_a\lesssim 0.02\,$eV~\cite{Arik:2008mq}.

As photon-ALP mixing is energy-dependent, the resulting modification of the photon flux, or luminosity, will not be the same at different energies. One of the standard tools of astronomy is comparing the luminosity in different energy bands, which often reveals correlations that are specific for a given class of objects. Photon-ALP mixing will impact the scatter of such correlations and induce non-Gaussianity in the distribution, an example of which was found in AGN data of optical/UV and X-ray emission~\cite{Burrage:2009mj}. Besides the obvious point that conventional astrophysics may lead to a non-Gaussian scatter distribution~\cite{Pettinari:2010ay}, substantial uncertainty in the interpretation arises from the question where exactly the conversion to ALPs occurs. If that happens inside the AGN, one would deduce $m_a\ll10^{-7}\,$eV and
$f_a\simeq3\times10^8\,$GeV. Most AGNs reside in galaxy clusters, and it is easy to see that 
photon-ALP conversion may also proceed in the $\mu$G-level magnetic field in the intracluster medium. In that case the appropriate parameters would be $m_a\ll10^{-12}\,$eV and
$f_a\lesssim 10^{10}\,$GeV. 

Another possible signature for photon-ALP mixing is a modification of intergalactic $\gamma\gamma$ absorption~\cite{SanchezConde:2009wu,Meyer:2013pny,2014ApJ...785L..16A}. Observations of AGNs at various redshifts suggest that the absorption of TeV-band $\gamma$-rays by pair production with the extragalactic infrared background~\cite{Aharonian:2005gh} is weaker than estimated. One possible explanation is the conversion of $\gamma$-rays into ALPs at the source and reconversion into photons in the galactic magnetic
field. Alternatively, conversion and reconversion may be caused by the much weaker extragalactic magnetic field, in which case one would deduce a completely different viable range for the parameters $m_a$ and $f_a$. Besides, conventional astrophysics can offer explanations for weak $\gamma$-ray absorption as well, for example if the sources actually produce cosmic rays of very high energy that along the entire line of sight produce photons through pion production in $p\gamma$ collisions~\cite{Essey:2009ju,Essey:2010er,Aharonian:2012fu}. A requirement is a weak deflection of cosmic rays, implying a weak extragalactic magnetic field. This interpretation would be challenged if variability in the multi-TeV $\gamma$-ray flux from distant AGN was observed.

%% file: section5.tex
\section{LHC searches}
\label{sec:5}

Due to their lack of electromagnetic interactions, non-baryonic nature and therefore
(at most) weak Standard-Model interactions, dark-matter particles escape detection
at colliders in a similar way as neutrinos. As a consequence, they produce a
characteristic signal of missing energy. At hadron colliders such as the Large
Hadron Collider (LHC) at CERN, only the missing transverse energy, $\not{\!\!\!E}_T$,
can be observed, since the longitudinal momentum fractions of the incoming partons
in the colliding hadrons are unknown. The missing $E_T$ is then determined
from the recoiling observed objects such as jets, heavy quarks, photons and leptons.

\subsection{Model-dependent searches}
\label{sec:5.1}
Assuming a particular dark-matter particle and its interactions with Standard-Model particles, it is possible
to perform complete signal and background Monte Carlo simulations. Specific kinematic cuts can then be defined that largely eliminate the Standard-Model background and thus permit
confronting the model expectations with experimental data. An excess in the
kinematic distributions may then be attributed to the dark-matter particle in question. The background
contribution can alternatively be determined directly from the data
by extrapolating from a control region to the signal region. This procedure would, of course, not eliminate irreducible backgrounds from, e.g., neutrinos.

If the dark-matter particle is the lightest and only stable member of a family of new states, it
would be the only one having survived the evolution of the universe. In that case, the typical LHC
signal would consist of cascade decays of the heavier new states, thereby producing
multi-parton or multi-lepton final states together with $\not{\!\!E}_T$.
If the new spectrum includes strongly interacting particles, the corresponding
hadronic cross sections are high, while the Standard-Model background rates decrease
with the particle multiplicity in the final state. For this reason, searches for dark matter in
cascade decays were the first to be performed at the LHC.

An important example is the search for the lightest supersymmetric particle (LSP), typically
a neutralino in $R$-parity conserving supersymmetry, in the pair production and subsequent
cascade decay of squarks and gluinos to jets and two LSPs. Assuming a constrained
Minimal Supersymmetric Standard Model (cMSSM), it is then possible to exclude regions in the parameter
space of $m_0$ and $m_{1/2}$, the common scalar and fermion cMSSM mass
parameters that in general also determine the LSP mass \cite{Aad:2015mia,Chatrchyan:2014goa}.

An alternative is to search for direct production of dark-matter particles, which is observable only if it involves at least one visible
particle in so-called mono-object events. This object can be either a jet, a photon, or an
electroweak gauge or Higgs boson, and it is radiated either from the initial state or
from an intermediate particle. The dark-matter signal then directly appears as an excess in
the tail of its $E_T$ distribution.
A typical example is the associated production of a jet with a pair of gravitinos,
which are often the LSPs in gauge-mediated supersymmetry-breaking models \cite{Klasen:2006kb}.

\subsection{Effective field theory approach}
\label{sec:5.2}

A model-independent way to search for dark matter at the LHC is to analyze mono-object
events within an effective field theory (EFT). Here, the coupling of dark matter
to Standard-Model particles is parameterized with a set of non-renormalizable, higher-dimensional
operators. The operators containing two dark-matter (fermionic or scalar) particles can be
categorized and are shown in Tab.\ \ref{tab:eft} \cite{Goodman:2010ku}.
\begin{table}
 \caption{Operators coupling two dark-matter particles $X$ to Standard-Model quarks ($q$) and gluons ($g)$.
 The operator names beginning with D, C, and R  apply to Dirac fermions, complex and real scalars, respectively.}\vspace*{2mm}
 \label{tab:eft}
\begin{tabular}{|c|c|c|}
\hline
   Name    & Operator & Coefficient  \\
\hline
D1 & $\bar{X}X\bar{q} q$ & $m_q/M_*^3$   \\
D2 & $\bar{X}\gamma^5X\bar{q} q$ & $im_q/M_*^3$    \\
D3 & $\bar{X}X\bar{q}\gamma^5 q$ & $im_q/M_*^3$    \\
D4 & $\bar{X}\gamma^5X\bar{q}\gamma^5 q$ & $m_q/M_*^3$   \\
D5 & $\bar{X}\gamma^{\mu}X\bar{q}\gamma_{\mu} q$ & $1/M_*^2$   \\
D6 & $\bar{X}\gamma^{\mu}\gamma^5X\bar{q}\gamma_{\mu} q$ & $1/M_*^2$    \\
D7 & $\bar{X}\gamma^{\mu}X\bar{q}\gamma_{\mu}\gamma^5 q$ & $1/M_*^2$   \\
D8 & $\bar{X}\gamma^{\mu}\gamma^5X\bar{q}\gamma_{\mu}\gamma^5 q$ & $1/M_*^2$   \\
D9 & $\bar{X}\sigma^{\mu\nu}X\bar{q}\sigma_{\mu\nu} q$ & $1/M_*^2$   \\
D10 & $\bar{X}\sigma_{\mu\nu}\gamma^5X\bar{q}\sigma_{\alpha\beta}q$ & $i/M_*^2$  \\
D11 & $\bar{X}X G_{\mu\nu}G^{\mu\nu}$ & $\alpha_s/4M_*^3$   \\
D12 & $\bar{X}\gamma^5X G_{\mu\nu}G^{\mu\nu}$ & $i\alpha_s/4M_*^3$   \\
D13 & $\bar{X}X G_{\mu\nu}\tilde{G}^{\mu\nu}$ & $i\alpha_s/4M_*^3$   \\
D14 & $\bar{X}\gamma^5X G_{\mu\nu}\tilde{G}^{\mu\nu}$  & $\alpha_s/4M_*^3$ \\
\hline
\end{tabular}
\begin{tabular}{|c|c|c|}
\hline
   Name    & Operator & Coefficient   \\
\hline
C1 & $X^\dagger X\bar{q}q$ & $m_q/M_*^2$    \\
C2 & $X^\dagger X\bar{q}\gamma^5 q$ & $im_q/M_*^2$   \\
C3 &  $X^\dagger\partial_\mu X\bar{q}\gamma^\mu q$ & $1/M_*^2$   \\
C4 &  $X^\dagger\partial_\mu X\bar{q}\gamma^\mu\gamma^5q$ & $1/M_*^2$    \\
C5 & $X^\dagger X G_{\mu\nu}G^{\mu\nu}$  & $\alpha_s/4M_*^2$   \\
C6 & $X^\dagger X G_{\mu\nu}\tilde{G}^{\mu\nu}$  & $i\alpha_s/4M_*^2$   \\ \hline
R1 & $X^2\bar{q}q$ & $m_q/2M_*^2$    \\
R2 & $X^2\bar{q}\gamma^5 q$ & $im_q/2M_*^2$   \\
R3 & $X^2 G_{\mu\nu}G^{\mu\nu}$  & $\alpha_s/8M_*^2$   \\
R4 & $X^2 G_{\mu\nu}\tilde{G}^{\mu\nu}$  & $i\alpha_s/8M_*^2$   \\
\hline
\end{tabular}
\end{table}
As in Fermi's four-fermion weak-interaction model, the effective couplings of
mass dimension $(-2)$ for these operators are obtained by integrating out the propagator
of the mediator with mass $M$ and coupling strength $g$ from $g^2/(p^2-M^2)$ to $g^2/M^2=:
1/M_\ast^2$ at low momentum transfer, $p$. Some of the coefficients in Tab.\ \ref{tab:eft}
are proportional to the quark mass $m_q$ and are thus more accessible in heavy-quark jets.

In contrast to many explicit dark-matter models like supersymmetry, the EFT approach depends on
only very few parameters, i.e.\ the scale of the interaction, $M_\ast$, and the dark-matter
mass, $m_X$.
Due to the crossing symmetry of the underlying Feynman diagrams, it can also serve as a
common standard for the comparison of LHC results with those from direct and indirect
searches \cite{Fox:2011pm}. A necessary condition for the applicability of EFT is, however, that $M_\ast$ is
indeed sufficiently large with respect to the momentum transfer
\cite{Buchmueller:2013dya,Busoni:2013lha}. The impact of this condition can be very
different in direct detection and at the LHC. Second, if the mediator is light enough to be produced
at the LHC, the collider analysis should be modified. Conversely, if the
dark-matter particle in question constitutes
only a fraction of the local dark-matter density, the direct-detection limits change. Finally, the information
obtained with EFT is of course less complete than that of a dedicated analysis.

\subsection{Experimental results from monoobject searches}
\label{sec:5.3}

\begin{figure}[t]
 \includegraphics[width=0.5\textwidth]{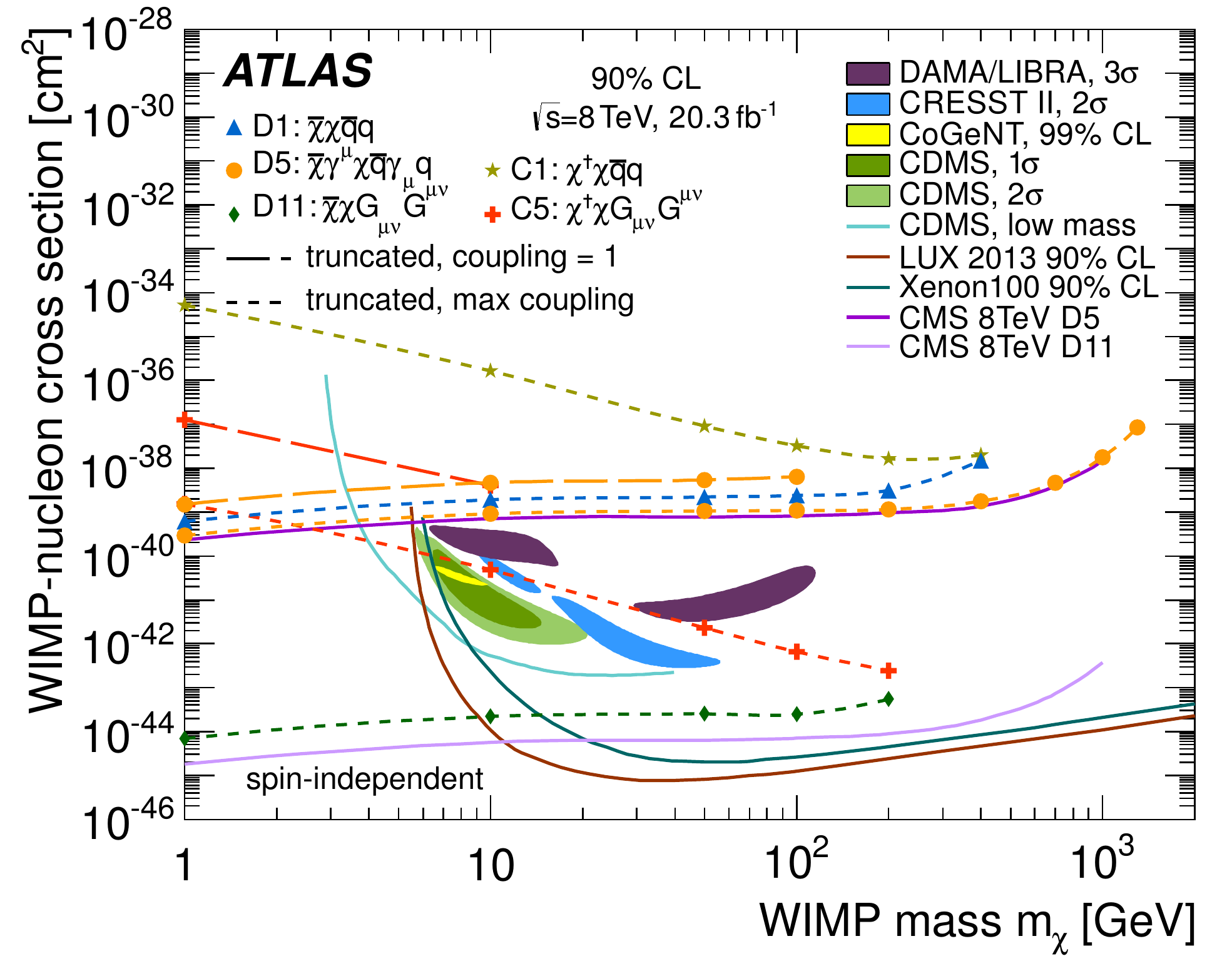}
 \includegraphics[width=0.5\textwidth]{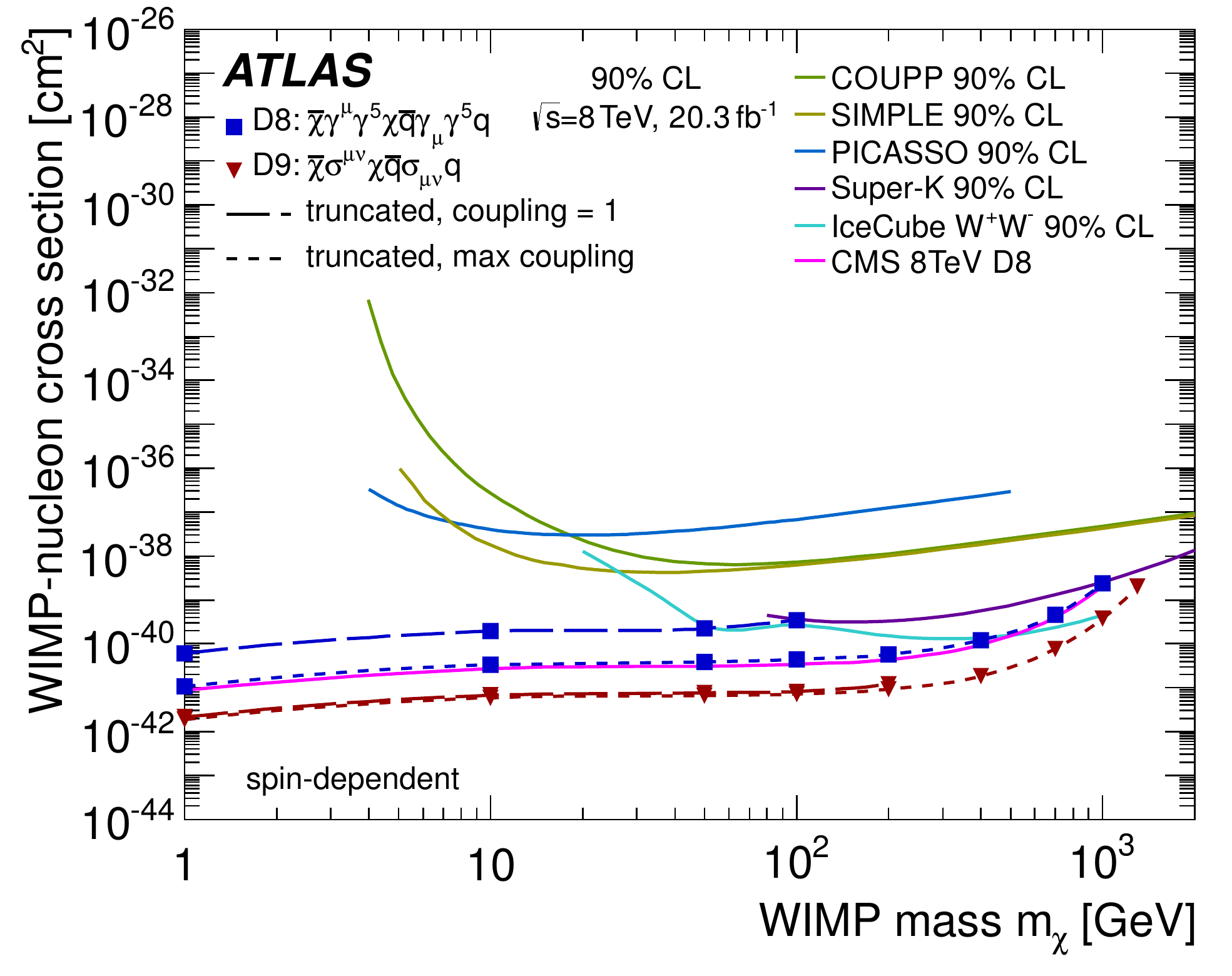}
 \caption{90\%-confidence upper limits obtained from a recent ATLAS monojet analysis
 at a center-of-mass energy of 8 TeV on the spin-independent (left) and spin-dependent (right)
 dark-matter/nucleon scattering cross section as a function of dark-matter mass $m_X$ for
 different operators. Results from direct-detection experiments for
 the spin-independent
 and spin-dependent
 cross section and the CMS results are shown
 for comparison \cite{Aad:2015zva}. 
  %
}
 \label{fig:atlassi2015}
\end{figure}

Recent monojet analyses by ATLAS \cite{Aad:2015zva} and CMS \cite{Khachatryan:2014rra}
demonstrate that the LHC is quite competitive with direct dark-matter searches, in particular for low masses.
An overview of results is presented in Fig.\ \ref{fig:atlassi2015}, focussing in particular on the operators
D1, D5, D11, C1 and C5 for the spin-independent case
and the models D8 and D9 for the spin-dependent case  \cite{Aad:2015zva}.

\begin{figure}[t]
 \includegraphics[width=0.5\textwidth]{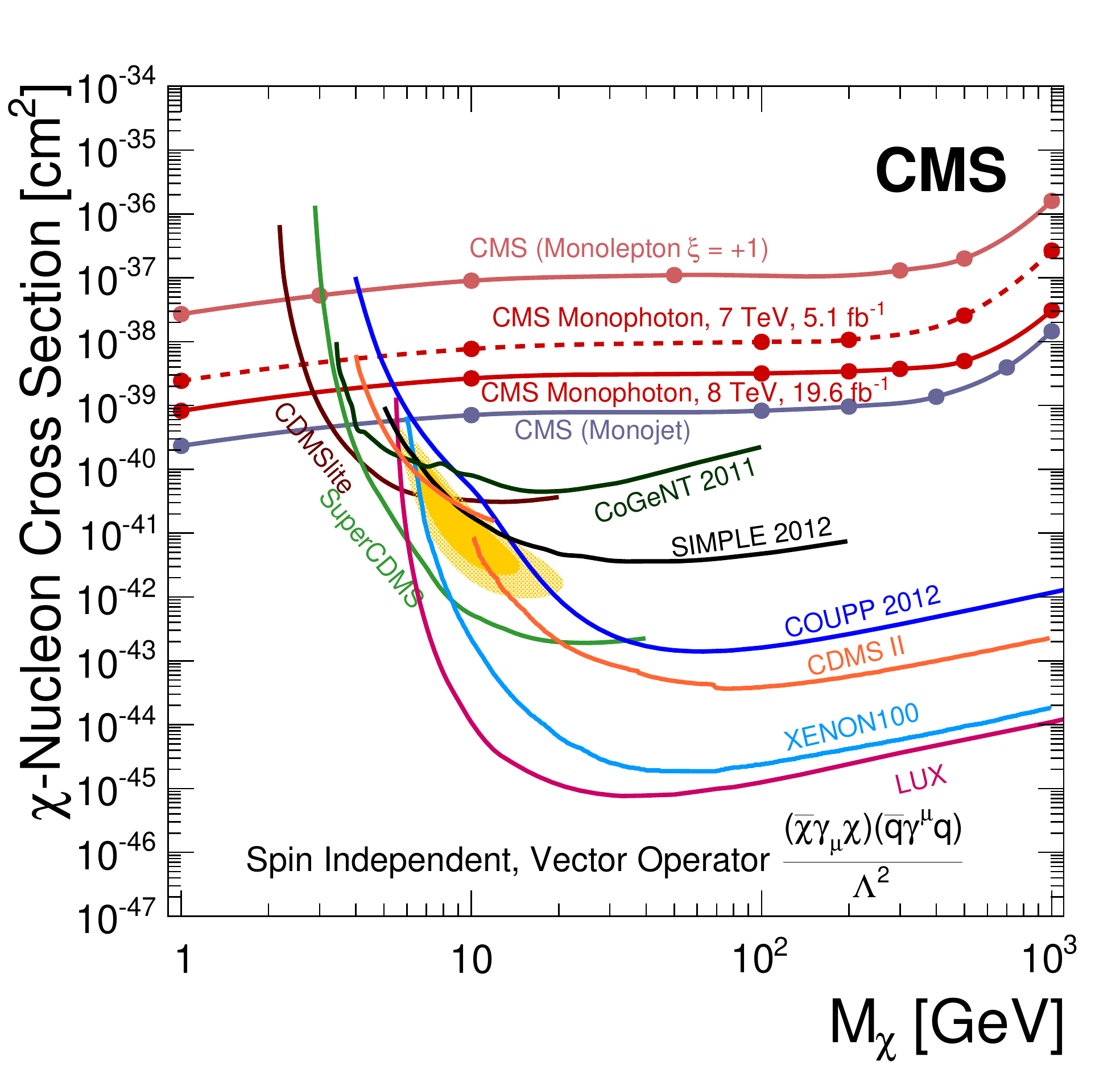}
 \includegraphics[width=0.5\textwidth]{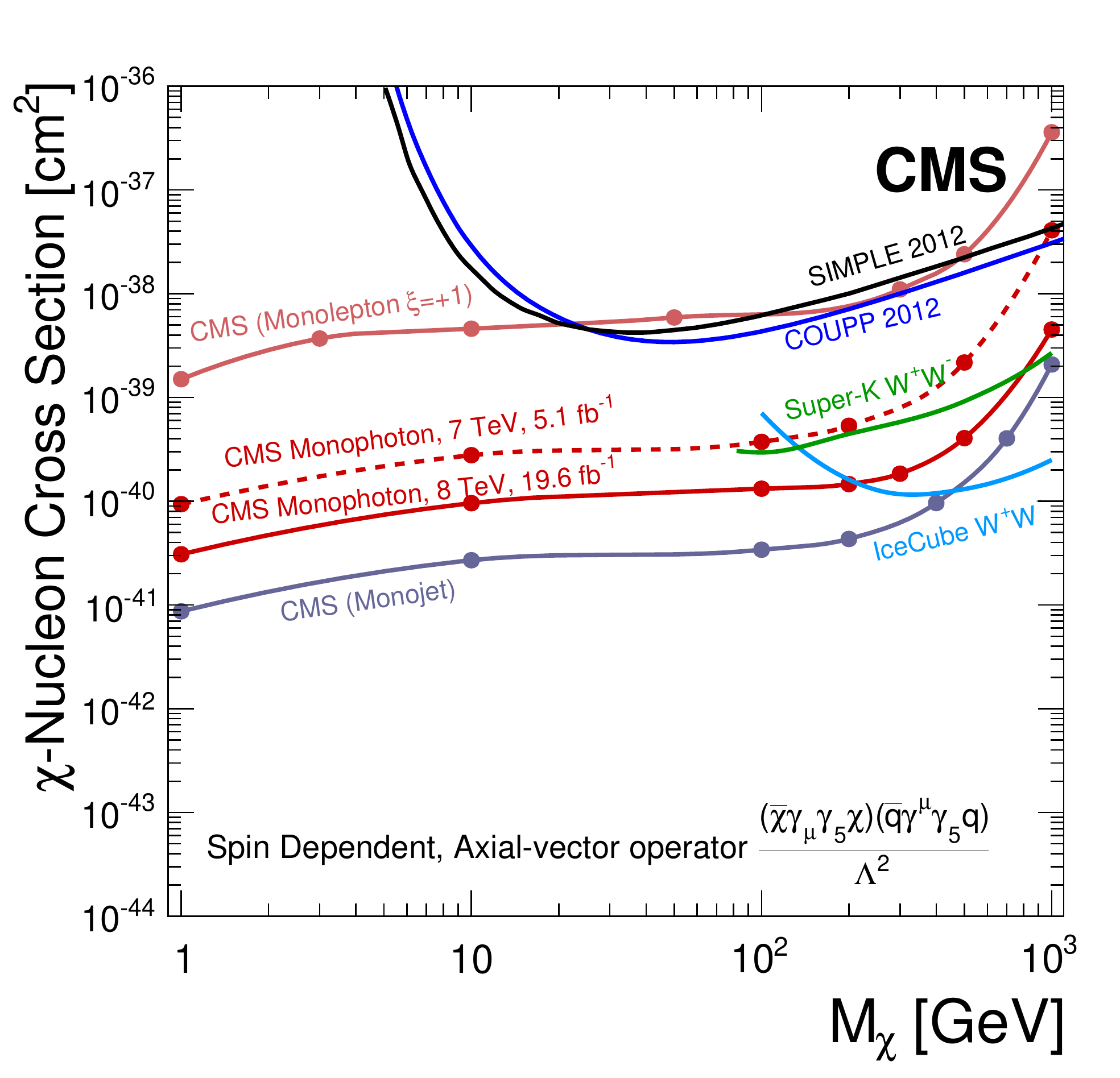}
 \caption{90\%-confidence upper limits  obtained from a recent CMS monophoton analysis
 at center-of-mass energies of 7 and 8 TeV 
 on the $X$-nucleon cross section as a function of the
 dark-matter particle mass, $M_{X}$, for spin-independent (left) and spin-dependent
 couplings (right)  \cite{Khachatryan:2014rwa}. Also shown are limits from CMS using monojet
  and
 monolepton
  signatures, including results for maximized interference parameter, $\xi= \pm 1$, as well as several published limits from direct-detection experiments.
  The solid/hatched yellow regions
 show the 68\% and 95\%-confidence contours respectively for a possible signal in
 CDMS.}
 \label{fig:cmssi2014}
\end{figure}

The monophoton analyses of ATLAS \cite{Aad:2014tda} and CMS \cite{Khachatryan:2014rwa}
are somewhat less competitive than the monojet and also the monolepton results, as Fig.~\ref{fig:cmssi2014} shows.
The monolepton channel may furthermore involve different couplings of the dark-matter particle
to up- and down-type quarks, and
so the ensuing interference effects have been parameterized in the couplings with an
additional factor $\xi=\pm1$.

In the future, measurements of correlations of visible particles, e.g.\ of the
azimuthal angle between two jets, may permit more discriminative cuts, in particular between the
signal and irreducible background coming from invisible
$Z$ decays to two neutrinos \cite{haisch}.
With search strategies adapted to the new pile-up and detector
conditions, the LHC experiments should surpass the previous limits from center-of-mass
energies of $\sqrt{s}=7$ and 8 TeV within one year of
data taking, reaching interaction scales $M_\ast$ of about 2 TeV at $\sqrt{s}=$14 TeV
with an integrated luminosity of about 100 fb$^{-1}$ \cite{doglioni}.

\subsection{Simplified models}
\label{sec:5.4}

\begin{figure}[t]
 \includegraphics[width=0.55\textwidth]{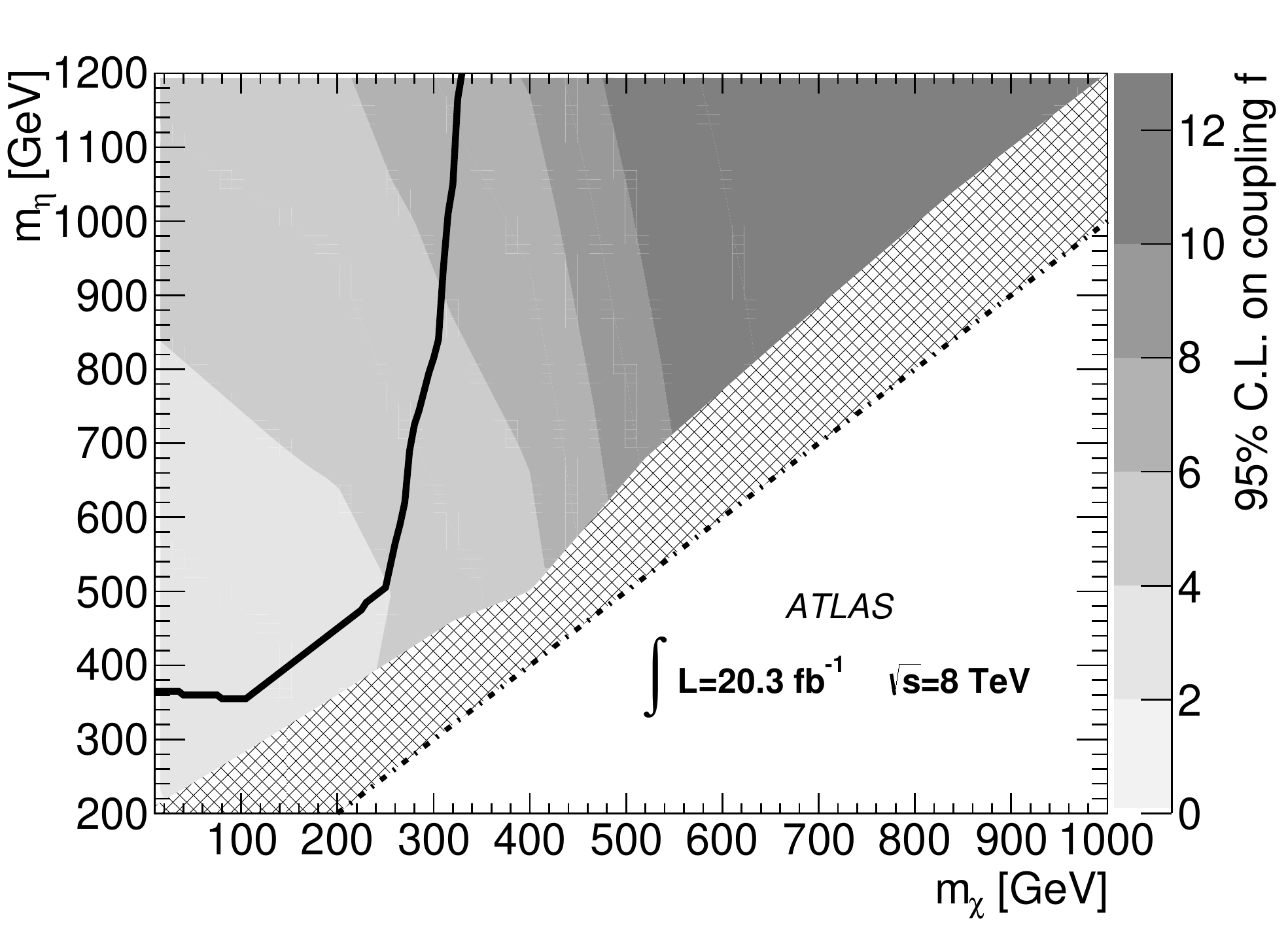}
 \includegraphics[width=0.41\textwidth]{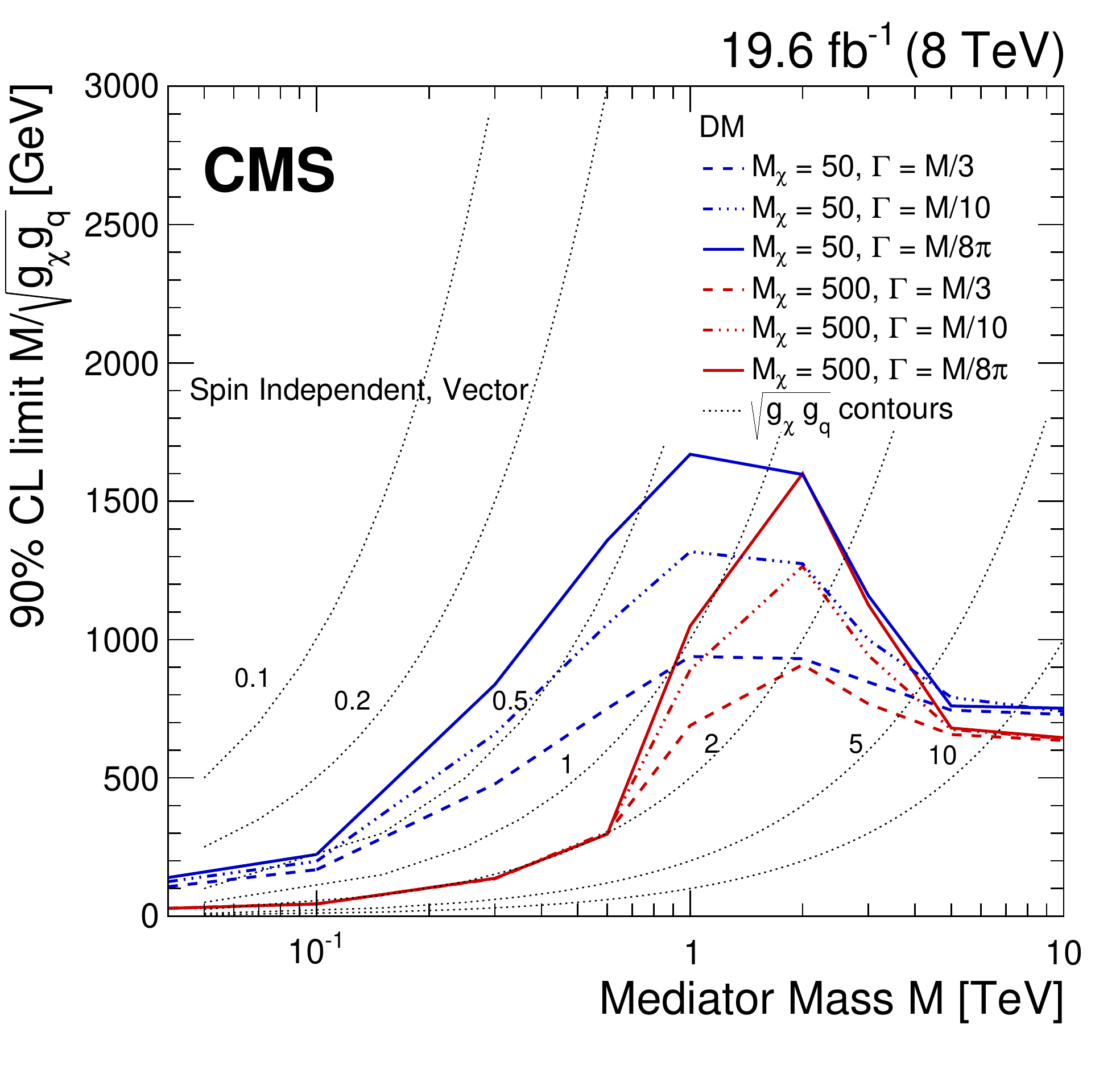}
\caption{Left: 95\%-confidence upper limits obtained from a recent ATLAS mono-$Z$-boson analysis
 at a center-of-mass energy of 8 TeV on the coupling constant, $f$, of the scalar-mediator theory
as a function of $m_{X}$ and the mediator mass, $m_{\eta}$.
The cross-hatching indicates the theoretically accessible region outside the range
covered by this analysis. The white region is phase space beyond the model's validity.
In the excluded region in the upper left-hand corner,
demarcated by the black line, the lower limit on $f$ from the relic-abundance
calculations
is larger than the upper limit measured in this analysis \cite{Aad:2014vka}.
Right: Observed limits obtained from a recent CMS monophoton analysis
 at a center-of-mass energy of 8 TeV on the dark-matter mediator mass divided by coupling, $M/\sqrt{
g_{X} g_{q}}$, as a function of the mediator mass, $M$, assuming vector interactions, for dark-matter particle masses of 50 GeV and 500 GeV. The width, $\Gamma$, of the mediator is varied between $M/8\pi$ and $M/3$. The dashed lines show contours of constant coupling \cite{Khachatryan:2014rwa}.}
 \label{fig:atlassimp2014}
\end{figure}

An intermediate level of interpretation of LHC dark-matter searches are simplified models.
They provide microscopic descriptions of interactions between dark matter and Standard-Model particles in a way that
is still independent of (ultraviolet) complete extensions of the Standard Model. They can be classified into four
basic types of messenger fields: scalar, pseudo-scalar, vector, or axial-vector. In this way, they
are still characterizable by a relatively small number of four to five parameters, including the mediator
mass and width, the dark-matter mass and one or two effective couplings \cite{Harris:2014hga}.

The ATLAS \cite{Aad:2014vka} and CMS \cite{Khachatryan:2014rwa} collaborations have
searched at $\sqrt{s}=8$ TeV for the production of single $Z$ bosons and single photons in association with missing
$E_T$, respectively. $Z$ bosons are identified by their leptonic (i.e.\
oppositely charged electron or muon) decay products. Since no excess above the Standard Model predictions
was observed, limits were first set on the mass scale of the contact interaction, $M_\ast$ (see
above), but also on the coupling and mediator mass in simplified models.

In the ATLAS analysis, only scalar mediators were considered. Figure~\ref{fig:atlassimp2014} (left) shows as a function of coupling, $f$, the permitted regions
in the parameter plane spanned by the dark-matter mass, $m_X$, and the scalar-mediator mass, $m_\eta$. Also included are limits in the $m_X$-$m_\eta$ plane 
obtained by imposing the observed dark matter relic abundance (thick black line), which in
the analysis was possible for dark matter and mediator masses up to 325~GeV and 1200~GeV, respectively.

In the CMS analysis, only vector mediators were considered. Figure~\ref{fig:atlassimp2014} (right) shows,
as a function of the mediator mass, limits
on the mediator mass, $M$, divided by the geometric average of the coupling to dark matter, $g_X$, and that to quarks, $g_q$. 

For Run II of the LHC, simplified models will play a prominent role. Experimental signatures of the types and characteristics of
mediators, e.g., their couplings to dark matter and Standard-Model particles, will be used as
building blocks for a common model implementation, benchmark definitions, and assessments of
the validity of the EFT approach. For the monojet signature, for example, resonant $s$-channel vector
and non-resonant $t$-channel scalar mediators coupling to quarks will be considered together
with scalar mediators coupling to gluons through heavy (in particular top) quark loops \cite{doglioni}. 

A rationale for this choice are models such as the inert-Higgs doublet \cite{LopezHonorez:2006gr}, which is
motivated by the recent Higgs discovery and the possibility that dark matter primarily couples to Higgs bosons.
The advantage of minimal models lies in their small parameter space and complete testability, not only at the LHC, but also in direct-detection experiments \cite{Klasen:2013btp}.
Signatures including a top quark will also be considered on account of its large coupling to Higgs bosons
and its important role in electroweak symmetry breaking.

%% file: section6.tex
\section{Towards global analyses}
\label{sec:6}

In this article, we have reviewed the observational evidence for the presence of
dark matter in our universe, covering a vast range of astrophysical scales, and possible
candidates from elementary particle physics together with their theoretical
relations to various proposed extensions of the Standard Model. We have then
summarized the three main roads followed in dark-matter searches: direct detection,
based on interactions of the local dark matter in dedicated laboratory experiments,
indirect detection, based on the search for decay and annihilation products of
dark matter on astrophysical and cosmological scales with multiple messengers, and
searches at the LHC. Special emphasis was placed on the complementarity of these
three detection modes. All three have specific advantages in that they can provide
critical information on specific particles that might constitute dark matter.
Finding and measuring the properties of GeV- to TeV-scale new particles at the LHC
is a unique channel for exploring the interactions with Standard-Model particles.
Scattering of dark-matter particles off various materials in direct-detection
experiments can provide insight into the local density and velocity distribution
of dark matter. Indirect detection through astrophysical measurements permits
tracing the distribution of dark matter in the sky.

Finding new particles that constitute dark matter would be a major breakthrough in
physics. As extraordinary new findings require extraordinary evidence, the hurdles
are high. Various unusual experimental results of the last decade have been interpreted
in terms of dark matter, but all of them could also be the result of a misunderstood
background or other effects. The problem is probably most obvious for the indirect
detection in astrophysical messengers, where conventional astrophysical processes can
lead to surprising observational features. It is therefore mandatory that all three
strategies of dark-matter detection be continued. There is complementarity in the
detection modes, but there is also overlap in the sense that they sample similar parts
of the parameter space with independent methods. The best option for the future appears
to be a global approach, in which the three detection modes are jointly pursued and
the parameter space of effective operators, classes of simplified models, and eventually
specific models are tested. Communication between the now often separately operating
research initiatives will be a key for success. Finding and identifying dark matter
requires thus a world-wide concerted effort.

We have also briefly discussed alternatives to dark matter, that may explain part of
the observational indications of excess gravitational action. Although the evidence for
dark matter composed of individual particles is huge, in science one is well advised to
keep an open eye to alternative or additional physical processes that may be at play.

\section*{Acknowledgements}
This work was supported by the Deutsche Forschungsgemeinschaft through the collaborative research centre SFB 676
and by the Helmholtz Alliance for Astroparticle Physics (HAP) funded by the Initiative and Networking Fund of the Helmholtz Association.